\documentclass{aims}
\usepackage{hyperref,amssymb,graphicx,float,subcaption}
\usepackage{amsmath,bm, amsthm}
\usepackage[ruled,vlined]{algorithm2e}
\SetKwComment{Comment}{$\triangleright$\ }{}
\usepackage{color}
\usepackage{geometry}
\usepackage{pdflscape}
\usepackage[pagewise]{lineno}
 
\usepackage{cite}
\hypersetup{urlcolor=blue, citecolor=red}

\def\currentvolume{X}
 \def\currentissue{X}
  \def\currentyear{200X}
   \def\currentmonth{XX}
    \def\ppages{X--XX}

\newtheorem{theorem}{Theorem}[section]

\theoremstyle{definition}
\newtheorem{definition}[theorem]{Definition}
\newtheorem{remark}{Remark}

\newcommand{\etal}{\textit{et al}.}

\newcommand{\norm}[1]{\left\lVert #1 \right\rVert}

\DeclareMathOperator*{\argmin}{arg\,min}
\def\Eqref#1{Eq.~(\ref{#1})}
\makeatletter
\newcommand\xleftrightarrow[2][]{%
  \ext@arrow 9999{\longleftrightarrowfill@}{#1}{#2}}
\newcommand\longleftrightarrowfill@{%
  \arrowfill@\leftarrow\relbar\rightarrow}
\makeatother

\title[Generative Deep Learning Turbulent Flows] 
      {Multi-fidelity Generative Deep Learning Turbulent Flows}

\author[Nicholas Geneva and Nicholas Zabaras]{}


\email{ngeneva@nd.edu}
\email{nzabaras@gmail.com}

\thanks{$^*$ Corresponding author: Nicholas Zabaras}

\begin{document}
\maketitle

\centerline{\scshape Nicholas Geneva}
\medskip
{\footnotesize
    \centerline{Scientific Computing and Artificial Intelligence (SCAI) Laboratory}
   \centerline{University of Notre Dame, 311 Cushing Hall}
   \centerline{Notre Dame, IN 46556, USA}
} 

\medskip

\centerline{\scshape Nicholas Zabaras$^*$}
\medskip
{\footnotesize
    \centerline{Scientific Computing and Artificial Intelligence (SCAI) Laboratory}
   \centerline{University of Notre Dame, 311 Cushing Hall}
   \centerline{Notre Dame, IN 46556, USA}
}

\bigskip

 \centerline{(Communicated by the associate editor name)}

\begin{abstract}
    In computational fluid dynamics, there is an inevitable trade off between accuracy and computational cost.
    In this work, a novel multi-fidelity deep generative model is introduced for the surrogate modeling of high-fidelity turbulent flow fields given the solution of a computationally inexpensive but inaccurate low-fidelity solver. 
    The resulting surrogate is able to generate physically accurate turbulent realizations at a computational cost magnitudes lower than that of a high-fidelity simulation.
    The deep generative model developed is a conditional invertible neural network, built with normalizing flows, with recurrent LSTM connections that allow for stable training of transient systems with high predictive accuracy.
    The model is trained with a variational loss that combines both data-driven and physics-constrained learning.
    This deep generative model is applied to non-trivial high Reynolds number flows governed by the Navier-Stokes equations including turbulent flow over a backwards facing step at different Reynolds numbers and turbulent wake behind an array of bluff bodies.
    For both of these examples, the model is able to generate unique yet physically accurate turbulent fluid flows conditioned on an inexpensive low-fidelity solution.
\end{abstract}

\section{Introduction}
\noindent
The numerical simulation and analysis of turbulent fluid flows is of great importance to many scientific and engineering domains.
Over the past several decades computational fluid dynamics (CFD) has become an integral component of academia and industry.
However high-accuracy fluid simulation remains a computationally demanding task particularly at high Reynolds numbers for which the flow is turbulent.
This has led to a hierarchy of simulation models to predict fluid flow ranging from the fast but typically inaccurate Reynolds-Averaged Navier-Stokes (RANS) to the fully resolved but super-computer demanding direct numerical simulation (DNS)~\cite{pope2001turbulent}.
Large-eddy simulation (LES) has become a work horse method for scientific and industrial analysis since it can achieve both reasonable accuracy and computational requirement.
Often only a section of the entire simulation domain is of interest or requires a greater degree of accuracy.
Such examples include boundary layers, turbulent wakes behind a suspended or wall mounted objects, the interface between two fluids, shock boundaries, etc.
This principle that different physical scales are of interest in different locations of the simulation domain has led to the development of various multiscale/multilevel methods~\cite{sagaut2013multiscale}.
Multiscale methods typically combine simulations at different resolutions to increase the accuracy of the simulation with minimal computational overhead.

Multiscale computational fluid dynamic methods constitute a rich and well developed field that encompasses many different methodologies that approach the multiscale aspect through different philosophies.
Of particular interest are adaptive multilevel methods which focus on resolving  different scales based on the complexity of the fluid flow~\cite{sagaut2013multiscale}.
Such approaches use a hierarchy of grids at various resolutions to resolve particular areas of the simulation domain  at various levels of accuracy.
This includes methods that use self-adaptive meshes in which the discretization of the simulation domain is evolved to meet specific resolution criteria~\cite{mitran2001comparison, terracol2001multilevel, hoffman2006approach}, global hybrid methods such as very large eddy simulation (VLES)~\cite{speziale1997computing} or detached eddy simulation (DES)~\cite{travin2002physical} and zonal methods for which prespecified regions of the flow domain are resolved with higher accuracy  to capture relevant physics~\cite{quemere2002zonal, schluter2004large, terracol2005hybrid}.
We take inspiration from these multiscale models to develop a deep learning model that takes advantage of simulations ran at multiple scales to predict high-fidelity turbulent fluid flow.

Machine learning in CFD, specifically the modeling of the N-S equations, has gained a growing interest in recent years with a wide variety of methods ranging from Kalman filters to deep neural networks.
These applications can be broken down into several major categories including: RANS turbulence modeling, LES sub-scale grid modeling, flow control and direct flow prediction.
Machine learning based turbulence modeling for RANS simulation seeks to approximate the Reynolds-Stress term in the RANS equations at an accuracy that 
is higher than the traditionally used closure models through the incorporation of prior physical knowledge and high-fidelity information~\cite{ling2016reynolds, xiao2016quantifying, wang2017physics, geneva2019quantifying, taghizadeh2020turbulence}.
Similarly, machine learning LES models seek to achieve the same goal of providing a sub-scale grid model that predicts the contribution of neglected turbulent length scales 
at a higher accuracy than the traditional methods~\cite{wang2018investigations, lapeyre2019training, maulik2019subgrid}.
These approaches are both very promising, however still rely on pre-existing physical assumptions, approximations and resolutions which fundamentally limit their predictive capability~\cite{wang2019reynolds}.
Another area of interest has been the use of machine learning models to build a controller to yield a particular fluid response~\cite{rabault2019artificial, bieker2019deep}.

The final category we discuss is direct fluid flow prediction where the machine learning model is used to predict the state variables of the fluid flow directly.
This includes the use of machine learning to approximate fluid flows for graphical simulations~\cite{tompson2017accelerating, wiewel2019latent, kim2019deep}, prediction of steady-state flows~\cite{guo2016convolutional, sun2020surrogate}, 
prediction of oscillating/unsteady flows~\cite{bieker2019deep, raissi2019deep, raissi2019physics,renkun2019novel}, and the super-resolution, compression or reproduction of various fluid systems~\cite{hennigh2017lat, mohan2019compressed, werhahn2019multi,subramaniam2020turbulence}.
While machine learning has become a popular tool to predict the behavior of fluids, we note that the majority of the test cases considered are focused on simple non-turbulent problems.
Many works that predict turbulent flows are largely focused on qualitative results (e.g. computer graphics).
This is expected due to the shear complexity of N-S turbulence that poses a challenging problem for even traditional numerical methods let alone machine learning models.
Given that the vast majority of fluid flows of interest are turbulent in nature, much work is still needed to push the application of machine learning to practical fluid flow problems of engineering concern.

In this work, we accelerate the prediction of high-fidelity turbulent flows given a computationally inexpensive  low-fidelity simulation through generative deep learning.
Although similar ideas have been presented in past literature, the proposed model differs in several respects.
First, we are interested in the prediction of physical turbulent fluid flow governed by the Navier-Stokes equations differing from the simpler inviscid Euler equations used in computer graphics~\cite{tompson2017accelerating, wiewel2019latent, kim2019deep}.
Second, in a similar spirit, we are interested in recovering accurate time-averaged and turbulent statistics as oppose to fluid flows that are just visually pleasing.
Third, in this work the input of our model is an inexpensive low-fidelity simulation that provides a coarse yet fairly inaccurate prediction.
This contrasts to many works in machine learning for turbulent applications where  compressed~\cite{hennigh2017lat,  mohan2019compressed} or sub-sampled~\cite{werhahn2019multi, subramaniam2020turbulence} fields of the high-fidelity target are used as the input.
Some auto-regressive models, such as the deep neural network (DNN) in~\cite{renkun2019novel}, are in fact even more dependent on a high-fidelity simulation which is needed to start the time-series prediction.
The reliance upon a direct/coarsened high-fidelity field as a model input contains much richer and more accurate information than a low-fidelity simulation since it is being sampled from a space for which the physics simulated is significantly more precise.
While this makes the machine learning problem significantly easier, it also results in models that need an expensive high-fidelity simulation to derive an input for making predictions. Thus the applicability of such models remains questionable.
Fourth, we are interested in developing a \textit{surrogate model} that can be used to predict multiple flows with different boundary conditions as opposed to just learning a single flow which is essential for justifying the model's training cost.
Lastly, in contrast to past deterministic approaches~\cite{hennigh2017lat, mohan2019compressed, subramaniam2020turbulence}, our generative model learns the probability distribution of high-fidelity flow fields conditioned on the low-fidelity simulation allowing for predictive probabilistic estimates. 

This paper makes the following novel contributions to the integration of deep learning with CFD: 
(a) A multi-fidelity deep generative model is proposed for the prediction of physical high-fidelity fluid flow from a low-fidelity solution. 
(b) A novel invertible neural network architecture is proposed to model the distribution of possible high-fidelity fluid flow solutions conditioned on the low-fidelity observation. 
(c) A backwards  Kullback–-Leibler (KL) divergence loss is used that allows for physics-constrained and standard data-driven training of the generative model.
(d) The model is deployed and evaluated for surrogate modeling of turbulent flows at different Reynolds numbers and  varying boundary conditions.

The remainder of this paper is structured as follows. In Section~\ref{sec:problem}, the problem of multi-fidelity generative modeling turbulent fluid flows is introduced and discussed. 
In Section~\ref{sec:tm-glow}, the generative invertible neural network architecture is introduced with details of each component of the model. 
Following in Section~\ref{sec:tm-glow-training}, the variational training of the generative model is outlined as well as the tuning of the model's hyper-parameters.
The first numerical example, in Section~\ref{sec:backwards-step}, investigates the surrogate modeling of turbulent flow over a backwards facing step at different Reynolds numbers.
The second numerical example, in Section~\ref{sec:cylinder-array}, focuses on the prediction of turbulent wake behind an array of bluff bodies in varying locations.
In Section~\ref{sec:computation}, the computational cost of both training and testing the proposed deep learning model is discussed.
Lastly, conclusions and discussion are provided in Section~\ref{sec:conclusion}.
All code, trained models and data used in this work is open-sourced for full reproducibility.\footnote{Code available at: \href{https://github.com/zabaras/deep-turbulence}{https://github.com/zabaras/deep-turbulence}.}

\section{Multi-fidelity Generative Modeling Fluid Flows}
\label{sec:problem}
\begin{figure}[H]
    \centering
    \begin{subfigure}{0.3\textwidth}
        \centering
        \includegraphics[height=2.5cm]{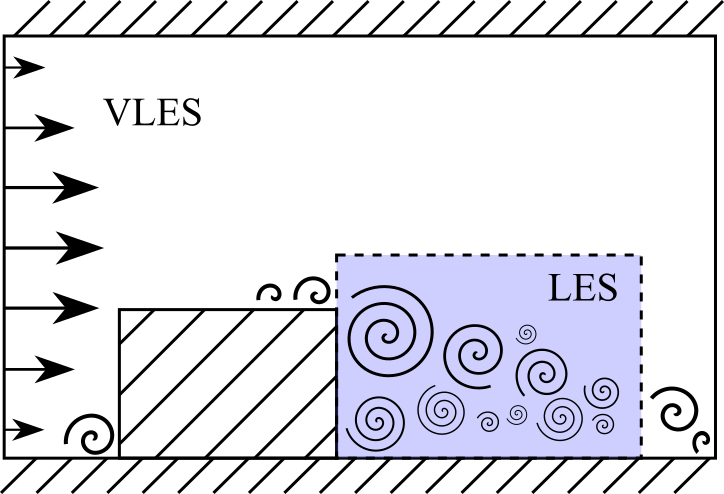}
        \caption{Hybrid VLES-LES.}
        \label{fig:zonal-LES}
    \end{subfigure}
    ~
    \begin{subfigure}{0.55\textwidth}
        \centering
        \includegraphics[height=2.5cm]{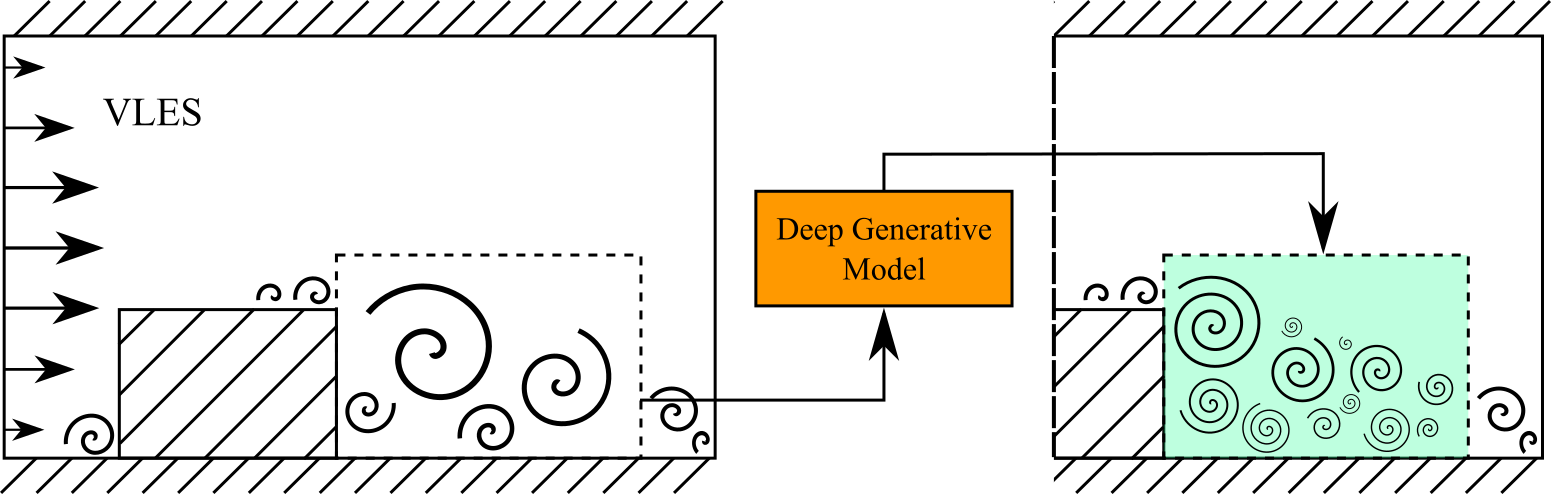}
        \caption{Multi-fidelity deep generative turbulence.}
        \label{fig:deep-generative-turbulence}
    \end{subfigure}
    \caption{Comparison between traditional hybrid VLES-LES simulation (left) and the proposed multi-fidelity deep generative turbulence model (right) for studying the wake behind a wall-mounted cube.}
\end{figure}

\noindent
Multiscale fluid simulation methods seek to strike an ideal balance between predictive accuracy and computational requirement.
In particular, zonal/hybrid methods couple a low-fidelity simulation with a high-fidelity simulation that is only evaluated in an area of interest.
This is most commonly done through the use of RANS or unsteady RANS in the low-fidelity region and LES in the high-fidelity region~\cite{holgate2019review}. 
Here, we consider the use of a very large eddy simulation (VLES) simulation (a LES simulation where the majority of the kinetic energy is unresolved due to a coarse grid) and a LES simulation on a finer mesh for the low- and high-fidelity areas, respectively.
As depicted in Fig.~\ref{fig:zonal-LES}, this results in two coupled simulations which are solved simultaneously with information being passed through the boundary of the high-fidelity simulation domain.
The objective in this work is to replace this high-fidelity simulation zone with a fast generative deep learning model which can quickly predict a high-fidelity realization given the low-fidelity simulation as illustrated in Fig.~\ref{fig:deep-generative-turbulence}.
We refer to this framework as a multi-fidelity generative model due to the distinct different physical scales resolved by the input and output.
While the scope of the numerical examples explored in this work is focused on the use of low-fidelity and high-fidelity LES simulations, everything discussed in this work can be extended to other multi-fidelity models using different coarse/fine simulation schemes.
We also  note that there is no limit on the size of the prediction area by the deep learning model, i.e. it can be the entire simulation domain if necessary. However,  in this work, we are motivated out of engineering needs where such zonal approaches are extremely applicable.

Simply learning a single solution of a PDE with a deep learning model has little practical benefit when a numerical solver exists due to the time and computational investment needed to tune and train the model.
Thus we are interested in surrogate modeling turbulence in multiple flows with varying boundary conditions (e.g. obstacle position or inlet velocity).
This is of particular interest for various engineering tasks including fluid-structure design/optimization, inverse modeling and uncertainty quantification.
To formalize the problem of interest consider an incompressible flow governed by the Navier-Stokes (N-S) equations:
\begin{equation}
\begin{gathered}
    \frac{\partial \overline{u}_{i}}{\partial t} + \overline{u}_{j}\frac{\partial \overline{u}_{i}}{\partial x_{j}} = -\frac{1}{\rho}\frac{\partial \overline{p}}{\partial x_{i}} + \frac{\partial}{\partial x_{j}}\left(\nu\frac{\partial \overline{u}_{i}}{\partial x_{j}} -  \overline{u'_{i}u'_{j}}\right), \\
    \bm{x}\in \Omega,\quad t\in[0,T], \quad \overline{u}_{i}(\bm{x},0)=u_{0}(\bm{x}), \quad \overline{p}(\bm{x},0)=p_{0}(\bm{x}),\\
    \mathcal{B}(\overline{u}_{i},\overline{p}) = \bm{b}(\bm{x},t),\quad \bm{x}\in\Gamma,
    \label{eq:ns-problem}
\end{gathered}
\end{equation}
where $\left\{\overline{u}_{j}, \overline{p}\right\}$ are filtered/averaged velocity and pressure, respectively, resolved within the spatial domain $\Omega$.
$\nu$ is the viscosity of the fluid and $\overline{u'_{i}u'_{j}}$ is the Reynolds stress.
$\Gamma$ denotes the boundary of the domain of interest for which the boundary operator $\mathcal{B}$ imposes the desired boundary conditions.
The initial state of the system is defined by $\left\{\bm{u}_{0}, \bm{p}_{0}\right\}$.

As depicted in Fig.~\ref{fig:deep-generative-turbulence}, we wish to build a deep generative model to infer from a low-fidelity flow field the corresponding high-fidelity realizations.
Due to their past success for modeling physical systems~\cite{zhu2018bayesian, zhu2019physics, geneva2019modeling}, we will choose to use a convolution based generative model with learnable parameters $\bm{\theta}$.
The use of convolutions implies that the data is placed onto a structured Euclidean grid, akin to that of pixels in images.
In the numerical examples used to test the proposed model, the flow in the $z$ direction is periodic and is not of any interest.
Thus we will be predicting a two--dimensional slice which for time-step $n$ would have a low-fidelity input $\bm{x}^{n}=\left\{\bm{u}_{l},\bm{p}_{l}\right\}\in\mathbb{R}^{3,d_{l1},d_{l2}}$ and a high-fidelity output $\bm{y}^{n}=\left\{\bm{u}_{h},\bm{p}_{h}\right\}\in\mathbb{R}^{3,d_{h1},d_{h2}}$ both of which span the same domain $\Omega'\in\Omega$ as depicted by the dashed boxes in Fig.~\ref{fig:deep-generative-turbulence}.
Although omitted, this is easily extendable to one- and three-dimensional fluid flows as well.
Given that $d_{l1},d_{l2} < d_{h1}, d_{h1}$, this requires the model to predict length scales not recovered by the coarse simulation making this problem ill-posed and motivating the need of a generative probabilistic model to predict the density $p(\bm{y}^{n}|\bm{x}^{n})$ as opposed to a single deterministic solution.
\begin{remark}
The inclusion of a low-fidelity simulator as an input to the deep learning surrogate allows for important information regarding the boundary conditions of the flow and approximate flow properties to be provided to the model.
This simplifies the learning task significantly by providing a physical coarse estimate of the flow, which is important for the prediction of the solution of the highly non-linear N-S equations at high Reynolds numbers.
While we have shown in our past work that deep learning surrogates can successfully model many complex physical systems independently~\cite{zhu2018bayesian, zhu2019physics, geneva2019modeling}, the systems of interest in these works are far less complex than the turbulent N-S equations. 
\end{remark}
Given that turbulence is a transient phenomenon, predicting at a single time-step is not sufficient, thus we wish to predict an entire time-series high-fidelity $\bm{Y}=\left\{\bm{y}^{1}, \bm{y}^{2},\ldots,\bm{y}^{N}\right\}$ given the respective low-fidelity observations $\bm{X}=\left\{\bm{x}^{1}, \bm{x}^{2},\ldots,\bm{x}^{N}\right\}$.
Although extensions can be made to simplify the model presented, we will assume that the time-step size, $\Delta t$, of the low-fidelity input and high-fidelity output is the same (i.e. for each input there is one output).
The objective for this surrogate is for fluid flow applications for which the boundary conditions are stochastic such that $\bm{b}\left(\hat{\bm{x}},t\right)\sim p(\bm{b})$, where $p(\bm{b})$  is an empirical, analytically known or an unknown probability distribution.
This spans problems including the modeling of a flow at different Reynolds numbers, different domain boundary conditions, flow through varying geometries or different initial conditions making this relevant to a vast number of fluid mechanics research studies.
Given that we wish to predict an entire time-series of high-fidelity realizations, $\bm{Y}$,
we pose the following definition for the generative multi-fidelity surrogate for flows with a stochastic boundary.

\theoremstyle{definition}
\begin{definition}{\bf Generative Surrogate for Flows with Stochastic Boundary Conditions.}
    Consider a low- and high-fidelity simulators that compute fluid flow governed by the N-S equations. For a given finite set of boundary conditions $\left\{\bm{b}\left(\hat{\bm{x}},t\right)_{i}\right\}_{i=1}^{M} \sim p(\bm{b})$, these simulators are used to collect a training set of  
     low- and high-fidelity simulation data $\mathcal{D}=\left\{\bm{X}_{i},\bm{Y}_{i}\right\}_{i=1}^{M}$ in the time interval $t\in[0,T]$. The problem of interest is training 
     a generative surrogate to learn $p_{\bm{\theta}}\left(\bm{Y}|\bm{X}\right)$ and compute the predictive conditional density $p_{\bm{\theta}}\left(\bm{Y}^{*}|\bm{X}^{*}, \mathcal{D}\right)$ of the high-fidelity flow field $\bm{Y}^{*}$ for \textit{any} low-fidelity flow time-series 
    $\bm{X}^{*}$ for  a given boundary condition $\bm{b}^{*}\left(\bm{x},t\right)\sim p(\bm{b})$.
\end{definition}

\section{Transient Multi-fidelity Glow}
\label{sec:tm-glow}
\subsection{Generative Normalizing Flows}
\label{sec:norm-flows}
\noindent
Deep generative models provide a flexible probabilistic framework with the most fundamental formulation centered around the use of random latent variables, $\bm{z}$, in a deep learning model (i.e. a neural network) to allow for the likelihood of the model's output, $\bm{y}$, to be expressed as the following marginal:
\begin{equation}
    p_{\bm{\theta}}(\bm{y})=\int p_{\bm{\theta}}\left(\bm{y}|\bm{z}\right) p_{\bm{\theta}}\left(\bm{z}\right) d\bm{z}, 
\end{equation}
in which $\bm{\theta}$ denotes the model's parameters.
In this work, the model's output, $\bm{y}$, is the high-fidelity flow field we wish to predict, however in this particular section $\bm{y}$ should be interpreted as a much more abstract output encompassing a wide variety of machine learning problems.
The latent variables are specifically designed such that their distribution is simple for sampling.
However, this marginal is typically not practical to train due to the large number of samples needed from $p_{\bm{\theta}}\left(\bm{z}\right)$ to approximate the marginalization.
Hence, generative models such as variational auto-encoders (VAEs)~\cite{kingma2013auto} as well as generative adversarial networks (GANs)~\cite{goodfellow2014generative} approximate this likelihood through variational inference or by a min-max adversarial game, respectively.

\begin{figure}[H]
    \centering
    \begin{subfigure}{0.22\textwidth}
        \centering
        \includegraphics[height=2.25cm]{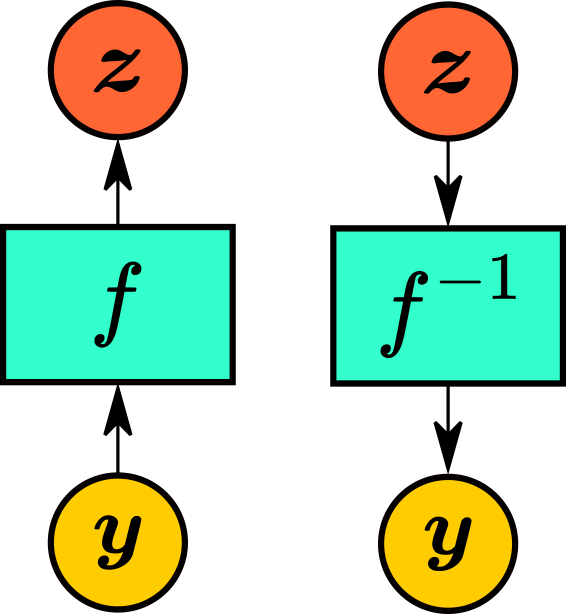}
        \caption{INN}
        \label{fig:inn}
    \end{subfigure}
    \begin{subfigure}{0.3\textwidth}
        \centering
        \includegraphics[height=2.25cm]{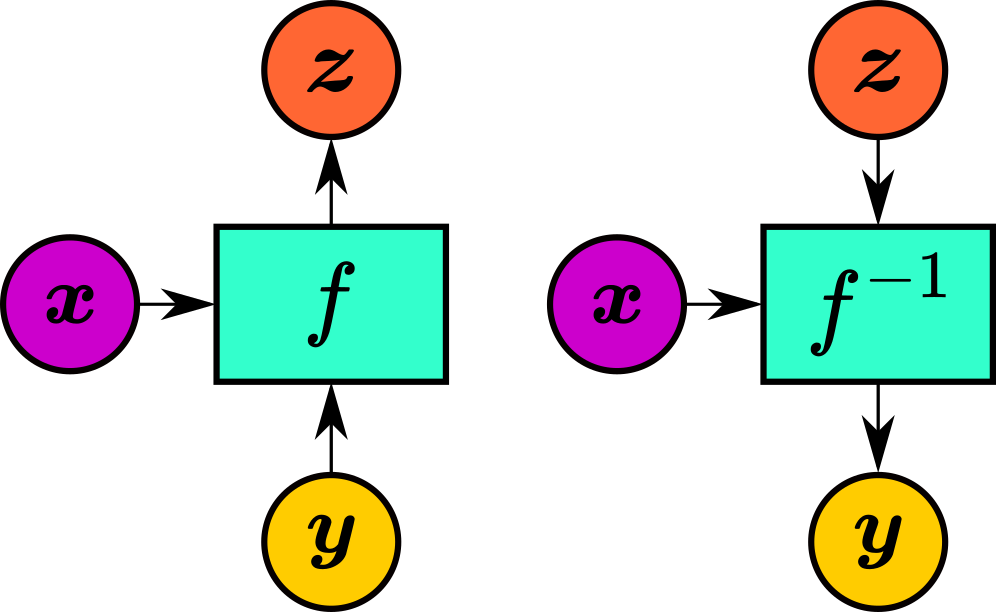}
        \caption{CINN}
        \label{fig:cinn}
    \end{subfigure}
    \begin{subfigure}{0.45\textwidth}
        \centering
        \includegraphics[height=2.25cm]{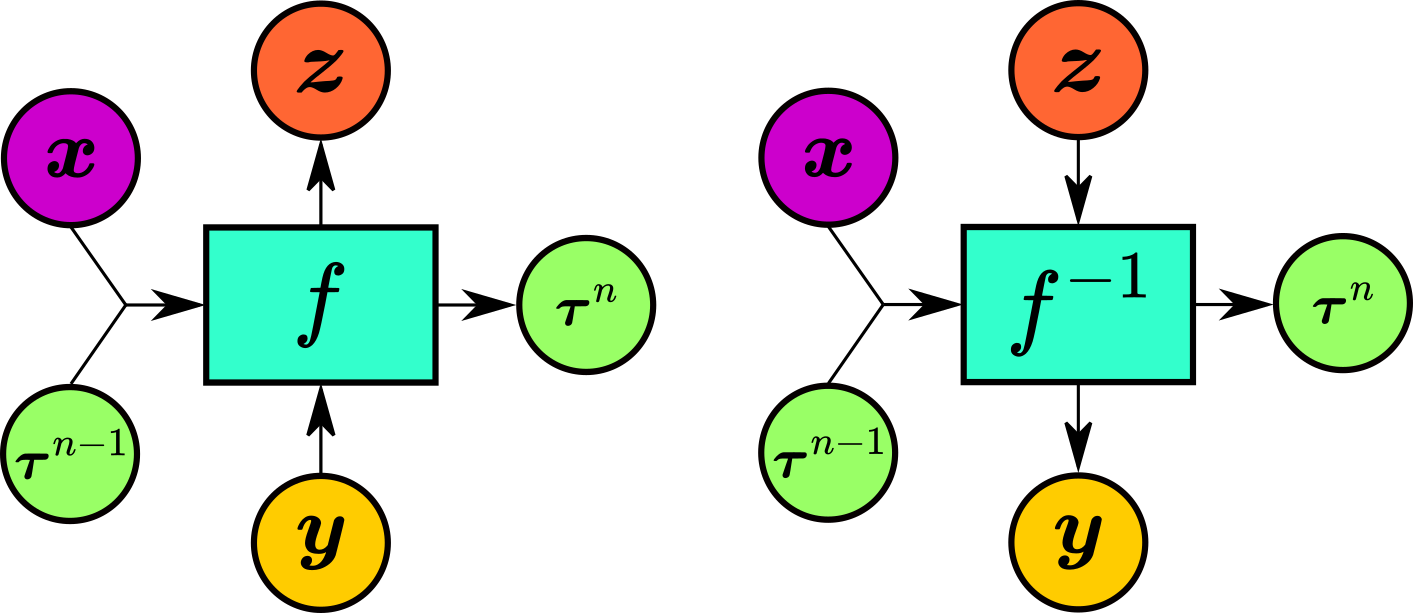}
        \caption{TM-Glow}
    \end{subfigure}
    
    \caption{Comparison of the forward and backward passes of various INN structures including (left to right) the standard INN, conditional INN (CINN)~\cite{zhu2019physics} and transient multi-fidelity Glow (TM-Glow) introduced in Section~\ref{sec:tm-glow-model}.}
\end{figure}

In this work, we will utilize normalizing flows which have gained increasing attention due to their extension to invertible neural networks (INNs) for tasks such as variational inference and generative modeling~\cite{dinh2014nice, dinh2016density, kingma2018glow, jacobsen2018revnet, kumar2019videoflow}.
Generative normalizing flows provide a bijective mapping between an unknown likelihood density of the observations $p_{\bm{\theta}}\left(\bm{y}\right)$  and a known latent density $p_{\bm{\theta}}\left(\bm{z}\right)$.
Typically, $p_{\bm{\theta}}\left(\bm{y}\right)$ can be viewed as the unknown likelihood of a system for which we have a finite number of observations, i.e. training data.
Let us consider   a mapping with a tractable Jacobian determinant, henceforth referred as the Jacobian,  which allows for the likelihood to be expressed w.r.t. the latent density as follows:
\begin{equation}
    p_{\bm{\theta}}\left(\bm{y}\right)=p_{\bm{\theta}}\left(\bm{z}\right)\left|\textrm{det}\left(\frac{\partial \bm{z}}{\partial \bm{y}}\right)\right|,
    \label{eq:change-of-vars}
\end{equation}
which is nothing more than the change of variables formula.
This implies that the model can be trained by maximizing the likelihood of $p_{\bm{\theta}}(\bm{y})$ (unknown) through the latent variables assigned a simple distribution $p_{\bm{\theta}}\left(\bm{z}\right)$ \textit{a-priori} (typically Gaussian).
As depicted in Fig.~\ref{fig:inn}, we use $f_{\bm{\theta}}\left(\cdot\right)$ to denote the learnable function, with a tractable Jacobian, that transforms  observation to  latent variables.
To generate samples of $\bm{y}_{i}\sim p_{\bm{\theta}}\left(\bm{y}\right)$, samples are drawn from the latent distribution $\bm{z}_{i}\sim p_{\bm{\theta}}\left(\bm{z}\right)$ which are then transformed using the inverse of the model $f^{-1}_{\bm{\theta}}\left(\cdot\right)$.

However, the requirement for a tractable Jacobian as well as a function that can efficiently be inverted for sampling is not trivial.
Normalizing flows address this challenge by using a series of change of variable transformations~\cite{tabak2010density, tabak2013family},
\begin{equation}
    \bm{y} 	\xleftrightarrow{f_{\bm{\theta}_{1}}} \bm{h}_{1} \xleftrightarrow{f_{\bm{\theta}_{2}}} \bm{h}_{2} \ldots \xleftrightarrow{f_{\bm{\theta}_{K}}} \bm{z},
    \label{eq:normalizing-flow}
\end{equation}
each of which has a tractable Jacobian and is invertible.
This allows for the log of the likelihood to be written as a summation of Jacobians:
\begin{equation}
    \log{p_{\bm{\theta}}\left(\bm{y}\right)} = \log{p_{\bm{\theta}}\left(\bm{z}\right)} + \sum_{k=1}^{K}\log{\left|\textrm{det}\left(\frac{\partial \bm{h}_{k}}{\partial \bm{h}_{k-1}}\right)\right|},
    \label{eq:log-likelihood}
\end{equation}
in which $\bm{h}_{0} \equiv \bm{y}$ and $\bm{h}_{K} \equiv \bm{z}$.

The core ideas of normalizing flows can be extended to constructing generative deep neural network models.
A particular subset of flow-based deep learning models we are interested in are coupling layer normalizing flows first proposed in NICE~\cite{dinh2014nice} and Real NVP~\cite{dinh2016density}.
A coupling layer is a carefully designed function such that the inverse mapping and the Jacobian can be easily calculated.
These layers can then be stacked, just like layers of a neural network to form an expressive model with a tractable Jacobian and inverse.
To increase the expressive capabilities of normalizing flow models, various transformations have been proposed to envelope coupling layers such as $1\times 1$ convolutions in the generative flow (Glow) model~\cite{kingma2018glow}.
Such flow-based models have the unique advantage of not needing an auto-encoder structure or discriminator, greatly simplifying the hyper-parameter search and increasing robustness against mode collapse.
As depicted in Fig.~\ref{fig:cinn}, further extensions can be made to learn conditional likelihoods, $p_{\bm{\theta}}\left(\bm{y}|\bm{x}\right)$,  for standard machine learning problems and surrogate modeling of  physical systems~\cite{zhu2019physics,ardizzone2019guided}.

\subsection{Transient Multi-fidelity Glow}
\label{sec:tm-glow-model}
\begin{figure}[H]
    \centering
    \includegraphics[width=0.6\textwidth]{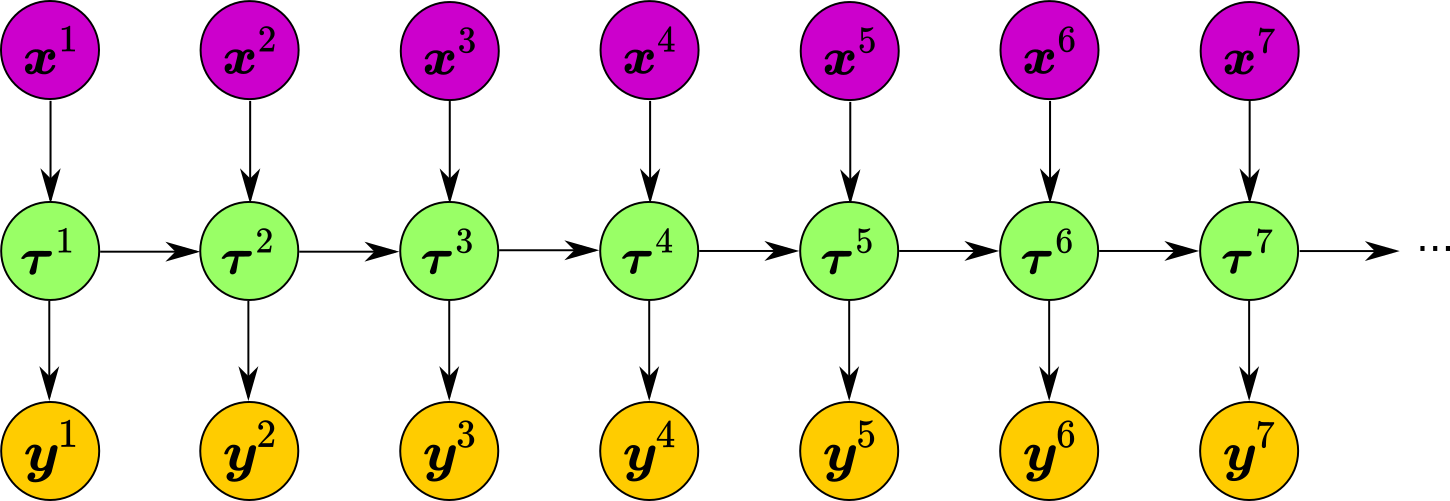}
    \caption{Unfolded computational graph of a recurrent neural network model for which the arrows show functional dependence.}
    \label{fig:nn-unfold}
\end{figure}

\begin{figure}[H]
    \centering
    \begin{subfigure}[b]{0.42\textwidth}
        \includegraphics[width=\textwidth]{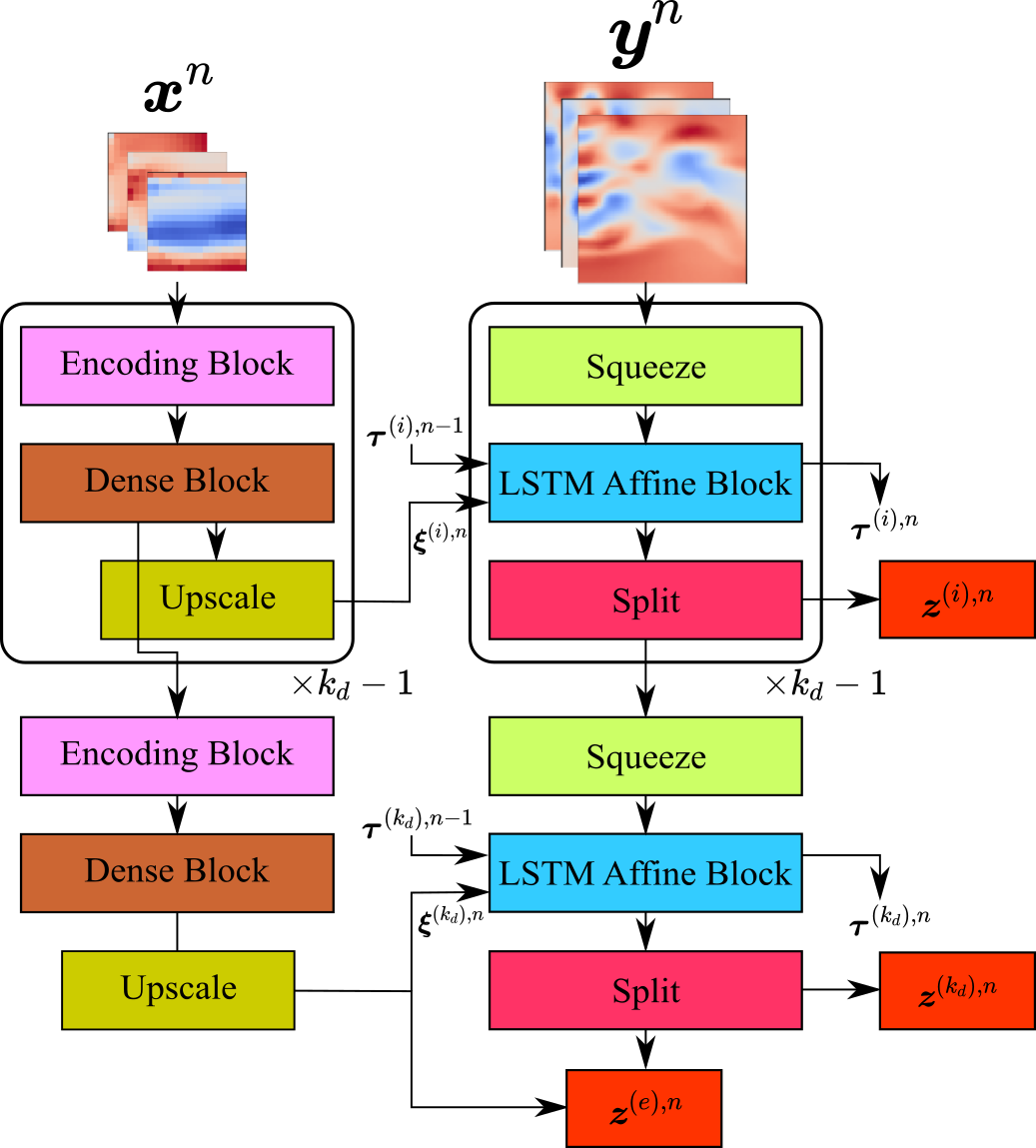}
        \caption{TM-Glow model schematic.}
        \label{fig:glowtm-schem}
    \end{subfigure}
    ~
    \begin{subfigure}[b]{0.55\textwidth}
        \includegraphics[width=\textwidth]{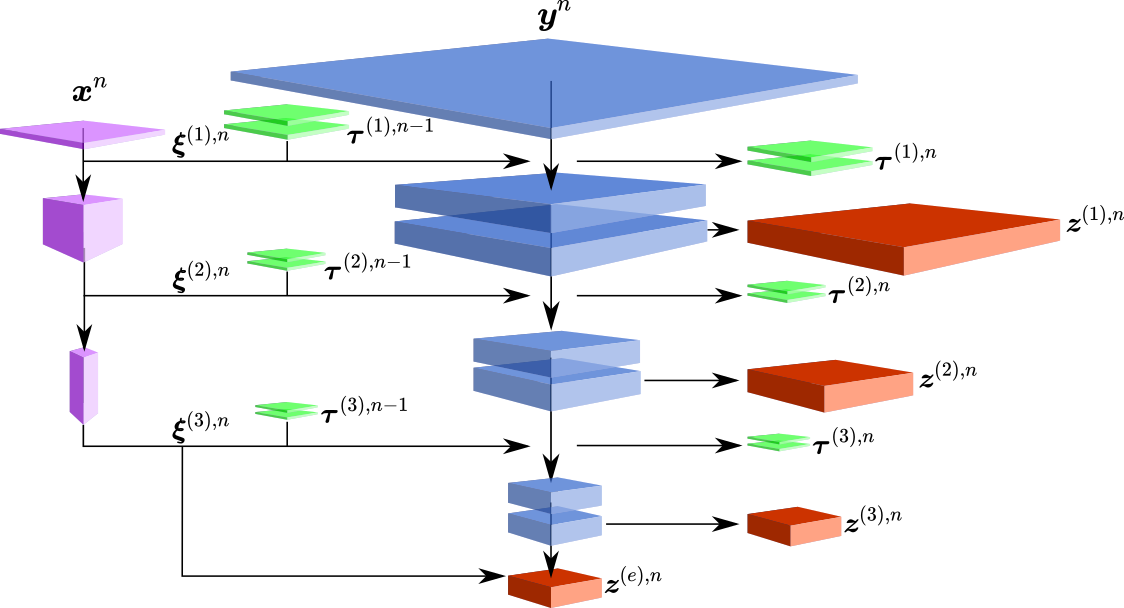}
        \caption{Dimensionality representation of TM-Glow with a model depth of $k_{d}=3$.}
        \label{fig:glowtm-dims}
    \end{subfigure}
    \caption{TM-Glow model. This model is comprised of a low-fidelity encoder that conditions a generative flow model to produce samples of high-fidelity field snapshots. LSTM affine blocks are introduced to pass information between time-steps using recurrent connections. Boxes with rounded corners in (a) indicate a stack of the elements inside and should not be confused with plate notation.
    Arrows illustrate the forward pass of the INN.
    (For interpretation of the colors in the figure(s), the reader is referred to the web version of this article.)}
    \label{fig:glowtm-model}
\end{figure}

\begin{figure}[H]
    \centering
    \includegraphics[width=0.6\textwidth]{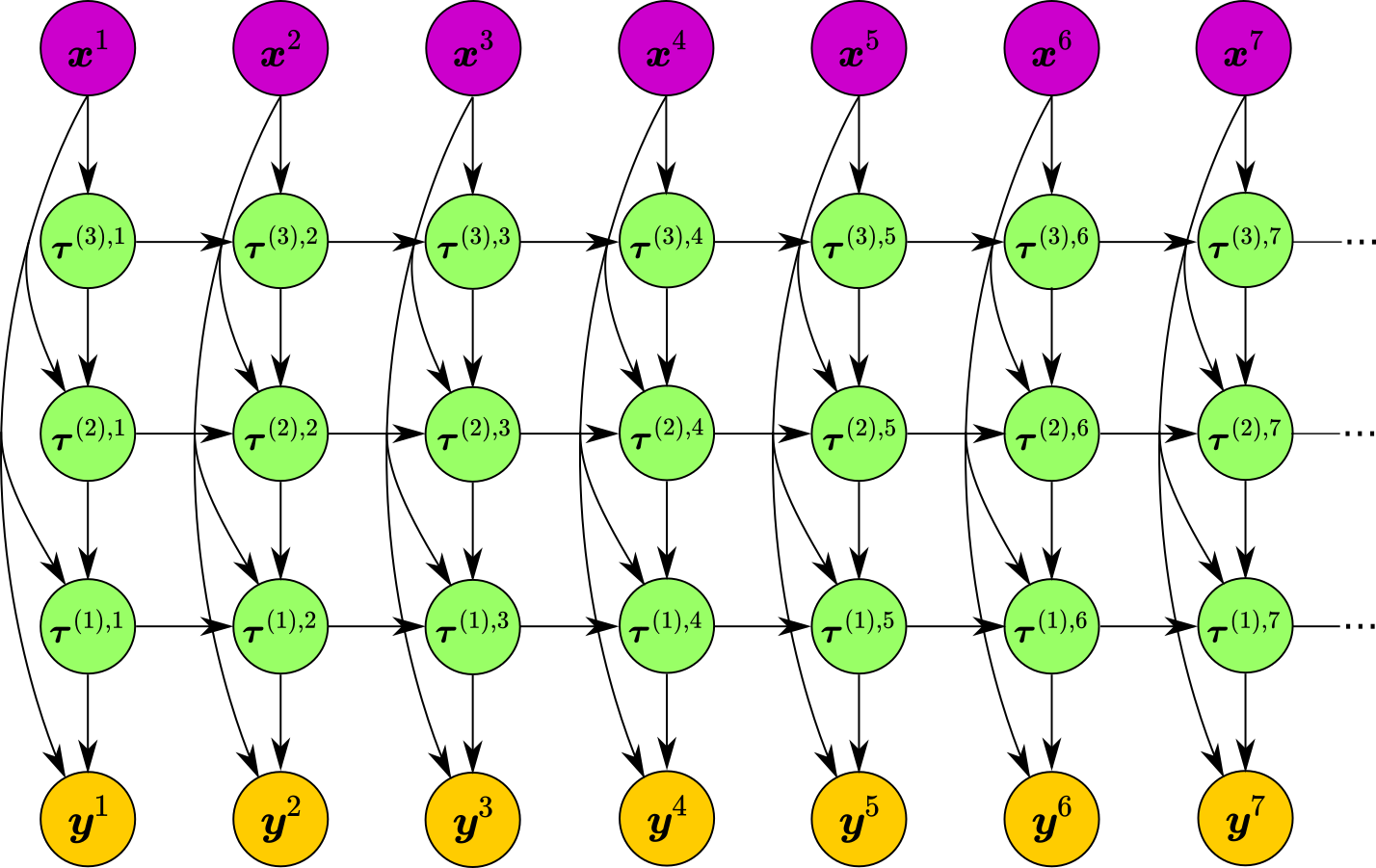}
    \caption{The unrolled computational graph of the TM-Glow model for a model depth of $k_{d}=3$.}
    \label{fig:glowtm-pgm}
\end{figure}

\noindent
As discussed in Section~\ref{sec:problem}, we are interested in the prediction of a high-fidelity flow, $\bm{Y}=\left\{\bm{y}^{1}, \bm{y}^{2},\ldots,\bm{y}^{N}\right\}$, given the corresponding solution of a low-fidelity simulation, $\bm{X}=\left\{\bm{x}^{1},\bm{x}^{2},\ldots, \bm{x}^{N}\right\}$.
In our past work~\cite{geneva2019modeling}, we formulated a deep convolutional auto-regressive model for modeling the evolution of a transient PDE as a  Markov chain.
To increase the predictive capability of our model and integrate the low-fidelity observations, in this work we will use a deep recurrent neural network (RNN) formulation which is a standard approach for time-series predictions in deep learning~\cite{goodfellow2016deep}.
While our model will still predict a single time-step at a time, latent information is passed between time-steps that the model can learn.
The computational graph of this RNN with recurrent features $\bm{\tau}^{n}$ is depicted in Fig.~\ref{fig:nn-unfold}.
The likelihood for the entire time-series can be decomposed as follows:
\begin{equation}
    \begin{aligned}
        p_{\bm{\theta}}\left(\bm{Y}|\bm{X}\right) = \prod_{n=1}^{N}p_{\bm{\theta}}\left(\bm{y}^{n}|\bm{x}^{1:n}\right) = \prod_{n=1}^{N}p_{\bm{\theta}}\left(\bm{y}^{n}|\bm{x}^{n},\bm{\tau}^{n-1}\right),
    \end{aligned}
    \label{eq:rnn-likelihood}
\end{equation}
in which the recurrent features, $\bm{\tau}^{n-1}$, carry information from past time-steps $\bm{x}^{1:n-1}$~\cite{goodfellow2016deep}.
This requires some initialization for $\bm{\tau}^{0}$ to be defined.
In this work, these states are made random with a known distribution as discussed in Section~\ref{sec:lstm-affine}, however there are many alternatives in RNN literature  such as making them constant (i.e. delta function density function) or making them learnable parameters.
Implementing the RNN framework, our goal is to develop a generative model that can learn the conditional density, $p_{\bm{\theta}}(\bm{y}^{n}|\bm{x}^{n},\bm{\tau}^{n-1})$, in a single high-fidelity time-step.
This poses the  following three design requirements: a) the core of our model must be generative allowing for probabilistic modeling of the likelihood, b) we must formulate a method for encoding the low-fidelity inputs $\bm{x}$ into features that can condition the generator and c) recurrent connections need to be integrated into the heart of the generative model to condition it on temporal information.
To this end, we present a novel Transient Multi-fidelity Glow (TM-Glow) model for probabilistic surrogate modeling of dynamical systems illustrated in Fig.~\ref{fig:glowtm-model}.
TM-Glow is built around the Glow model proposed by Kingma~\etal~\cite{kingma2018glow} which will be the core generative INN for modeling the conditional likelihood.
This model is depicted in the right column of Fig.~\ref{fig:glowtm-schem} and the blue boxes in Fig.~\ref{fig:glowtm-dims}.
Glow is designed to provide a multiscale encoding of the high-fidelity fields, $\bm{y}^{n}$, into a set of random latent variables, $\bm{z}^{n}$, represented by the orange boxes in Fig.~\ref{fig:glowtm-model}.
To address the second design requirement, we use the convolutional conditional encoder proposed by Zhu~\etal~\cite{zhu2019physics} which conditions Glow model on the low-fidelity input, $\bm{x}^{n}$, through a set of learnable features.
This conditional encoder is shown in the left column of Fig.~\ref{fig:glowtm-schem} and the pink boxes in Fig.~\ref{fig:glowtm-dims}.
Lastly, to allow for temporal conditioning of the Glow model, recurrent connections are integrated into novel LSTM affine coupling blocks discussed in Section~\ref{sec:lstm-affine}.
These LSTM based operations allow for recurrent features to flow in and out of the generator illustrated by the green boxes in Fig.~\ref{fig:glowtm-dims}.
As shown in Fig.~\ref{fig:glowtm-schem}, the TM-Glow core component is the multiscale Glow model comprised of squeeze, LSTM Affine Block and split operations discussed in detail in Section~\ref{sec:multi-scale-glow}.
Superscript numbers enclosed by parenthesis are used to denote variables at different TM-Glow model levels.
We emphasize that TM-Glow is an INN, thus this model can evaluate the conditional likelihood \textit{exactly} through the change of variables:
\begin{equation}
    \log{p_{\bm{\theta}}\left(\bm{Y}|\bm{X}\right)} = \sum_{n=1}^{N}\log{p_{\bm{\theta}}\left(\bm{z}^{n}|\bm{x}^{n},\bm{\tau}^{n-1}\right)} + \sum_{k=1}^{K}\log{\left|\textrm{det}\left(\frac{\partial \bm{h}^{n}_{k}}{\partial \bm{h}^{n}_{k-1}}\right)\right|},
    \label{eq:cond-log-likelihood}
\end{equation}
in which $\left\{\bm{h}^{n}_{k}\right\}^{K}_{k=1}$ is used to denote the hidden layers of TM-Glow that are the inputs/outputs of the various invertible operations discussed in Section~\ref{sec:multi-scale-glow} and specifically in Table~\ref{tab:flow-operations}.
The forward pass of the model, $f_{\bm{\theta}}(\cdot)$, encodes the high-fidelity observation $\bm{y}^{n}$ into a set of random latent variables $\bm{z}^{n}=\left\{\bm{z}^{(1),n},\bm{z}^{(2),n},\ldots,\bm{z}^{(k_{d}),n},\bm{z}^{(e),n}\right\}$.
The backward/inverse pass of the model, $f^{-1}_{\bm{\theta}}(\cdot)$, generates a sample of $\bm{y}^{n}$ by sampling each random latent variable.
The novel LSTM Affine block contains recurrent connections between time-steps conditioning the INN on the latent states of previous time-steps $ \bm{\tau}^{n-1}= \left\{\bm{\tau}^{(1),n-1},...,\bm{\tau}^{(k_{d}),n-1}\right\}$.
The dense convolutional encoder, detailed in Section~\ref{sec:tm-glow-conditioning}, encodes a low-fidelity input into conditional feature maps, $\bm{\xi}^{n}=\left\{\bm{\xi}^{(1),n},\bm{\xi}^{(2),n},\ldots,\bm{\xi}^{(k_{d}),n}\right\}$ that are injected into the multiscale glow at each dimensional level as depicted in Fig.~\ref{fig:glowtm-dims}.
The use of the recurrent connections in the LSTM block as well as the conditioning encoder results in the following directed graphical representation of the model in Fig.~\ref{fig:glowtm-pgm}.

\subsection{Multiscale Glow}
\label{sec:multi-scale-glow}
\noindent
Our model is centered around a multiscale structure to promote the discovery of low-dimensional representation of the physics that govern the system.
As seen in Fig.~\ref{fig:glowtm-model}, a multiscale Glow model originally proposed by Dinh~\etal~\cite{dinh2016density} is employed to generate a flow field realization.
In Fig.~\ref{fig:glowtm-schem}, this is the right column of the model comprised of Squeeze, LSTM Affine Block and Split operations each of which is invertible.
We remind the reader that the goal of each of these operations is to provide a computationally efficient but descriptive mapping between the high-fidelity flow field and the random latent variables.
As previously discussed in~\Eqref{eq:normalizing-flow}, this is achieved through the series of transformations between the hidden layers $\left\{\bm{h}^{n}_{k}\right\}^{K}_{k=1}$.
These transformations are precisely the operations discussed in the subsequent sections and listed in Table~\ref{tab:flow-operations}.

\subsubsection{LSTM Affine Block}
\label{sec:lstm-affine}
\begin{figure}[H]
    \centering
    \includegraphics[width=\textwidth]{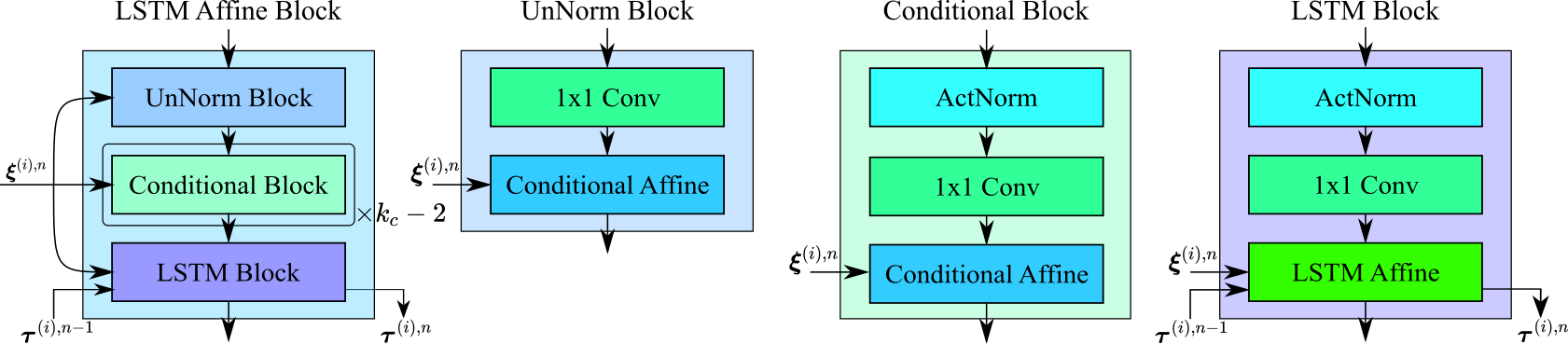}
    \caption{The LSTM affine block used in TM-Glow consisting of $k_{c}$ affine coupling layers including an unnormalized conditional affine block (UnNorm Block), a stack of conditional affine blocks (Conditional Block) and a conditional LSTM affine block (LSTM Block).}
    \label{fig:affine-blocks}
\end{figure}

\begin{figure}[H]
    \centering
    \begin{subfigure}{0.45\textwidth}
        \centering
        \includegraphics[width=0.7\textwidth]{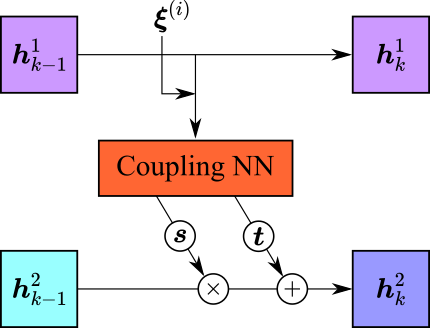}
        \caption{Conditional coupling layer.}
        \label{fig:affine-coupline-layer}
    \end{subfigure}
    ~
    \begin{subfigure}{0.45\textwidth}
        \centering
        \includegraphics[width=0.7\textwidth]{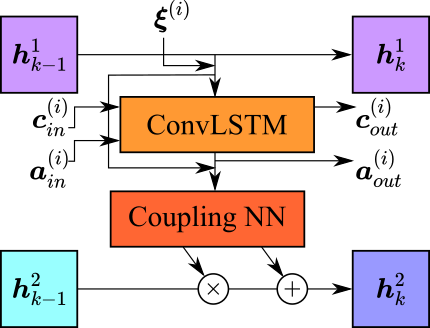}
        \caption{Conditional LSTM coupling layer.}
         \label{fig:lstm-coupline-layer}
    \end{subfigure}
    \caption{The two variants of affine coupling layers used in TM-Glow with an input and output denoted as $\bm{h}_{k-1} = \left\{\bm{h}_{k-1}^{1}, \bm{h}_{k-1}^{2}\right\}$ and $\bm{h}_{k} = \left\{\bm{h}_{k}^{1}, \bm{h}_{k}^{2}\right\}$, respectively. Time-step superscripts have been omitted for clarity of presentation.}
    \label{fig:coupling-layers}
\end{figure}

\noindent
The core component of the generative portion of the TM-Glow model is the LSTM affine block, which is a novel extension of the conditional affine coupling layers~\cite{zhu2019physics, ardizzone2019guided} designed specifically for transient time-series prediction.
The LSTM affine block is comprised of three different sub-components illustrated in Fig.~\ref{fig:affine-blocks}: an unnormalized conditional affine block, a stack of conditional affine blocks and a conditional LSTM affine block.
The core component of all these blocks are affine coupling layers~\cite{dinh2014nice}, a specially designed function that allows for an efficient inversion and Jacobian calculation.
As depicted in Fig.~\ref{fig:affine-coupline-layer}, half of the input, $\bm{h}^{2}_{k-1}$, is modified by the scale and translation parameters, $\bm{s}$ and $\bm{t}$, respectively, calculated from a coupling neural network (coupling NN).
As implemented in Zhu~\etal~\cite{zhu2019physics}, this coupling NN is a shallow dense convolutional network with an input of the other half original feature map, $\bm{h}^{1}_{k-1}$, and the conditional input $\bm{\xi}^{(i),n}$, which are simply concatenated together.
This coupling NN contains the learnable parameters that can be updated using any gradient decent method.
As detailed in Table~\ref{tab:flow-operations}, the retention of the input to the coupling NN allows for a simple inversion and Jacobian calculation.

This conditional affine coupling layer is further extended with a convolutional LSTM (convLSTM) depicted in Fig.~\ref{fig:lstm-coupline-layer} for transient problems.
ConvLSTM is a variation of the traditional LSTM structure that employs convolutional operations~\cite{xingjian2015convolutional}, making it better suited for convolutional models such as TM-Glow.
This input to the ConvLSTM is the same as the input of the coupling NN in the conditional coupling layer, which is conditioned on $\bm{\xi}^{(i),n}$.
Following the standard ConvLSTM formulation, the recurrent features have two states, $\bm{\tau}^{(i),n-1}=\left\{\bm{a}^{(i)}_{in},\bm{c}^{(i)}_{in}\right\}$, which correspond to the LSTM hidden and cell state, respectively.
The output of the LSTM, $\bm{\tau}^{(i),n}=\left\{\bm{a}^{(i)}_{out},\bm{c}^{(i)}_{out}\right\}$, is passed to the subsequent time-step and $\bm{a}^{(i)}_{out}$ is used as an input to the coupling NN of the affine layer.
The initial states of the hidden and cell states at the first time-step are assigned the following densities:
\begin{equation}
    \bm{\tau}^{(i),0}=\left\{\bm{a}^{(i)}_{in}\sim \mathcal{U}\left(-1,1\right),\bm{c}^{(i)}_{in}\sim \mathcal{N}\left(0,1\right)\right\}.
\end{equation}
The resulting coupling layer is conditioned on both the current low-fidelity input as well as past time-step states.
In the recent work of Kumar~\etal~\cite{kumar2019videoflow}, residual connections between generative flow models are also proposed which are implemented by simply using the previous latent variables as an input to the shallow neural network in the split operation, as detailed in Section~\ref{sec:split}.
The proposed use of ConvLSTM affine layer prevents a vanishing gradient and enables much more descriptive recurrent feature maps to be learned.
In the coupling layer blocks ActNorm is used which was originally proposed by Kingma and Dhariwal~\cite{kingma2018glow} as an alternative to batch-normalization.
ActNorm applies an invertible normalization to each feature channel, detailed in Table~\ref{tab:flow-operations}, that allows for smaller batch-sizes to be used without performance desegregation seen in traditional batch-normalization.
The last essential component of the affine coupling blocks is the $1 \times 1$ convolution also originally proposed in the Glow model~\cite{kingma2018glow}.
Due to this convolutional operation being $1\times 1$, it can be efficiently inverted and has a trivial Jacobian as detailed in Table~\ref{tab:flow-operations}.
The purpose of this convolution is to permute the feature maps between coupling layers.
Since, the coupling layers used in Fig.~\ref{fig:coupling-layers} only operate on half of the input data, permutation between layers is essential to increase the expressibility of the model.

\renewcommand{\arraystretch}{1.25}
\begin{table}[H]
\caption{Invertible operations used in the generative normalizing flow method of TM-Glow. Being consistent with the notation in~\cite{kingma2018glow}, we assume the inputs and outputs of each operation are of dimension $\bm{h}_{k-1},\bm{h}_{k} \in \mathbb{R}^{c\times h \times w}$ with $c$ channels and a feature map size of $\left[h \times w\right]$. Indexes over the spatial domain of the feature map are denoted by $\bm{h}(x,y)\in \mathbb{R}^{c}$. The coupling neural network and convolutional LSTM are abbreviated as $NN$ and $LSTM$, respectively. Time-step superscripts have been neglected for clarity of presentation.}
\label{tab:flow-operations}
\resizebox{\textwidth}{!}{%
\begin{tabular}{c|l|l|l}
\hline
Operation & Forward & Inverse & Log Jacobian \\ \hline

\begin{tabular}[c]{@{}c@{}} Conditional\\ Affine Layer\end{tabular} & 
$\begin{aligned}
    &\left\{\bm{h}_{k-1}^{1}, \bm{h}_{k-1}^{2}\right\} = \bm{h}_{k-1}\\
    &(\log \bm{s}, \bm{t}) = NN(\bm{h}_{i-1}^{1}, \bm{\xi}^{(i)})\\
    &\bm{h}_{k}^{2}=\exp\left(\log \bm{s}\right)\odot \bm{h}_{k-1}^{2} + \bm{t}\\
    &\bm{h}_{k}^{1} = \bm{h}_{k-1}^{1}\\
    &\bm{h}_{k} = \left\{\bm{h}_{k}^{1}, \bm{h}_{k}^{2}\right\}
\end{aligned}$ & 
$\begin{aligned}
    &\left\{\bm{h}_{k}^{1}, \bm{h}_{k}^{2}\right\} = \bm{h}_{k} \\
    &(\log \bm{s}, \bm{t}) = NN(\bm{h}_{k}^{1}, \bm{\xi}^{(i)})\\
    &\bm{h}_{k-1}^{2}= \left(\bm{h}_{k}^{2} - \bm{t}\right)/\exp\left(\log \bm{s}\right)\\
    & \bm{h}_{k-1}^{1} = \bm{h}_{k}^{1}\\
    &\bm{h}_{k-1} = \left\{\bm{h}_{k-1}^{1}, \bm{h}_{k-1}^{2}\right\}
\end{aligned}$
 &   $\textrm{sum}\left(\log \left|\bm{s}\right|\right)$ \\ \hline
 
\begin{tabular}[c]{@{}c@{}} LSTM Affine \\ Layer \end{tabular} &  
$\begin{aligned}
    &\left\{\bm{h}_{k-1}^{1}, \bm{h}_{k-1}^{2}\right\} = \bm{h}_{i-1}\\
    & \bm{a}^{(i)}_{out}, \bm{c}^{(i)}_{out} = LSTM\left(\bm{h}_{k-1}^{1}, \bm{\xi}^{(i)}, \bm{a}^{(i)}_{in}, \bm{c}^{(i)}_{in}\right)\\
    &(\log \bm{s}, \bm{t}) = NN(\bm{h}_{k-1}^{1}, \bm{\xi}^{(i)}, \bm{a}^{(i)}_{out})\\
    &\bm{h}_{k}^{2}=\exp\left(\log \bm{s}\right)\odot \bm{h}_{k-1}^{2} + \bm{t}\\
    &\bm{h}_{k}^{1} = \bm{h}_{k-1}^{1}\\
    &\bm{h}_{k} = \left\{\bm{h}_{k}^{1}, \bm{h}_{k}^{2}\right\}
\end{aligned}$ & 
$\begin{aligned}
    &\left\{\bm{h}_{k}^{1}, \bm{h}_{k}^{2}\right\} = \bm{h}_{k} \\
    &\bm{a}^{(i)}_{out}, \bm{c}^{(i)}_{out} = LSTM\left(\bm{h}_{k}^{1}, \bm{\xi}^{(i)}, \bm{a}^{(i)}_{in}, \bm{c}^{(i)}_{in}\right)\\
    &(\log \bm{s}, \bm{t}) = NN(\bm{h}_{k}^{1}, \bm{\xi}^{(i)}, \bm{a}^{(i)}_{out})\\
     &\bm{h}_{k-1}^{2}= \left(\bm{h}_{k}^{2} - \bm{t}\right)/\exp\left(\log \bm{s}\right)\\
    & \bm{h}_{k-1}^{1} = \bm{h}_{k}^{1}\\
    &\bm{h}_{k-1} = \left\{\bm{h}_{k-1}^{1}, \bm{h}_{k-1}^{2}\right\}
\end{aligned}$ &  $\textrm{sum}\left(\log \left|\bm{s}\right|\right)$ \\ \hline

ActNorm & $\forall x,y\quad \bm{h}_{k}(x,y)=\bm{s}\odot \bm{h}_{k-1}(x,y) + \bm{b}$  & $\forall x,y\quad \bm{h}_{k-1}(x,y)=(\bm{h}_{k}(x,y)-\bm{b})/\bm{s}$  & $h\cdot w \cdot \textrm{sum}\left(\log \left|\bm{s}\right|\right)$ \\ \hline 

$1\times 1$ Convolution & $\forall x,y\quad \bm{h}_{k}(x,y)=\bm{W}\bm{h}_{k-1}(x,y) \quad \bm{W}\in\mathbb{R}^{c\times c}$  & $\forall x,y\quad \bm{h}_{k-1}(x,y) =\bm{W}^{-1}\bm{h}_{k}(x,y)$  & $h\cdot w \cdot \log\left(\det \left|\bm{W}\right|\right)$ \\ \hline

Split & 
$\begin{aligned}
    &\left\{\bm{h}_{k-1}^{1}, \bm{h}_{k-1}^{2}\right\} = \bm{h}_{k-1}\\
    &\left(\bm{\mu},\bm{\sigma}\right) = NN\left(\bm{h}_{k-1}^{1}\right)\\
    &p_{\bm{\theta}}(\bm{z}_{k}) = \mathcal{N}\left(\bm{h}_{k-1}^{2}| \bm{\mu}, \bm{\sigma} \right)\\
    &\bm{h}_{k} = \bm{h}_{k-1}^{1}
\end{aligned}$ 
& 
$\begin{aligned}
    &\bm{h}_{k-1}^{1} = \bm{h}_{k}\\
    &\left(\bm{\mu},\bm{\sigma}\right) = NN\left(\bm{h}_{k-1}^{1}\right)\\
    &\bm{h}_{k-1}^{2} \sim \mathcal{N}\left(\bm{\mu},\bm{\sigma} \right)\\
    &\bm{h}_{k-1} = \left\{\bm{h}_{k-1}^{1}, \bm{h}_{k-1}^{2}\right\}
\end{aligned}$ 
& N/A
\end{tabular}}
\end{table}

\subsubsection{Squeeze}
\begin{figure}[H]
    \centering
    \begin{subfigure}{0.48\textwidth}
        \centering
        \includegraphics[height=2.5cm]{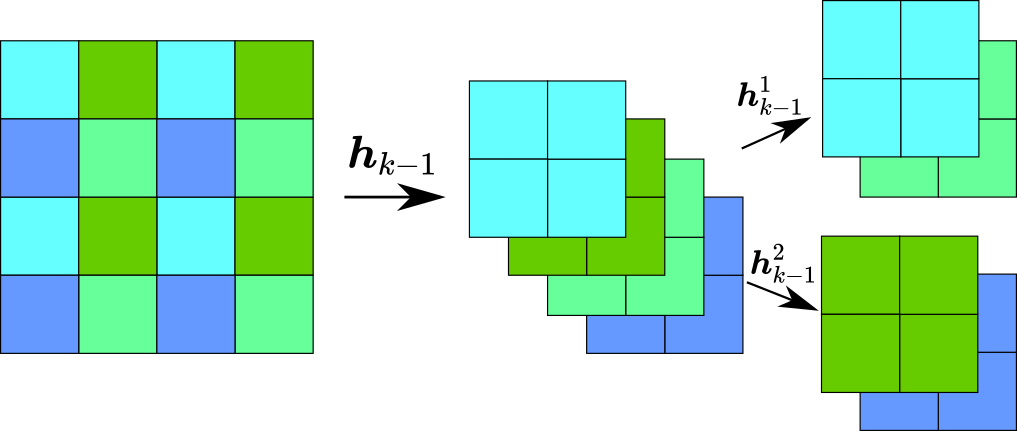}
        \caption{Squeeze operation.}
        \label{fig:squeeze}
    \end{subfigure}
    ~
    \begin{subfigure}{0.48\textwidth}
        \centering
        \includegraphics[height=2.5cm]{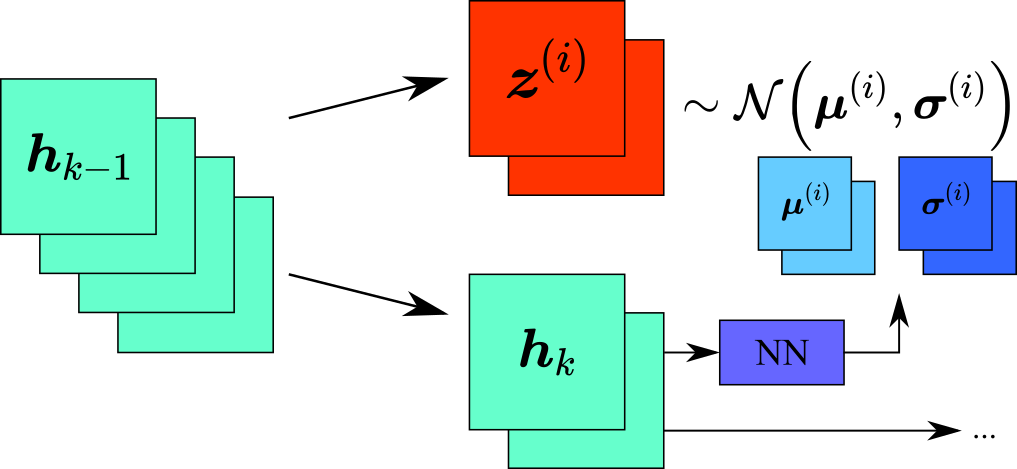}
        \caption{Split operation.}
        \label{fig:split}
    \end{subfigure}
    \caption{Squeeze and split forward operations used to manipulate the dimensionality of the features in TM-Glow. (Left) The squeeze operation compresses the input feature map $\bm{h}_{k-1}$ using a checkerboard pattern halving the spatial dimensionality and increasing the number of channels by four. (Right) The split operation factors out half of an input $\bm{h}_{k-1}$ which are then taken to be latent random variable $\bm{z}^{(i)}$. The remaining features, $\bm{h}_{k}$ are sent deeper in the network. Time-step 
    superscripts have been omitted for clarity of presentation.}
\end{figure}

\noindent
As seen in Fig.~\ref{fig:coupling-layers}, the affine coupling layer requires two inputs for which only one is modified to allow for efficient inversion.
To form these two inputs, a squeeze operation is applied to the feature maps which reduces the dimensions of the feature map by a half and increases the number of channels by a factor of two.
In this work, we use the squeeze method originally proposed by Dinh~\etal~\cite{dinh2016density} and also implemented in the Glow model~\cite{kingma2018glow}.
As depicted in Fig.~\ref{fig:squeeze}, the image is separated using a checkerboard pattern resulting in four sub-sampled versions.
\subsubsection{Split}
\label{sec:split}
\noindent
Unlike standard convolutional operations, the affine coupling layer is volume preserving meaning that the number of output elements must be the same as the input.
Retaining the total dimensionality input through all layers of the model is not ideal for a convolutional model since this increases the computational and memory cost of the model.
Thus we use the multiscale architecture proposed by Dinh~\etal~\cite{dinh2016density}, which is illustrated clearly in Fig.~\ref{fig:glowtm-dims}.
This multiscale flow model factors out half of the current feature maps at multiple intervals of the architecture which are then treated as random latent variables~\cite{kingma2018glow, zhu2019physics}.
A single split operation is illustrated in Fig.~\ref{fig:split} in which the density of these latent variables is taken to be a fully-factorizable Gaussian with mean and standard deviation governed from the remaining features using a shallow neural network.

When the split is executed in the inverse direction, the hyper-parameters are dependent on the features being provided from deeper within the model as seen in Table~\ref{tab:flow-operations}.
This dependence on deeper feature therefore conditions the random latent variables on both conditional features representing the coarse simulation input, $\bm{x}^{n}$, as well as the recurrent features $\bm{\tau}^{n-1}$.
As an example to illustrate this point, consider a TM-Glow with a model depth of $k_{d}=3$ as illustrated in Fig.~\ref{fig:glowtm-dims}.
Each of the four random latent variables and high-fidelity output for a single time-step can be described as the following conditional distributions:
\begin{equation}
\begin{gathered}
    \bm{z}^{(e),n}\sim p_{\bm{\theta}}\left(\bm{z}^{(e),n}|\bm{\xi}^{(3),n}\right), \quad \bm{z}^{(3),n}\sim p_{\bm{\theta}}\left(\bm{z}^{(3)}|\bm{z}^{(e)}\right), \\
    \bm{z}^{(2),n}\sim p_{\bm{\theta}}\left(\bm{z}^{(2),n}|\bm{z}^{(3),n},\bm{\xi}^{(3),n},\bm{\tau}^{(3),n-1}\right),
    \quad \bm{z}^{(1),n}\sim p_{\bm{\theta}}\left(\bm{z}^{(1,n)}|\bm{z}^{(2),n},\bm{\xi}^{(2),n},\bm{\tau}^{(2),n-1}\right),\\
    \bm{y}^{n}\sim p_{\bm{\theta}}\left(\bm{y}^{n}|\bm{z}^{(1),n},\bm{\xi}^{(1),n},\bm{\tau}^{(1),n-1}\right),
\end{gathered}
\end{equation}
which clearly is a hierarchical modeling of the distribution $\bm{y}^{n}\sim p\left(\bm{y}^{n}|\bm{x}^{n},\bm{\tau}^{n-1}\right)$ for which TM-Glow was designed to learn.

As discussed by Dinh~\etal~\cite{dinh2016density}, this multiscale architecture has multiple intrinsic benefits.
The first is that it results in the model learning intermediate representations of the output field, with deeper latent variables representing more global characteristics and shallower ones representing finer details.
Additionally this permutes the loss across multiple layers of the network which can improve training and predictive accuracy.
With respect to modeling physical systems, such a multiscale architecture is well suited as a vast number of physical phenomena are multiscale in nature.
Specifically in fluids, it is well known that turbulence occurs at multiple length and time scales making TM-Glow well suited for fluid flow prediction.
\begin{remark}
A particularly interesting attribute of the Glow model is the presence of random latent variables at multiple levels in the generator.
This characteristic is absent from traditional VAE or GANs models for which the random latent variables are only present at one level of the model, typically the lowest-dimensional.
This unique architecture arises out of necessity, but allows for the generative model to learn probabilistic densities at multiple scales.
In the context of physical systems, this could allow the model to learn stochastic phenomena at varying length scales which lends itself nicely to many multiscale systems.
\end{remark}

\subsection{Low-Fidelity Conditioning}
\label{sec:tm-glow-conditioning}
\begin{figure}[H]
    \centering
    \includegraphics[width=0.8\textwidth]{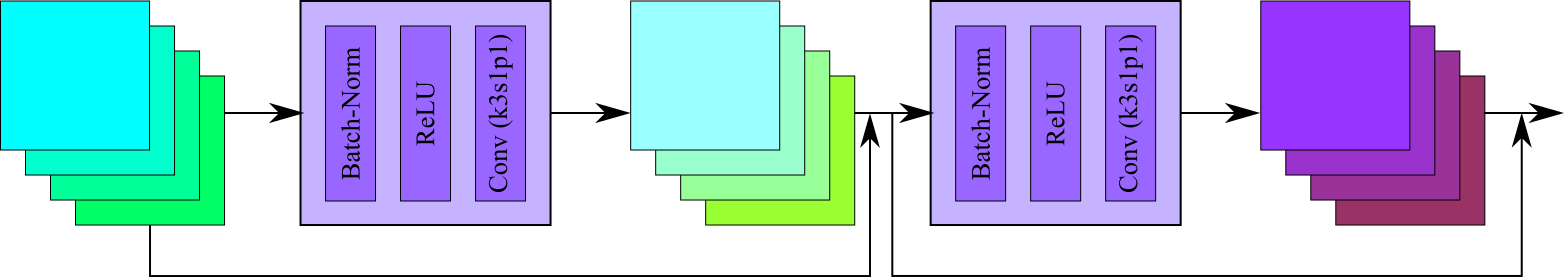}
    \caption{Dense block with a growth rate and length of $2$. Residual connections between convolutions progressively stack feature maps resulting in $12$ output channels in this schematic.  Standard batch-normalization~\cite{ioffe2015batch} and Rectified Linear Unit (ReLU) activation functions are used~\cite{glorot2011deep} in junction with the convolutional operations.  Convolutions are denoted by the kernel size $k$, stride $s$ and padding $p$.}
    \label{fig:dense-block}
\end{figure}

\noindent
To condition the generative model on the low-fidelity fluid field at multiple levels, a densely connected convolutional encoder is used.
This convolutional encoder, illustrated on the right side of Fig.~\ref{fig:glowtm-schem}, is comprised of encoding and dense blocks following the approach originally taken by Zhu~\etal~\cite{zhu2019physics}.
Examples of the encoding and dense blocks are illustrated in Fig.~\ref{fig:dense-block} which have been used successfully for modeling many physical systems in the past~\cite{zhu2018bayesian, zhu2019physics, geneva2019modeling, mo2019deep}.
The encoding blocks down-scale the feature maps forcing the model to learn low-dimensional representations while the densely connected blocks increase predictive accuracy of the model and have better performance than standard residual connections~\cite{huang2017densely}.
The feature maps are taken from multiple levels of  the convolutional encoder, up-scaled and passed to the affine coupling blocks conditioning the generator.
These are denoted by $\bm{\xi}^{(i),n}$ in Fig.~\ref{fig:glowtm-model}, passing detailed high-dimensional features towards the beginning of the encoder and global low-dimensional features towards the end.
%

\section{TM-Glow Training}
\label{sec:tm-glow-training}
\noindent
One of the key benefits of using INNs is the ability to calculate the likelihood of the data exactly with respect to the latent variables.
This makes data-driven training straight forward as one can simply pose the optimization as the minimization of the negative of the log likelihood in~\Eqref{eq:log-likelihood}~\cite{dinh2014nice, dinh2016density, kingma2018glow, grathwohl2018ffjord}.
However, since this encodes the output of the model to the latent parameters, this training does not allow physical-constraints to be imposed on the generated samples of the model.
In the work of Zhu~\etal~\cite{zhu2019physics}, in which physics-constrained learning is used in the absence of data, the reverse  Kullback-Leibler (KL) divergence is used as an optimization objective.
The reverse KL divergence poses the optimization though the generated samples of the INN, which allows for physical-constraints to be imposed on the produced realizations.

Due to the complex dynamics of the N-S equations at high-Reynolds numbers, physics-constrained learning turbulent fluid flows through a PDE based loss alone poses a difficult optimization objective.
Thus we will use a  semi-supervised extension of the reverse KL-divergence loss that allows both supervision with data as well as additional physics-constrained components.
Consider a training set of i.d.d. cases $\mathcal{D}=\left\{\bm{X}_{d},\bm{Y}_{d}\right\}_{d=1}^{D}$, then the loss is as follows:
\begin{equation}
\begin{aligned}
    \mathcal{L} &= \argmin_{\bm{\theta}}  \sum^{D}_{d=1} D_{KL}\left( p_{\bm{\theta}}\left(\bm{Y}_{d}|\bm{X}_{d}\right)|| p_{\bm{\beta}}\left(\bm{Y}_{d}|\bm{X}_{d}\right)\right) = \sum^{D}_{d=1}\mathbb{E}_{p_{\bm{\theta}}}\left[\log \frac{p_{\bm{\theta}}\left(\bm{Y}_{d}|\bm{X}_{d}\right)}{p_{\beta}\left(\bm{Y}_{d}|\bm{X}_{d}\right)}\right],
\end{aligned}
\end{equation}
in which we have made use of the fact that the KL divergence is additive for independent distributions.
$p_{\bm{\theta}}\left(\bm{Y}|\bm{X}\right)$ is the density of TM-Glow with parameters $\bm{\theta}$  for a single time-step.
$p_{\beta}\left(\bm{Y}|\bm{X}\right)$ is an energy based density function with a controllable parameter $\beta$ representing the true high-fidelity targets.
Note that the expectation is calculated using the samples of the generative model $\bm{y}\sim p_{\bm{\theta}}$, requiring a \textit{backward} pass of the INN which is the opposite direction of the standard maximum likelihood approach.

Currently this loss is posed across the entire time-series, however we desire it to be expressed in terms of single time-steps to make it computationally tractable with TM-Glow.
First, we pose the energy-based density as a product of independent distributions at each individual time-step $p_{\beta}\left(\bm{Y}|\bm{X}\right) = \prod_{n=1}^{N}p_{\beta}\left(\bm{y}^{n}|\bm{x}^{n}\right)$.
This is a similar form as the definition of the model's likelihood in~\Eqref{eq:rnn-likelihood}, $p_{\bm{\theta}}\left(\bm{Y}|\bm{X}\right) = \prod_{n=1}^{N}p_{\bm{\theta}}\left(\bm{y}^{n}|\bm{x}^{n},\bm{\tau}^{n-1}\right)$.
The loss for a time-series of $N$ time-steps can be written as:
\begin{equation}
    \mathcal{L} = \sum^{D}_{d=1}\sum_{n=1}^{N} \mathbb{E}_{p_{\bm{\theta}}}\left[\log{p_{\bm{\theta}}\left(\bm{y}^{n}_{d}|\bm{x}^{n}_{d},\bm{\tau}^{n-1}_{d}\right)} - \log{p_{\beta}\left(\bm{y}^{n}_{d}|\bm{x}^{n}_{d}\right)}\right].
\end{equation}
The first term is an entropy promoting term, $\log{p_{\bm{\theta}}\left(\bm{y}^{n}_{d}|\bm{x}^{n}_{d},\bm{\tau}^{n-1}_{d}\right)}$, encouraging diversity in the models samples and avoiding mode collapse.
A unique advantage using an INN is that the entropy, $H(\bm{y}^{n}_{d}|\bm{x}^{n}_{d},\bm{\tau}^{n-1}_{d}) = -\mathbb{E}_{ p_{\bm{\theta}}}\left[\log{p_{\bm{\theta}}\left(\bm{y}^{n}_{d}|\bm{x}^{n}_{d},\bm{\tau}^{n-1}_{d}\right)}\right]$, can be evaluated exactly though the change of variables in~\Eqref{eq:change-of-vars} as oppose to approximating~\cite{li2018learning, yang2019adversarial} or learning it~\cite{kumar2019maximum}.
The second term, the negative log energy density, $-\log{p_{\beta}\left(\bm{y}|\bm{x}\right)}$, encourages consistency between the models generated samples and the specified physical-constraints.
In this work, we use the Boltzmann distribution to model $p_{\beta}\left(\bm{y}^{n}_{d}|\bm{x}^{n}_{d}\right)$ which is standard in energy based models~\cite{lecun2006tutorial}:
\begin{equation}
    p_{\beta}\left(\bm{y}^{n}_{d}|\bm{x}^{n}_{d}\right) = \frac{\textrm{exp}\left(-\beta V_{PDE}(\cdot)\right)}{Z_{\beta}},
\end{equation}
in which $V_{PDE}(\cdot)$ is a PDE based potential discussed in further detail in Section~\ref{sec:pc-learning}.
$Z_{\beta}$ is a normalizing constant which does not impact the optimization and thus neglected.
$\beta$ is a tunable parameter that corresponds to the inverse temperature in the Boltzmann distribution that controls the strength of the potential in the backward KL loss.
The resulting form of the reverse KL divergence follows:
\begin{multline}
     \mathcal{L} = \sum^{D}_{d=1}\sum_{n=1}^{N}\mathbb{E}_{p_{\bm{\theta}}} \left[\log{p_{\bm{\theta}}\left(\bm{z}^{n}_{d}|\bm{x}^{n}_{d},\bm{\tau}^{n-1}_{d}\right)} + \sum_{k=1}^{K}\log{\left|\textrm{det}\left(\frac{\partial \bm{h}^{n}_{k,d}}{\partial \bm{h}^{n}_{k-1,d}}\right)\right|} + \beta V_{PDE}(\cdot)\right].
\end{multline}
In practice, the expectation in the KL divergences is taken as point estimate during training.
Due to the large number of times this loss is evaluated during the stochastic optimization of our model, the effects of such point estimates have been empirically shown to be minimal~\cite{kingma2013auto}. 

\subsection{Physics-Constrained Potential}
\label{sec:pc-learning}
\noindent
The potential $V_{PDE}$ represents imposed physical constraints one wishes to impose on the model's samples.
Similar to past physics-constrained literature~\cite{zhu2019physics, geneva2019modeling}, we will use the governing equations to aid the formulation of this potential.
Within this work, we pose $V_{PDE}$ in terms of three components:
\begin{gather}
    V_{PDE} = V_{Pres} + V_{Div} + V_{L2} + V_{RMS},\\
    V_{Pres} = \frac{v_{c}^{2}}{n_{s}}\norm{\frac{1}{\rho}\left(\frac{\partial^{2}\bm{p}^{n}}{\partial x^{2}} + \frac{\partial^{2}\bm{p}^{n}}{\partial y^{2}}\right) + \left(\frac{\partial \bm{u}_{x}^{n}}{\partial x}\right)^{2} + 2\frac{\partial \bm{u}_{x}^{n}}{\partial y}\frac{\partial \bm{u}_{y}^{n}}{\partial x} + \left(\frac{\partial \bm{u}_{y}^{n}}{\partial y}\right)^{2}}^{2}_{2},\\
    V_{Div} = \frac{v_{c}^{2}}{n_{s}}\norm{\frac{\partial \bm{u}_{x}^{n}}{\partial x} + \frac{\partial \bm{u}_{y}^{n}}{\partial y}}^{2}_{2}, \quad V_{L2} = \frac{1}{n_{s}}\norm{\bm{y}^{n} - \bm{y}_{HF}^{n}}_{2}^{2} \\
    V_{RMS} = \frac{1}{n_{s}}\norm{RMS\left(\bm{y}'\right) - RMS\left(\bm{y}'_{HF}\right) }_{2}^{2},
\end{gather}
which consists of the residual of the Poisson equation for pressure, the divergence free constraint for incompressible flow and two $L_2$ supervised learning terms. 
The first is between the predicted state variables of the model, denoted by $\bm{y}^{n}$, and the observed high-fidelity solution $\bm{y}^{n}_{HF}$.
The second is between the root-mean-square (RMS) of the fluctuation states predicted by the model, $\bm{y}'$, and the observed high-fidelity RMS values of $\bm{y}'_{HF}$.
This term can be interpreted as matching the turbulent intensity between the predicted time-series and the high-fidelity observables.
$n_{s}$ is the number of nodes in the predicted high-fidelity spatial domain.
Both residual loss terms are scaled by the cell volume, $v_{c} = \Delta x \cdot \Delta y$, to help balance each loss component.
While the potential resembles forms of other data and PDE based constrained loss functions~\cite{tompson2017accelerating, subramaniam2020turbulence}, ours is posed in probabilistic framework for learning the full distribution of solutions opposed to a single deterministic prediction.

The PDE residual terms are evaluated using the model's predictions and constrains the predictions to be physically realizable.
To evaluate the gradients we use the same methods successfully used in our past works for various physical systems~\cite{zhu2019physics, geneva2019modeling}.
Efficient finite difference based convolutions to approximate first-order gradients:
 \begin{equation}
     \frac{\partial \bm{u}^{n}}{\partial x} = \frac{1}{8\Delta x}\begin{bmatrix} 
    -1 & 0 & 1 \\
    -2 & 0 & 2 \\
    -1 & 0 & 1
    \end{bmatrix}\ast \bm{u}^{n}, \quad 
    \frac{\partial \bm{u}^{n}}{\partial y} =\frac{1}{8\Delta y}\begin{bmatrix} 
        -1 & -2 & -1 \\
        0 & 0 & 0 \\
        1 & 2 & 1
    \end{bmatrix}\ast \bm{u}^{n}, \\
 \end{equation}
 as well as second-order gradients:
 \begin{equation}
    \frac{\partial^{2} \bm{p}^{n}}{\partial x^{2}}=\frac{1}{4\Delta x^{2}}\begin{bmatrix} 
    1 & -2 & 1 \\
    2 & -4 & 2 \\
    1 & -2 & 1
    \end{bmatrix}\ast \bm{p}^{n}, \quad 
    \frac{\partial^{2} \bm{p}^{n}}{\partial y^{2}}=\frac{1}{4\Delta y^{2}}\begin{bmatrix} 
        1 & 2 & 1 \\
        -2 & -4 & -2 \\
        1 & 2 & 1
    \end{bmatrix}\ast \bm{p}^{n}. \\
 \end{equation}
These smoothed second-order accurate finite difference approximations are based on image processing filters such as the Sobel filter 2D convolutions~\cite{sobel19683x3} which have been found to improve training stability over pure finite-difference calculations.
The convolutional filter approach allows for efficient computation of these gradients during training that directly integrates itself into the computational graph for back-propagation.
In this work, since we are predicting a sub-domain for which we do not know the complete boundary conditions, we only compute the PDE constraint terms on the deep nodes of the predicted domain ignoring the boundary values. 
The supervised $L_2$ terms help stabilize the PDE based losses which can be unstable due to their gradients and encouraging turbulence in the predicted fluid flow.
Similar supervised losses are used in GAN models for time-series predictions to increase time-series accuracy and continuity~\cite{xiong2018learning, zhao2018learning}.
Pseudocode for the training process is outlined in Algorithm~\ref{algo:training} for a single training case but easily extends to a full training data-set.
The pseudocode for sampling of TM-Glow is also outlined in Algorithm~\ref{algo:sampling}, from which statistics are then computed in traditional Monte Carlo fashion.
\begin{algorithm}
    \caption{Training TM-Glow for a single training case.}
    \label{algo:training}
    \KwIn{TM-Glow model: $f_{\bm{\theta}}$; Low-fidelity and high-fidelity time-series data $\left\{\bm{X},\bm{Y}\right\}=\left\{\bm{x}^{n},\bm{y}_{HF}^{n}\right\}^{N}_{n=1}$ of length $N$; Number of epochs: $M$; Back-propagation through time interval: $p$; Learning rate: $\eta$}
    $\overline{\bm{y}} \approx (1/n)\sum_{n=1}^{N} \bm{y}^{n}_{HF}$ \Comment*[r]{Approx. mean flow field}
    
    \For{$\textrm{epoch} = 1$ \KwTo $M$}{
        $\bm{\tau}^{0} \sim p\left(\bm{\tau}^{0}\right)$ \Comment*[r]{Sample initial recurrent state}
        \For{$n = 1$ \KwTo $N$}{
            $\bm{y}^{n}, \bm{\tau}^{n}, \log{p(\bm{y}^{n}|\bm{x}^{n}, \bm{\tau}^{n-1})} \leftarrow f_{\bm{\theta}}^{-1}\left(\bm{x}^{n}, \bm{\tau}^{n-1}\right)$ \Comment*[r]{Sample TM-Glow}
            $V_{Pres}(\bm{y}^{n}) = (v_{c}^{2}/n_{s})\norm{\triangle \bm{p}^{n} + \nabla\cdot\left((\bm{u}^{n}\cdot\nabla)\bm{u}^{n}\right)}_{2}^{2}$ \Comment*[r]{Poisson residual}
            $V_{Div}(\bm{y}^{n}) = 	(v_{c}^{2}/n_{s})\norm{\nabla \cdot \bm{u}^{n}}_{2}^{2}$ \Comment*[r]{Divergence residual}
            $V_{L2} = (1/n_{s})\norm{\bm{y}^{n} - \bm{y}_{HF}^{n}}^{2}_{2}$ \Comment*[r]{L2 Loss}
            $\mathcal{L} \mathrel{+}= \log{p(\bm{y}^{n}|\bm{x}^{n}, \bm{\tau}^{n-1})} + \beta\left(V_{Pres}+V_{Div}+V_{L2}\right)$ \Comment*[r]{Backward KL}
            
            \If{Mod(n,p)=0}{ 
                $\bm{y}' = \bm{y}^{n-p:n} - \overline{\bm{y}}$ \Comment*[r]{Approx. TM-Glow fluctuation fields}
                $\mathcal{L} \mathrel{+}=  (\beta p/n_{s})\norm{RMS\left(\bm{y}'\right) - RMS\left(\bm{y}'_{HF}\right)}_{2}^{2}$ \Comment*[r]{RMS loss}
                $\nabla \bm{\theta} \leftarrow  \textrm{Backprop}(\mathcal{L})$ \Comment*[r]{Back-propagation}
                $\bm{\theta} \leftarrow \bm{\theta} - \eta \nabla \bm{\theta}$ \Comment*[r]{Gradient Descent}
                 $\mathcal{L} = 0$ \Comment*[r]{Zero loss}
            }
        }
        $\overline{\bm{y}} = (1/n)\sum_{n=1}^{N} \bm{y}^{n}$ \Comment*[r]{Update mean flow field estimate}
    }
    \KwOut{Trained TM-Glow model $f_{\bm{\theta}}$;}
\end{algorithm}
\begin{algorithm}
    \caption{Sampling TM-Glow high-fidelity time-series.}
    \label{algo:sampling}
    \KwIn{Trained TM-Glow model: $f_{\bm{\theta}}$; Low-fidelity time-series data $\left\{\bm{X}\right\}=\left\{\bm{x}^{n}\right\}^{N}_{n=1}$ of length $N$; Number of samples: $M$}
    
    \For{$m = 1$ \KwTo $M$}{
        $\bm{\tau}^{0} \sim p\left(\bm{\tau}^{0}\right)$ \Comment*[r]{Sample initial recurrent state}
        \For{$n = 1$ \KwTo $N$}{
            $\bm{y}^{n}, \bm{\tau}^{n} \leftarrow f_{\bm{\theta}}^{-1}\left(\bm{x}^{n}, \bm{\tau}^{n-1}\right)$ \Comment*[r]{Sample time-step from TM-Glow}
        }
        $\bm{Y}^{m} = \left\{\bm{y}^{1},\bm{y}^{2},...,\bm{y}^{N}\right\}$ \Comment*[r]{Store sampled time-series}
    }
    \KwOut{High-fidelity flow samples: $\bm{Y}^{1:M}$}
\end{algorithm}

\subsection{Hyper-parameter Tuning}
\label{sec:hyper-tuning}
\begin{figure}[H]
    \centering
    \includegraphics[width=0.8\textwidth]{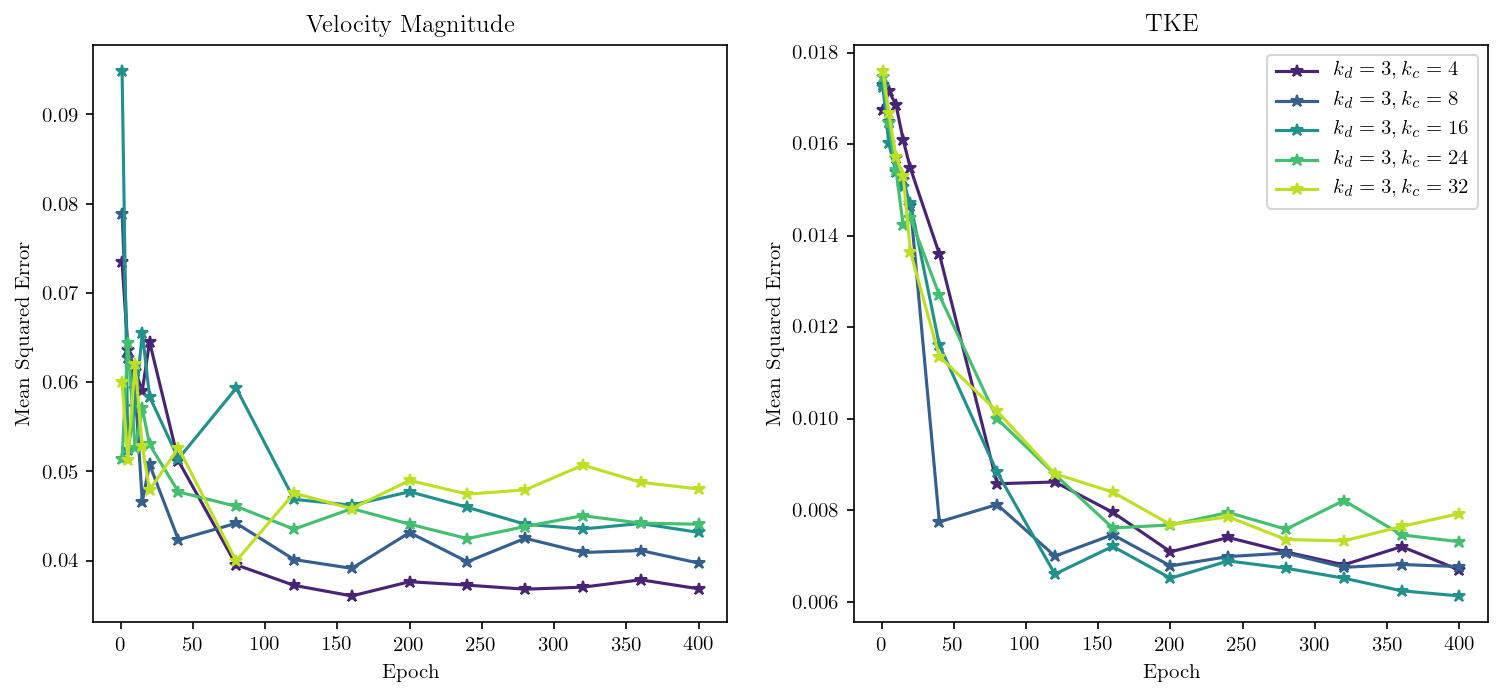}
    \caption{(Left to right) Velocity magnitude MSE and turbulent kinetic energy (TKE) test MSE for TM-Glow models containing $k_{d}\cdot k_{c}$ affine coupling layers.}
    \label{fig:model-depth}
\end{figure}

\noindent
TM-Glow contains a large set of hyper-parameters including model depth, the number of affine coupling layers, coupling neural network depth, learning rate, mini-batch size, etc. which are all coupled together making an extensive hyper-parameter search extremely difficult.
While automated methods exist to aid this search, we opted to take a simpler approach by empirically finding a reasonable model architecture that fits the desired needs of the problems of interest (e.g. predictive capability, memory consumption of the model, stability, etc.) and do a more extensive search on ones deemed more important.
First various model depths are tested by adjusting the number of affine coupling layers in each LSTM affine block, denoted by $k_{c}$ in Fig.~\ref{fig:affine-blocks}.
Each model is trained on a small data-set ($32$ flows from the second numerical example in Section~\ref{sec:cylinder-array}) to try to keep the computational cost of the hyperparameter search reasonable.
To quantify the accuracy of each model, the following time-averaged prediction mean squared errors (MSE) are used for a validation set of $n_{test}=16$ flows:
\begin{gather}
    MSE_{Mag} = \frac{1}{n_{s}n_{test}}\norm{\mathbb{E}_{p_{\bm{\theta}}}\left[\left|\overline{\bm{u}}\right|\right] - \left|\overline{\bm{u}}_{HF}\right|}_{2}^{2}, \quad \left|\overline{\bm{u}}\right| = \frac{1}{T}\int_{0}^{T} \sqrt{\bm{u}^{2}_{x}(t) + \bm{u}^{2}_{y}(t)} dt, \label{eq:mse-mag}\\
    \begin{gathered}MSE_{TKE} = \frac{1}{n_{s}n_{test}}\norm{\mathbb{E}_{p_{\bm{\theta}}}\left[\bm{k}\right] - \bm{k}_{HF}}_{2}^{2}, \quad  \bm{k} = \frac{1}{2}\left(\overline{\left(\bm{u}^{'}_{x}\right)^{2}} + \overline{\left(\bm{u}^{'}_{y}\right)^{2}}\right), \label{eq:mse-tke}\end{gathered}
\end{gather}
in which the expected value of the model's prediction is estimated using $20$ model samples.
The first error assesses the accuracy of the mean flow magnitude, the second assesses the accuracy of the predicted turbulent kinetic energy (TKE).
The test error of the models considered are plotted in Fig.~\ref{fig:model-depth}.
We find that there is a trade off between average velocity and turbulent energy accuracy and larger models begin to over fit on this small training data-set.
Based on these results, we select a TM-Glow model with $k_{d}=3$ and $k_{c}=16$.
Essential TM-Glow and training hyper-parameters are outlined in Table~\ref{tab:tm-glow-paramters}.
The resulting model contains $1.7$ million learnable parameters.
Back-propagation through time (BPTT)~\cite{goodfellow2016deep} occurs at $10$ time-step intervals due to the memory constraints of the GPU used to train the model.
Although ideally the loss should be calculated for the entire time-series with a single back-propagation, this results in a large computational graph that requires significant memory on the GPU making it not practical.
An observed side effect of using smaller BPTT intervals is the convergence of the recurrent states for later time-steps resulting in predicted fields converging as well to a similar prediction for all samples.
To prevent this the initial recurrent state, $\bm{\tau}^{0}$, is averaged with the current recurrent state after every BPTT.
This borrows the idea of using various recurrent features at different timescales in hierarchical RNNs for neural language processing~\cite{liu2015multi, chung2016hierarchical}.
Although not necessary, this algorithmic heuristic helps prevent the information from the initial state being lost which was found to improve the model's accuracy and sample diversity.
For additional details, we direct the reader to the source code.
\begin{table}[H]
\centering
\caption{TM-Glow model and training parameters used for both numerical test cases. For the parameters that vary between test cases the superscript $\dagger$ and $\ddagger$ to denote numerical examples in Sections~\ref{sec:backwards-step} and~\ref{sec:cylinder-array}, respectively. Hyper-parameter differences are due to memory constraints imposed from the varying predictive domain sizes.}
\label{tab:tm-glow-paramters}
\begin{tabular}{llll}
TM-Glow                   & \multicolumn{1}{l|}{}          & Training        &                                                                      \\ \hline
Model Depth, $k_{d}$               & \multicolumn{1}{l|}{$3$}     & Optimizer       & ADAM~\cite{kingma2014adam}                                         \\
Conditional Features, $\bm{\xi}^{(i)}$      & \multicolumn{1}{l|}{$32$}      & Weight Decay    & $1e-6$                                                               \\
Recurrent Features, $\bm{a}^{(i)}_{in}, \bm{c}^{(i)}_{in}$        & \multicolumn{1}{l|}{$64, 64$}      & Epochs          & $400$                                                                \\
Affine Coupling Layers, $k_{c}$ & \multicolumn{1}{l|}{$16$} & Mini-batch Size & $32^{\dagger}, 64^{\ddagger}$                      \\
Coupling NN Layers          & \multicolumn{1}{l|}{$2$}       & BPTT  & $10$ time-steps   \\
                          &  \multicolumn{1}{l|}{}  &  Inverse Temp.,  $\beta$ & 200                      
\end{tabular}
\end{table}

\begin{figure}[H]
    \centering
    \includegraphics[width=\textwidth]{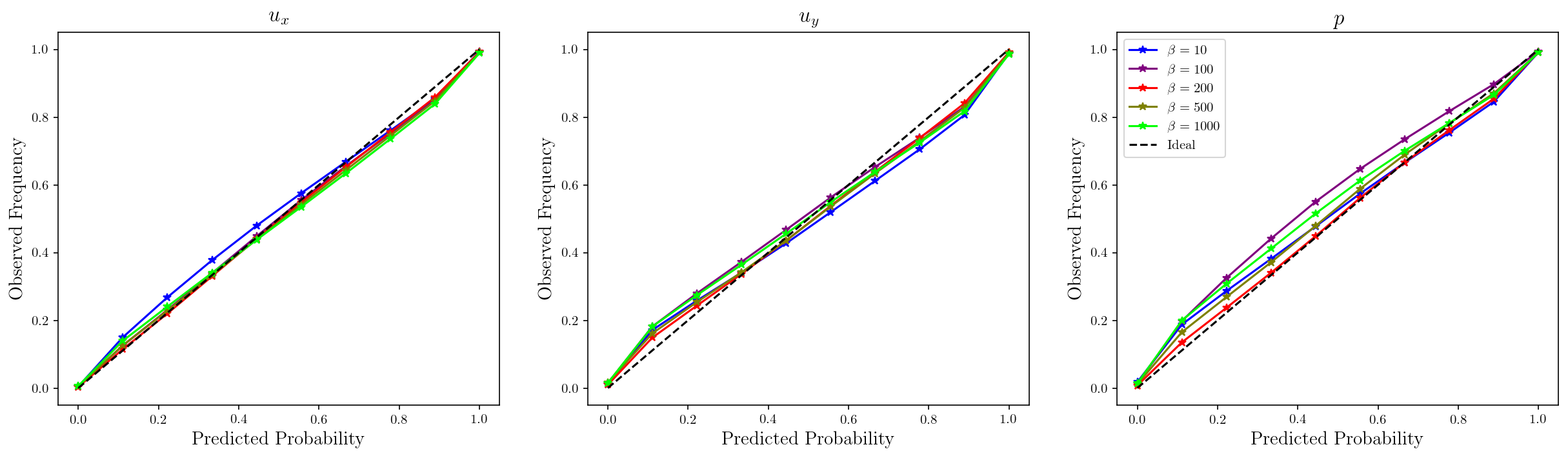}
    \caption{Reliability diagrams of the predicted x-velocity, y-velocity and pressure fields predicted with TM-Glow evaluated over $12000$ model predictions. The black dashed line indicates matching empirical distributions between the model's samples and observed validation data.}
    \label{fig:hyper-reliability}
\end{figure}

The inverse temperature parameter, $\beta$, in the energy density controls the balance between the model satisfying the physics-based potential and the model's entropy.
Given this parameter close relation to the model's probabilistic nature, reliability diagrams of the model's predictions are used to assess the predictive uncertainty quality.
Several models with different $\beta$ values are trained on the same small data-set of $32$ flows from the second numerical example in Section~\ref{sec:cylinder-array} used when calibrating the model depth.
For a small validation data-set of $16$ flows, for each model we compute the empirical density function for each of the model's output fields over all samples, time-steps and validation cases at each spatial location independently.
The values of the predicted density function at several quantiles are then compared to the empirical density function of the high-fidelity data which is then averaged over the spatial domain and plotted in Fig.~\ref{fig:hyper-reliability} for each state variable.
Interestingly, unlike Zhu~\etal~\cite{zhu2019physics}, the predicted quantiles all match fairly well with the high-fidelity data with apparently little sensitivity to $\beta$.
Based on these results we selected $\beta=200$.
\begin{remark}
For each numerical example, additional fine tuning is certainly possible to obtain the highest level of accuracy.
However, in this work we will not be performing any case specific tuning to demonstrate that decent results can be obtained using TM-Glow for multiple problems of different nature and dimensionality.
\end{remark}

\subsection{Ablation Study}
\label{sec:ablation}
\noindent
An ablation study is performed to investigate the impact each loss component has on the model's predictive accuracy.
Additionally, the model is also trained using the standard maximum likelihood approach for INNs, by maximizing~\Eqref{eq:cond-log-likelihood}, to act as the traditional base-line.
The same training/validation data-set used for the accuracy and uncertainty calibration studies was also used here.
As listed in Table~\ref{tab:ablation}, we train several models using variants of the propose backward KL loss and compute the mean squared error of various flow-field quantities across the validation data-set.
Again, $20$ model samples are used to compute the expected value of each predicted flow quantity from which the error is computed.

First we note that training the model through the traditional maximum likelihood approach generally yields worse results than the backwards KL losses with the exception of  some of the time-averaged mean flow quantities.
Additionally, the large residual errors for the maximum likelihood training indicated that the instantaneous flow fields are non-physical.
Interestingly, the proposed loss does not produce the most accurate mean flow or turbulent statistics.
This appears to be due to the inclusion of the Poisson pressure residual loss, which enforces physical coupling of the output fields.
Without this PDE loss, the model has more freedom and can achieve greater accuracy of the flow statistics.
However, this comes at the cost of having nonphysical instantaneous flow field realizations which is indicated by the increase in the pressure residual.
Given that we are interested in predicting physical fluid flow, we believe that inclusion of the Poisson residual is essential even at the sacrifice of the time-average statistics.

\newcommand{\xmark}{\ding{55}}%
\begin{table}[H]
\caption{Ablation study of the impact of different parts of the backward KL loss. As a base-line we also train TM-Glow using the standard maximum likelihood estimation (MLE) approach. The mean square error (MSE) of various flow field quantities for various loss formulations are listed. The lowest values for each error are bolded. }
\label{tab:ablation}
\resizebox{\textwidth}{!}{%
\begin{tabular}{ccccccccccccc}
\multicolumn{1}{l}{MLE} & \multicolumn{1}{l}{$V_{Pres}$} & \multicolumn{1}{l}{$V_{Div}$} & \multicolumn{1}{l}{$V_{L2}$} & \multicolumn{1}{l|}{$V_{RMS}$}  & $MSE\left(\overline{\bm{u}}_{x}\right)$ & $MSE\left(\overline{\bm{u}}_{y}\right)$ & $MSE\left(\overline{\bm{p}}\right)$ & $MSE\left(\sqrt{\overline{\left(\bm{u}^{'}_{x}\right)^{2}}}\right)$ & $MSE\left(\sqrt{\overline{\left(\bm{u}^{'}_{y}\right)^{2}}}\right)$ & $MSE\left(\sqrt{\overline{\left(\bm{p}^{'}\right)^{2}}}\right)$ & $\overline{V_{Div}}$ & $\overline{V_{Pres}}$ \\ \hline
\checkmark  & \xmark & \xmark & \xmark  & \multicolumn{1}{c|}{\xmark} & 0.0589 & 0.0085 &  \textbf{0.0135} & 0.0204 & 0.0486 & 0.0137 & 0.0019 & 0.0615\\
\xmark  & \checkmark & \checkmark & \checkmark  & \multicolumn{1}{c|}{\checkmark} & 0.0490 & 0.0115 &  0.0188 & 0.0168 & 0.0292 & 0.0125 & 0.0012 & \textbf{0.0192}\\
\xmark  & \xmark & \checkmark & \checkmark  & \multicolumn{1}{c|}{\checkmark} & \textbf{0.0390} & \textbf{0.0078} &  0.0189 & \textbf{0.0162} & \textbf{0.0251} & \textbf{0.0106} & 0.0013 & 0.0402\\
\xmark  & \xmark & \xmark & \checkmark  & \multicolumn{1}{c|}{\checkmark} & 0.0463 & 0.0113 & 0.0158 & 0.0166 & 0.0256 & 0.0129 & 0.0012 & 0.0424 \\
\xmark  & \xmark & \xmark & \checkmark  & \multicolumn{1}{c|}{\xmark} & 0.0435 & 0.0089 &  0.0140 & 0.0168 & 0.0272 & 0.0131 & 0.0012 & 0.0366 
\end{tabular}}
\end{table}

\section{Turbulent Flow over a Backwards Step}
\label{sec:backwards-step}
\begin{figure}[H]
    \centering
    \includegraphics[width=\textwidth]{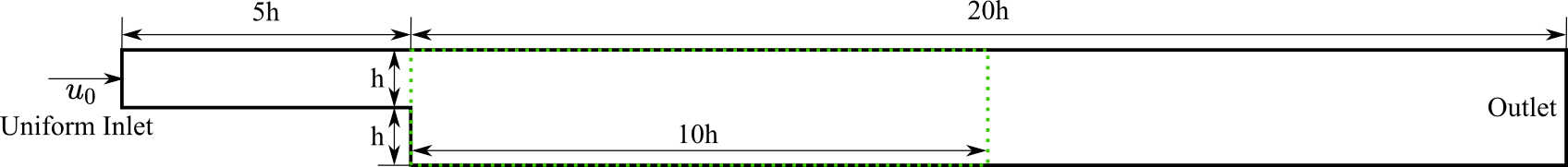}
    \caption{Flow over a backwards step. The green region indicates the recirculation region TM-Glow will be used to predict. All domain boundaries are no-slip with the exceptions of the uniform inlet and zero gradient outlet. The total outlet simulation length is made to be double that of the prediction range to negate effects of the boundary condition on this zone.}
    \label{fig:backward-step-schematic}
\end{figure}

\begin{figure}[H]
    \centering
    \begin{subfigure}{0.48\textwidth}
        \includegraphics[trim={0 10cm 0 10cm},clip, width=\textwidth]{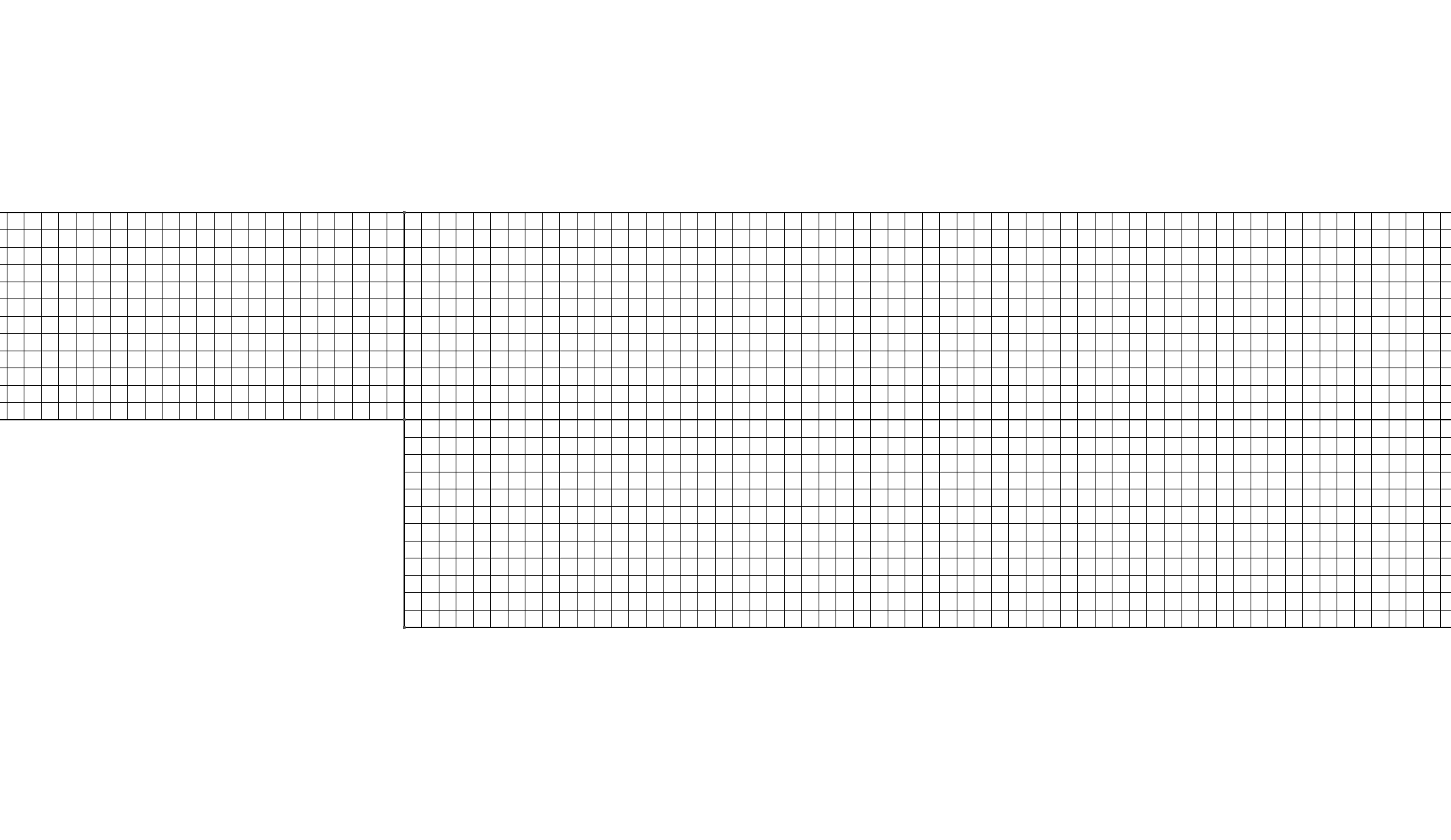}
        \caption{Low-fidelity}
    \end{subfigure}
    ~
    \begin{subfigure}{0.48\textwidth}
        \includegraphics[trim={0 10cm 0 10cm},clip, width=\textwidth]{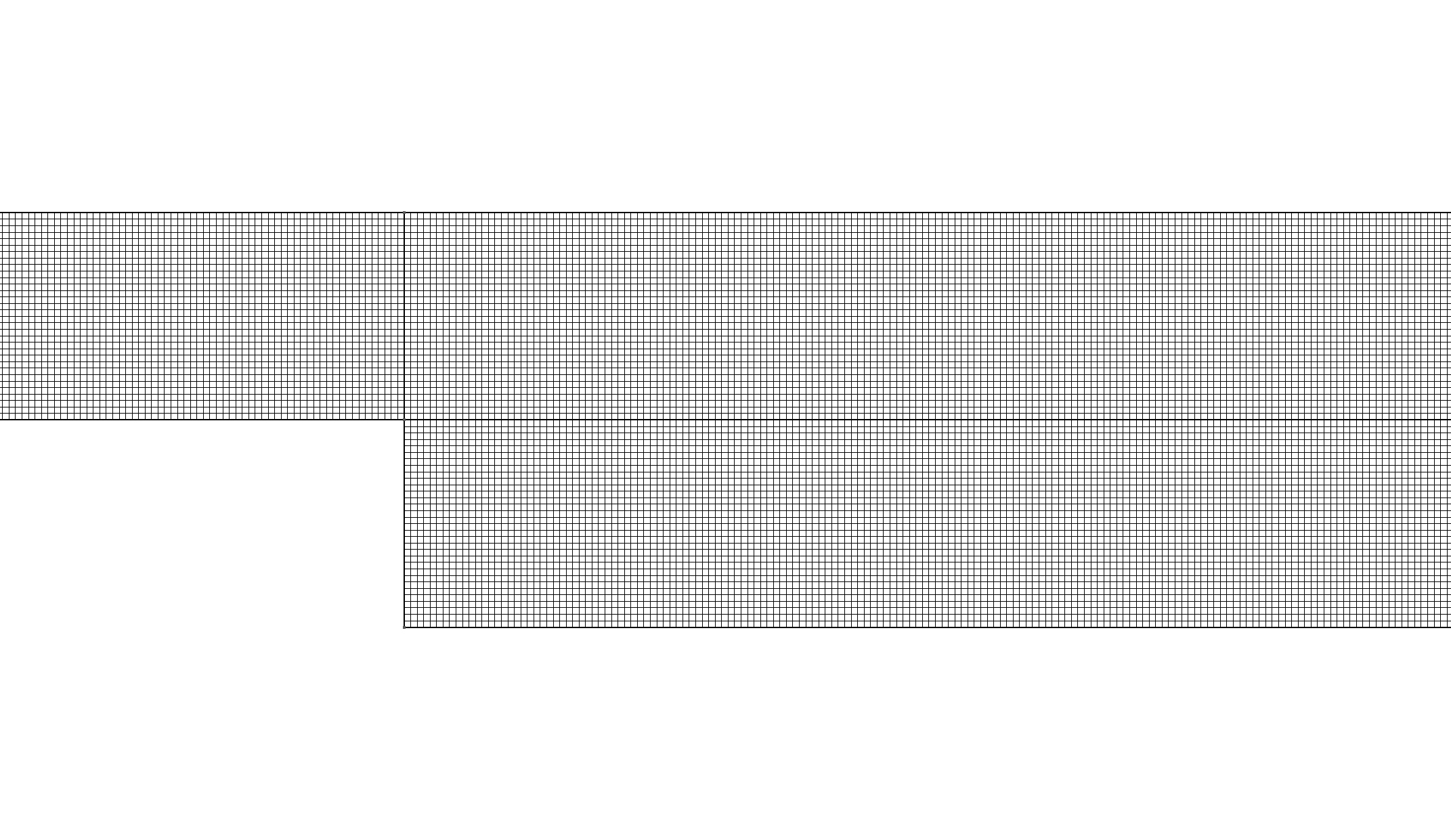}
        \caption{High-fidelity}
    \end{subfigure}
    \caption{Computational mesh around the backwards step used for the low- and high-fidelity CFD simulations solved with OpenFOAM~\cite{jasak2007openfoam}.}
    \label{fig:backward-step-mesh}
\end{figure}

\noindent
We first apply the proposed model to surrogate modeling turbulent flow over a backward step at different Reynolds numbers, a classical benchmark problem in computational fluids.
As illustrated in Fig.~\ref{fig:backward-step-schematic}, the feature of interest is the flow separation that occurs following the step.
Such phenomena can be found in a surprisingly large number of systems including heat exchanges, flow around buildings, combustion engines and aerodynamic elements~\cite{erturk2008numerical, chen2018review}.
The Reynolds number of the flow is governed by the inlet velocity $u_{0}$, viscosity $\nu=0.0002$ and the height of the step $h=1$. 
In this benchmark, the inlet boundary condition is varying in magnitude and thus varying the Reynolds number of the flow.
Here we are interested in predicting the recirculation region, marked by the green box in Fig.~\ref{fig:backward-step-schematic}, for different Reynolds number.
This region is the typical area of study for this flow due the presence of flow separation, Kevin-Helmholtz instability and turbulent flow with various eddy formations.
The low-fidelity simulator that will be the input of the model has a mesh characteristic resolution of $l_{c} = h/12$ and the target high-fidelity field has a resolution of $l_{c} = h/32$ as shown in Fig.~\ref{fig:backward-step-mesh}.
Both the velocity and pressure fields are normalized by the inlet velocity and an additional constant input field set to the inlet velocity is added.
The resulting model input for a single time-step is $\bm{x}^{n}=\left\{\bm{u}^{n}_{l}/u_{0}, \bm{p}_{l}^{n}/u^{2}_{0}, \bm{u}_{0}\right\} \in \mathbb{R}^{4,24,120}$ with an output $\bm{y}^{n}=\left\{\bm{u}^{n}_{h}, \bm{p}_{h}^{n}\right\} \in \mathbb{R}^{3,64,320}$ with a time-step size of $\Delta t =0.5$.
The full training data set consists of fluid flows evenly distributed between Reynolds number $5000$ to $50000$ each consisting of $80$ time-steps.
Simulations were performed using the OpenFOAM finite volume solver using standard Smagorinsky LES sub-grid scale models~\cite{jasak2007openfoam}.
During training we augment these time-series by splitting them in half into two time-series of $40$ time-steps to artificially create more flow training cases.
Further details on the computational cost of the low-fidelity and high-fidelity simulations along with the training of TM-Glow are discussed in Section~\ref{sec:computation}.

\begin{figure}[H]
    \centering
    \includegraphics[width=0.7\textwidth]{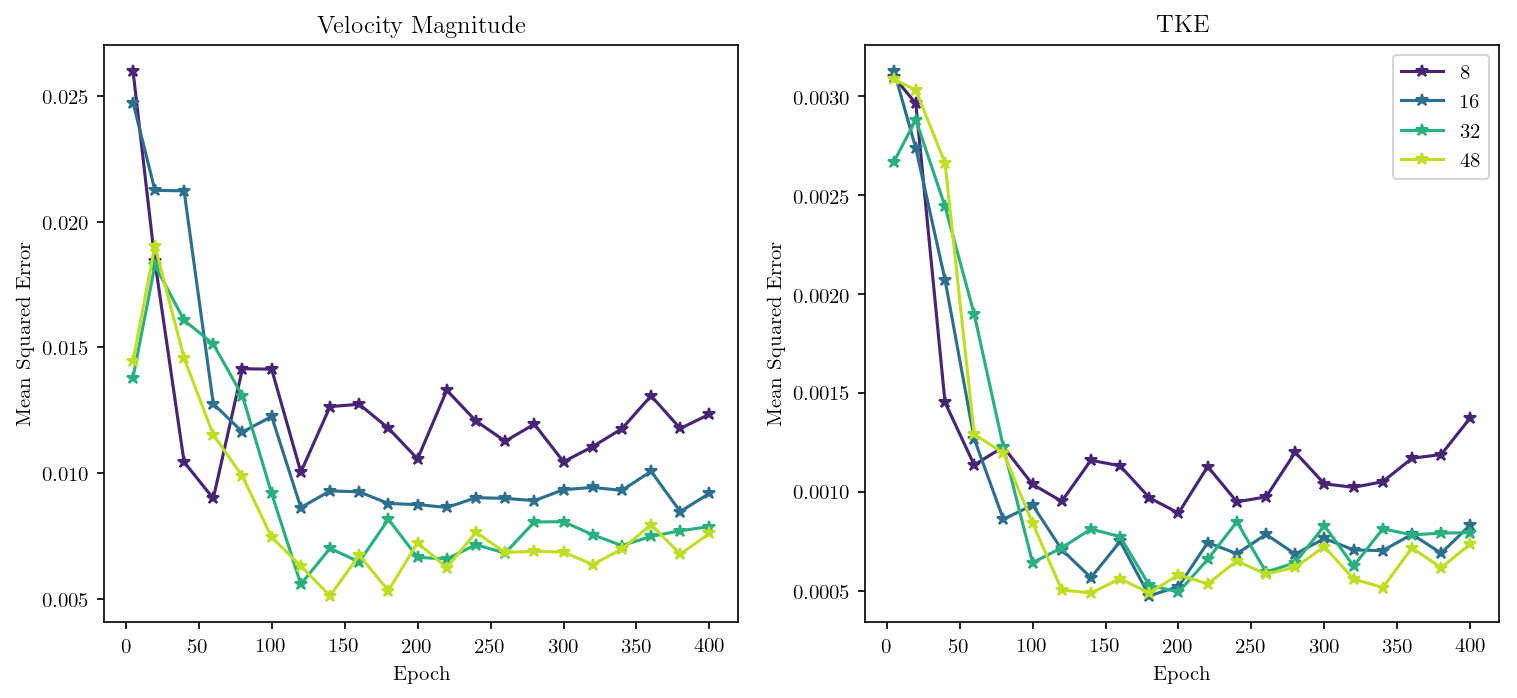}
    \caption{(Left to right) Flow over backwards step velocity magnitude and turbulent kinetic energy (TKE) error during training of TM-Glow on different data set sizes. Error values were average over five model samples.}
    \label{fig:bstep-training}
\end{figure}

A test set of $17$ flows with evenly spaced Reynolds numbers between $[7500,47500]$ are used to evaluate the performance of TM-Glow.
Four models are trained on $8$, $16$, $32$ and $48$ flows.
The test MSE error of the velocity magnitude and TKE, defined in~\Eqref{eq:mse-mag} and~\Eqref{eq:mse-tke}, during training is plotted in Fig.~\ref{fig:bstep-training}.
The test errors of various flow field quantities are listed in Table~\ref{tab:bstep-error} along with the error obtained from naively interpolating the low-fidelity solution to the high-fidelity mesh.
TM-Glow is able to produce time-average statistics that are far more accurate than the low-fidelity solution as expected.
\begin{table}[H]
\caption{Backwards step test error of various normalized time-averaged flow field quantities of the low-fidelity solution interpolated to the high-fidelity mesh and TM-Glow trained on various training data set sizes. Lower is better. TM-Glow errors were averaged over $20$ samples from the model. The training wall-clock (WC) time of each data set size is also listed.}
\label{tab:bstep-error}
\resizebox{\textwidth}{!}{%
\begin{tabular}{cccccccc}
  & $MSE\left(\overline{\bm{u}}_{x}/u_{0}\right)$ & $MSE\left(\overline{\bm{u}}_{y}/u_{0}\right)$ & $MSE\left(\overline{\bm{p}}/u^{2}_{0}\right)$ & $MSE\left(\sqrt{\overline{\left(\bm{u}^{'}_{x}\right)^{2}}}/u_{0}\right)$ & $MSE\left(\sqrt{\overline{\left(\bm{u}^{'}_{y}\right)^{2}}}/u_{0}\right)$ & $MSE\left(\sqrt{\overline{\left(\bm{p}^{'}\right)^{2}}}/u^{2}_{0}\right)$ & WC Hrs. \\ \hline
\multicolumn{1}{c|}{Low-Fidelity} & 0.1212 & 0.0224 & 0.0199 & 0.0237 & 0.0177 & 0.0124 & - \\
\multicolumn{1}{c|}{$8$ Flows} & 0.0182 & 0.0036 & 0.0023 & 0.0053 & 0.0059 & 0.0034 & 6.5 \\
\multicolumn{1}{c|}{$16$ Flows} & 0.0185 & 0.0031 & 0.0021 & 0.0030 & 0.0033 & 0.0023 & 10.0 \\
\multicolumn{1}{c|}{$32$ Flows} & 0.0091 & 0.0019 & 0.0014 & 0.0022 & 0.0022 & 0.0014 & 12.1 \\
\multicolumn{1}{c|}{$48$ Flows} & 0.0074 & 0.0017 & 0.0014 & 0.0021 & 0.0022 & 0.0013 & 16.6 
\end{tabular}}
\end{table}

\begin{figure}[H]
    \centering
    \begin{subfigure}{\textwidth}
        \includegraphics[width=\textwidth]{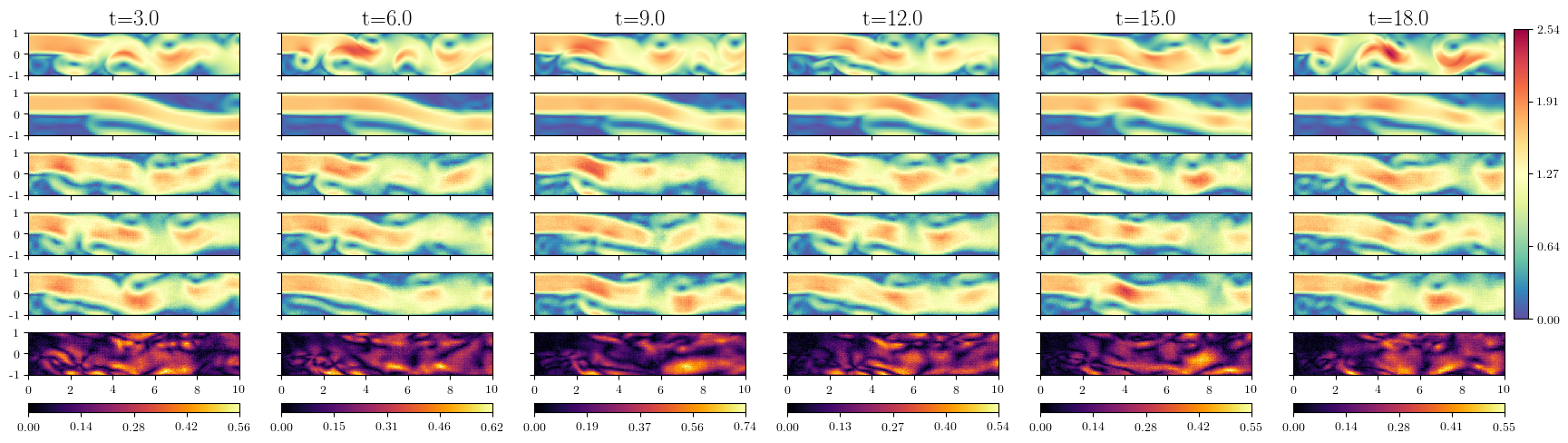}
        \caption{$Re=7500$}
    \end{subfigure}\\
    \begin{subfigure}{\textwidth}
        \includegraphics[width=\textwidth]{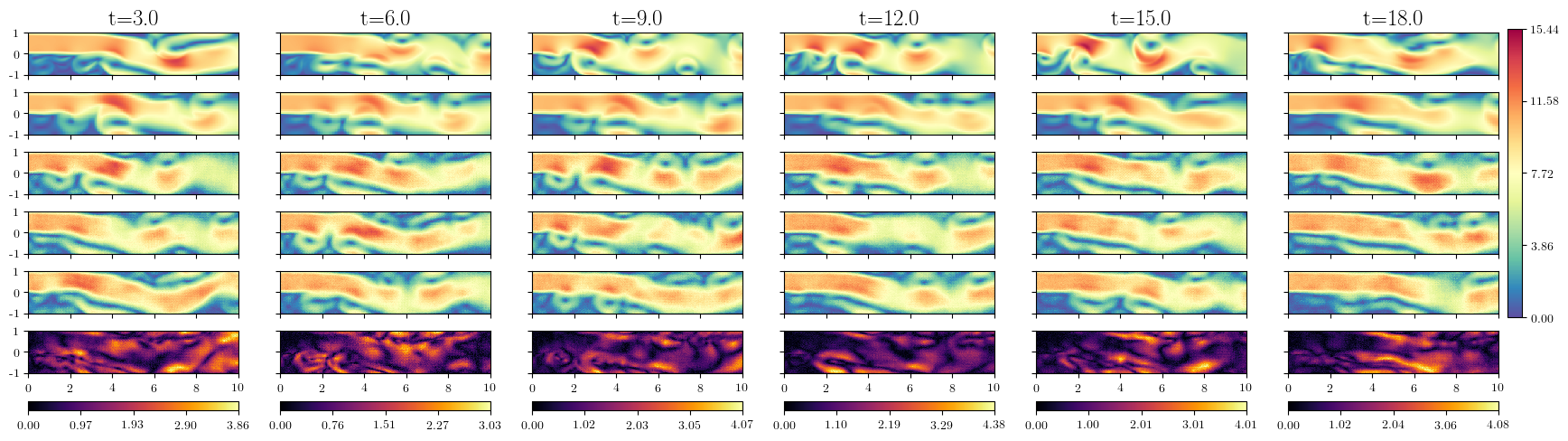}
        \caption{$Re=47500$}
    \end{subfigure}
    \caption{(Top to bottom) Velocity magnitude of the high-fidelity target, low-fidelity input, $3$ TM-Glow samples and standard deviation for two test flows.}
    \label{fig:bstep-vmag-1}
\end{figure}

\begin{figure}[H]
    \centering
    \begin{subfigure}{\textwidth}
        \includegraphics[width=\textwidth]{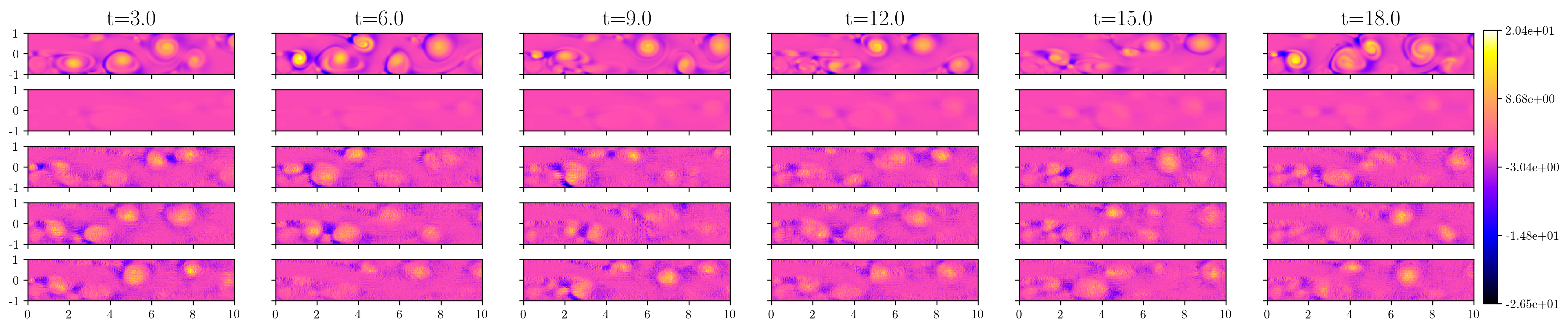}
        \caption{$Re=7500$}
    \end{subfigure}\\
    \begin{subfigure}{\textwidth}
        \includegraphics[width=\textwidth]{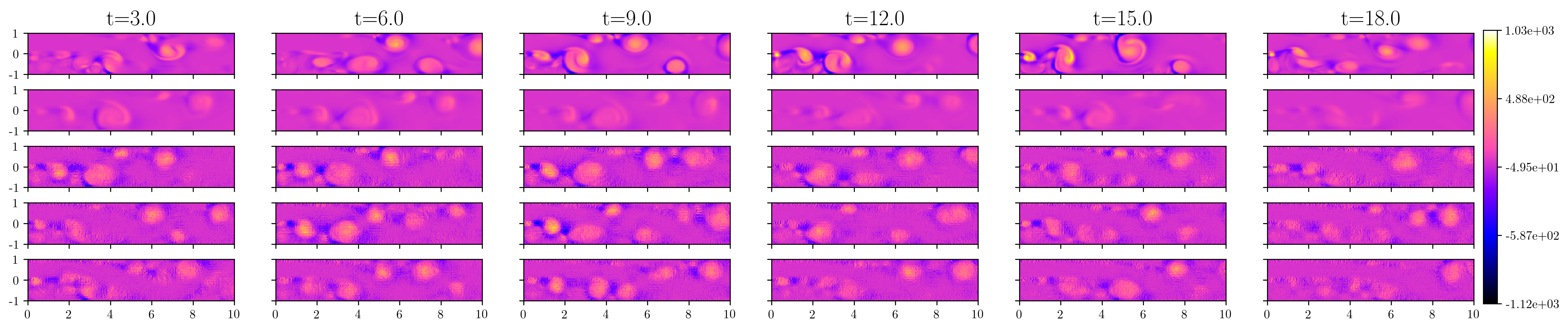}
        \caption{$Re=47500$}
    \end{subfigure}
    \caption{(Top to bottom) Q-criterion of the high-fidelity target, low-fidelity input and three TM-Glow samples for two test flows.}
    \label{fig:bstep-qcrit-1}
\end{figure}

\begin{figure}[H]
    \centering
    \begin{subfigure}{0.9\textwidth}
        \centering
        \includegraphics[width=\textwidth]{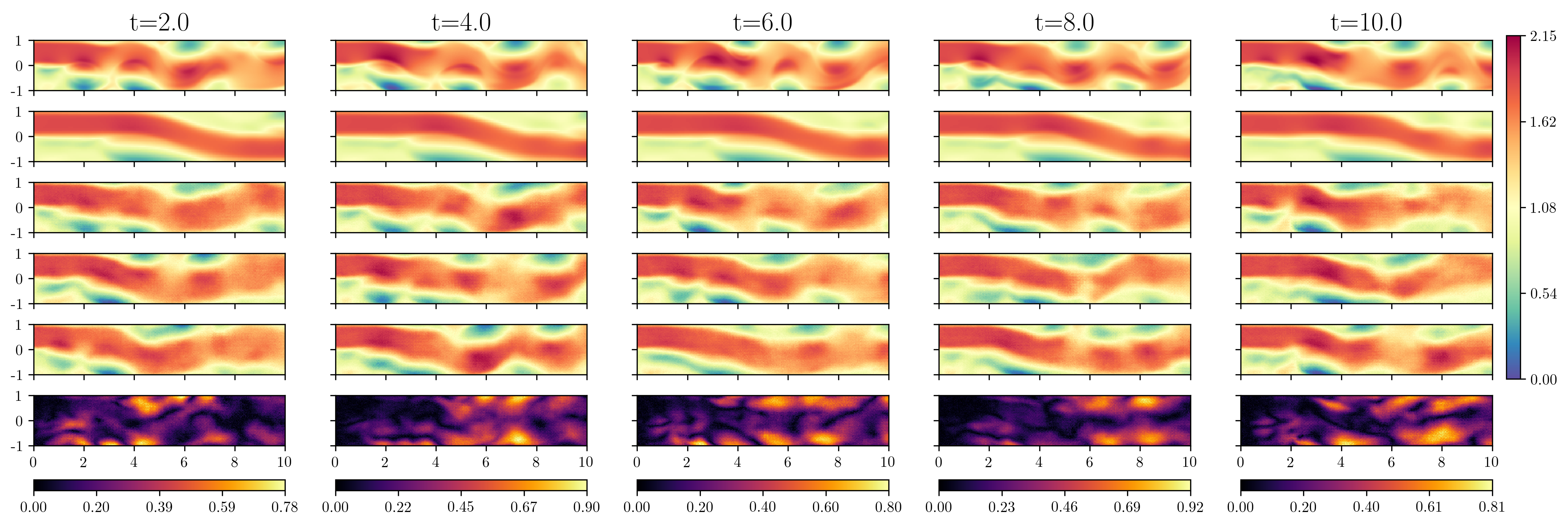}
        \caption{X-velocity}
    \end{subfigure}\\
    \begin{subfigure}{0.9\textwidth}
        \centering
        \includegraphics[width=\textwidth]{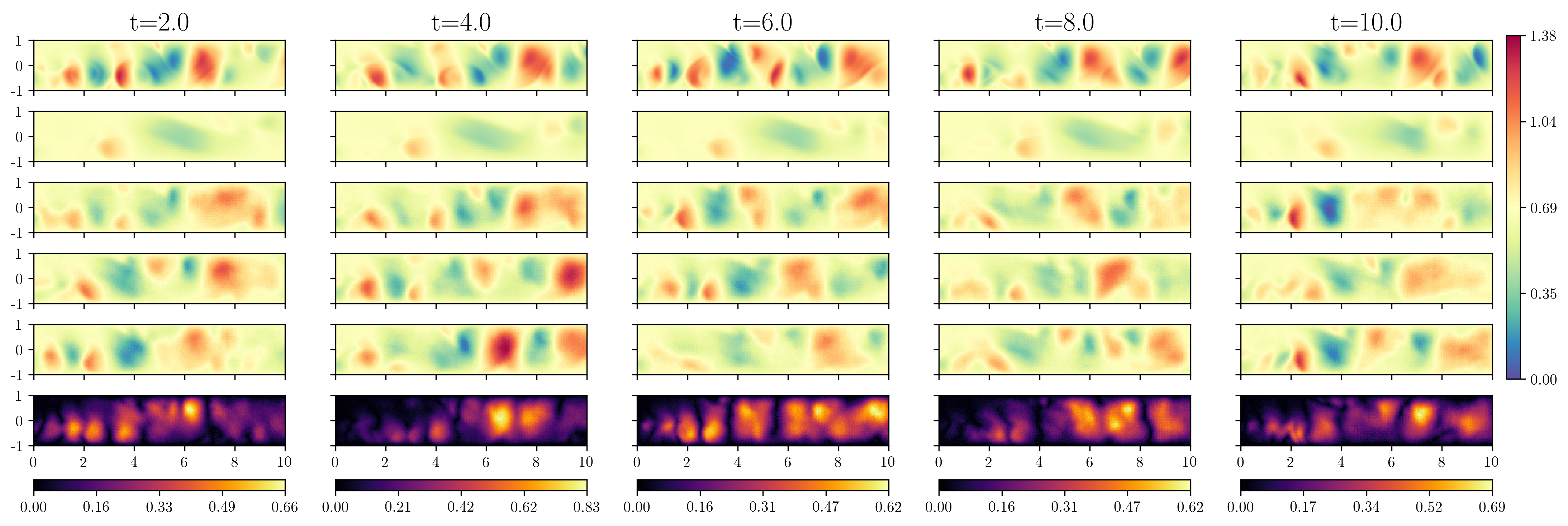}
        \caption{Y-velocity}
    \end{subfigure}\\
    \begin{subfigure}{0.9\textwidth}
        \centering
        \includegraphics[width=\textwidth]{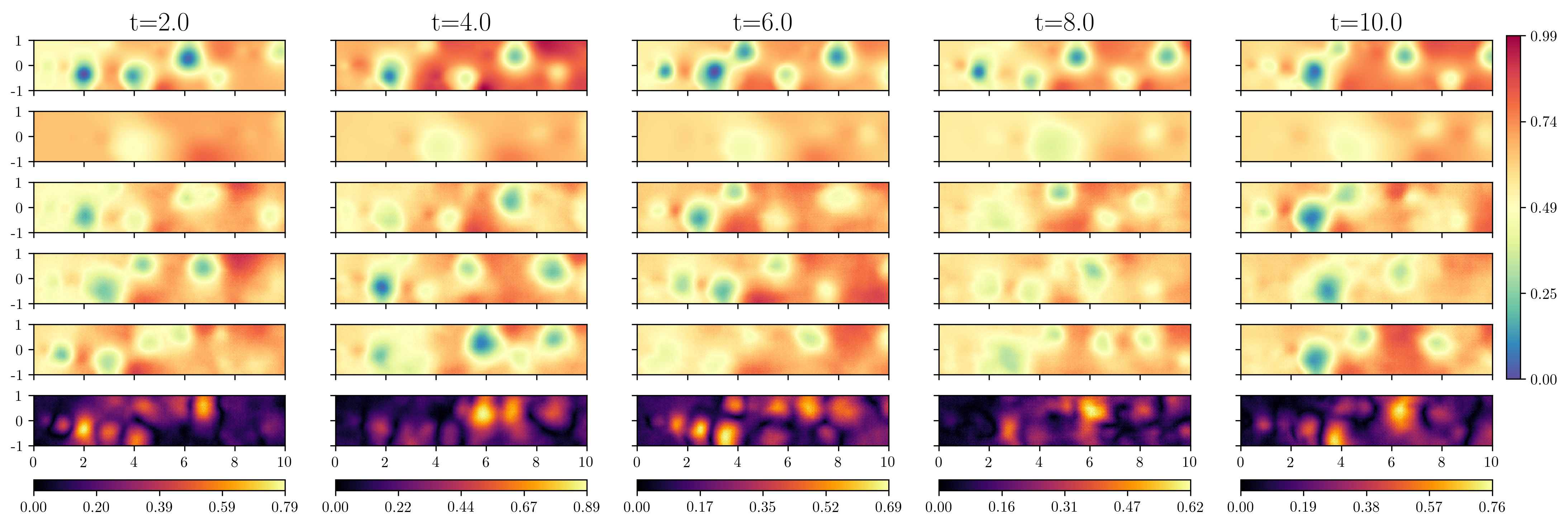}
        \caption{Pressure}
    \end{subfigure}
    \caption{TM-Glow time-series samples of $x-$velocity, $y-$velocity and pressure fields for a backwards step test case at $Re=7500$. For each field (top to bottom) the high-fidelity ground truth, low-fidelity input, three TM-Glow samples and the resulting standard deviation are plotted.}
    \label{fig:bstep-field-sample-1}
\end{figure}

\begin{figure}[H]
    \centering
    \includegraphics[width=\textwidth]{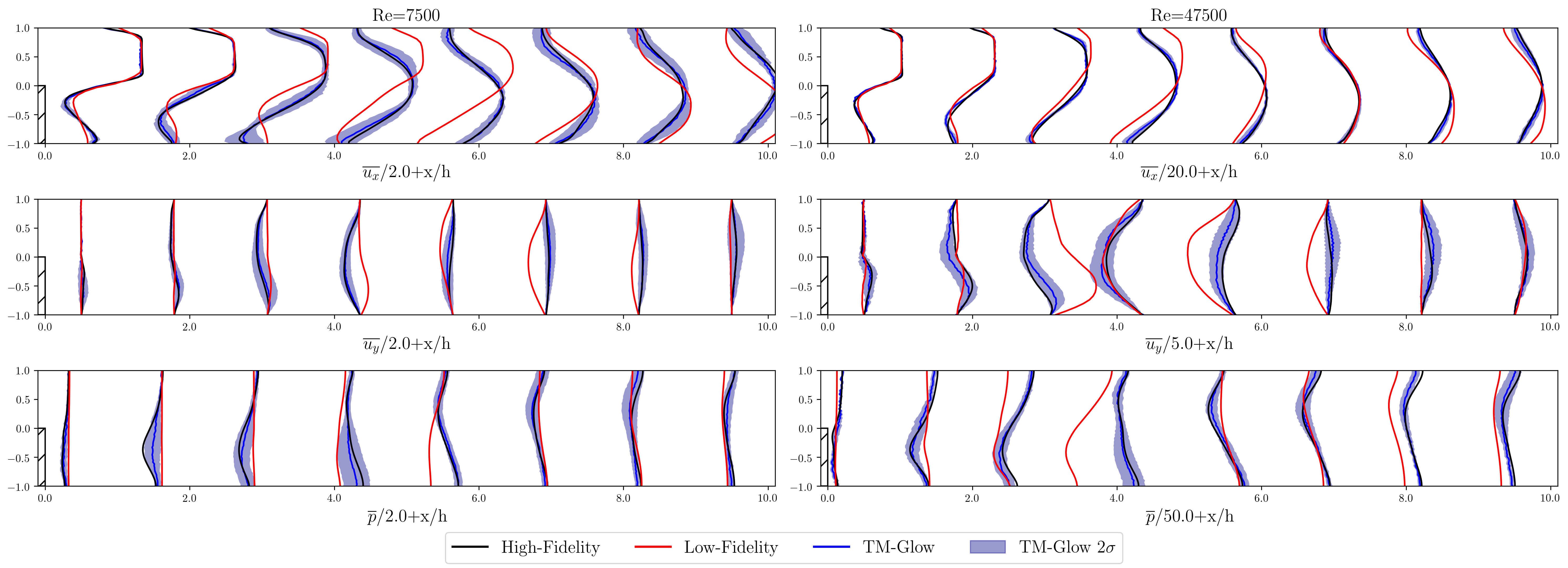}
    \caption{(Top to bottom) Time averaged x-velocity, y-velocity and pressure profiles for two different test cases at (left to right) $Re=7500$ and $Re=47500$. TM-Glow expectation (TM-Glow) and confidence interval (TM-Glow $2\sigma$) are computed using $20$ time-series samples.}
    \label{fig:bstep-mean-profiles-1}
\end{figure}

\begin{figure}[H]
    \centering
    \includegraphics[width=\textwidth]{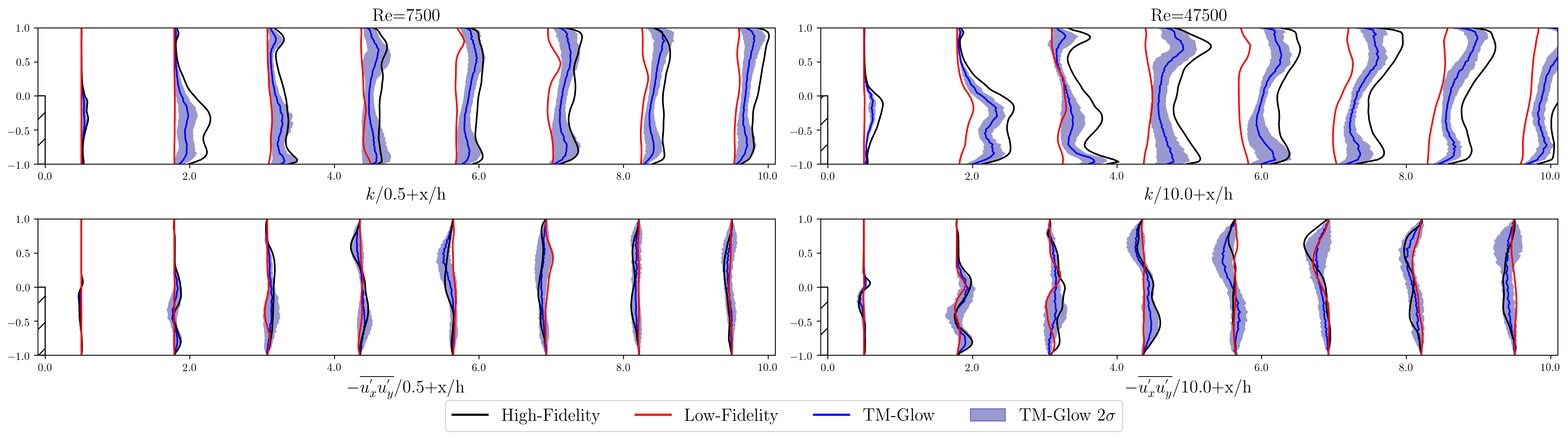}
    \caption{(Top to bottom) Turbulent kinetic energy and Reynolds shear stress profiles for two different test cases at (left to right) $Re=7500$ and $Re=47500$. TM-Glow expectation (TM-Glow) and confidence interval (TM-Glow $2\sigma$) are computed using $20$ time-series samples. }
    \label{fig:bstep-tke-profiles-1}
\end{figure}

The model trained of $48$ flows is used for the rest of this section to further analyze TM-Glow's predictions.
Several time-steps of the velocity magnitude are plotted for a few model samples in Fig.~\ref{fig:bstep-vmag-1} as well as the Q-criterion (also known as the elliptic Okubo-Weiss criterion for 2D flows)~\cite{hunt1988eddies, haller2005objective} in Fig.~\ref{fig:bstep-qcrit-1}.
Samples of each state variable for this numerical test case are illustrated in Fig.~\ref{fig:bstep-field-sample-1}.
TM-Glow clearly generates diverse fluid flow samples that are far closer to the high-fidelity solution compared to the low-fidelity simulation both in the magnitude of the fluid velocity and predicted vortex structure.
In general, the model's samples produce accurate fluid flow and turbulent statistics as illustrated in Figs.~\ref{fig:bstep-mean-profiles-1} and~\ref{fig:bstep-tke-profiles-1}.
In Fig.~\ref{fig:bstep-mean-profiles-1}, the mean flow profiles are plotted of the state variables along with the predicted uncertainty for two test flows.
Following in Fig.~\ref{fig:bstep-tke-profiles-1}, the turbulent kinetic energy and Reynolds shear stress profiles are illustrated.
TM-Glow is able to make dramatic improvements to the flow statistics for turbulent flows differing in Reynolds numbers by almost an order of magnitude.

%

\section{Turbulent Flow around an Array of Cylinders}
\label{sec:cylinder-array}
\noindent
While the prediction of a flow at different Reynolds numbers is a practical test case, the reality is that the underlying flow structures have a relatively similar form.
Thus for our second numerical example, we wish to stress this model further by investigating the prediction of a flow where the underlying flow structures are varying dramatically between test cases.
A classical fluid mechanics benchmark is the flow around a cylinder, however in its traditional form its not up to the complexity we are interested in.
Thus for a more challenging problem, we will consider the prediction of turbulent wake behind an array of cylinders with a stochastic location.

\begin{figure}[H]
    \centering
    \includegraphics[width=0.7\textwidth]{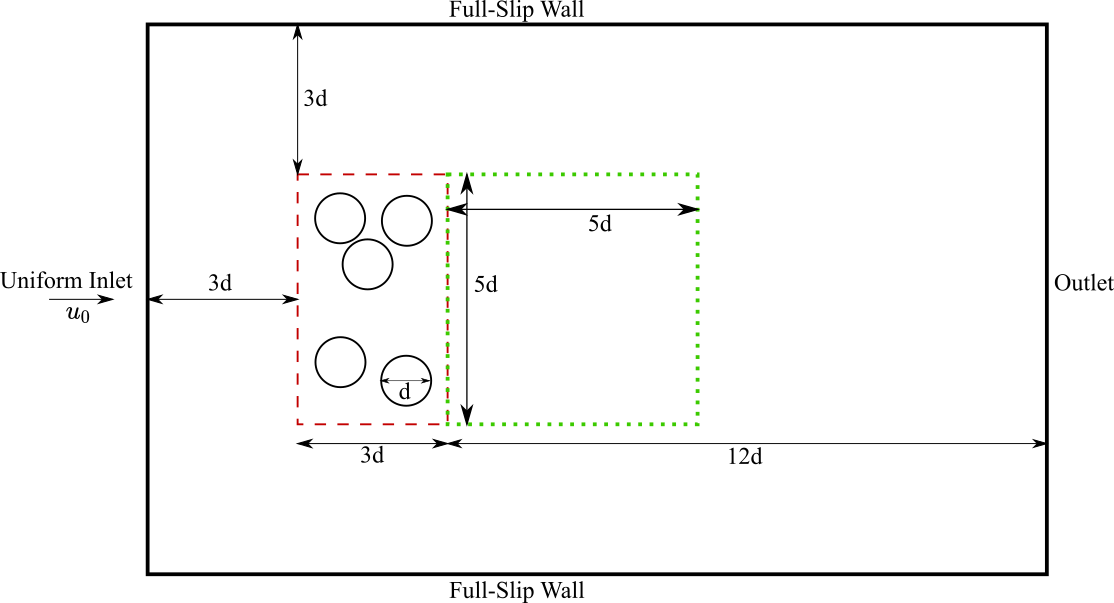}
    \caption{Flow around array of bluff bodies. The red region indicates the area for which the bodies can be placed randomly. The green region indicates the wake zone that we will use TM-Glow to predict a high-fidelity response from a low-fidelity simulation.}
    \label{fig:cylinder-schem}
\end{figure}

\begin{figure}[H]
    \centering
    \begin{subfigure}{0.1\textwidth}
        \includegraphics[height=1.75cm]{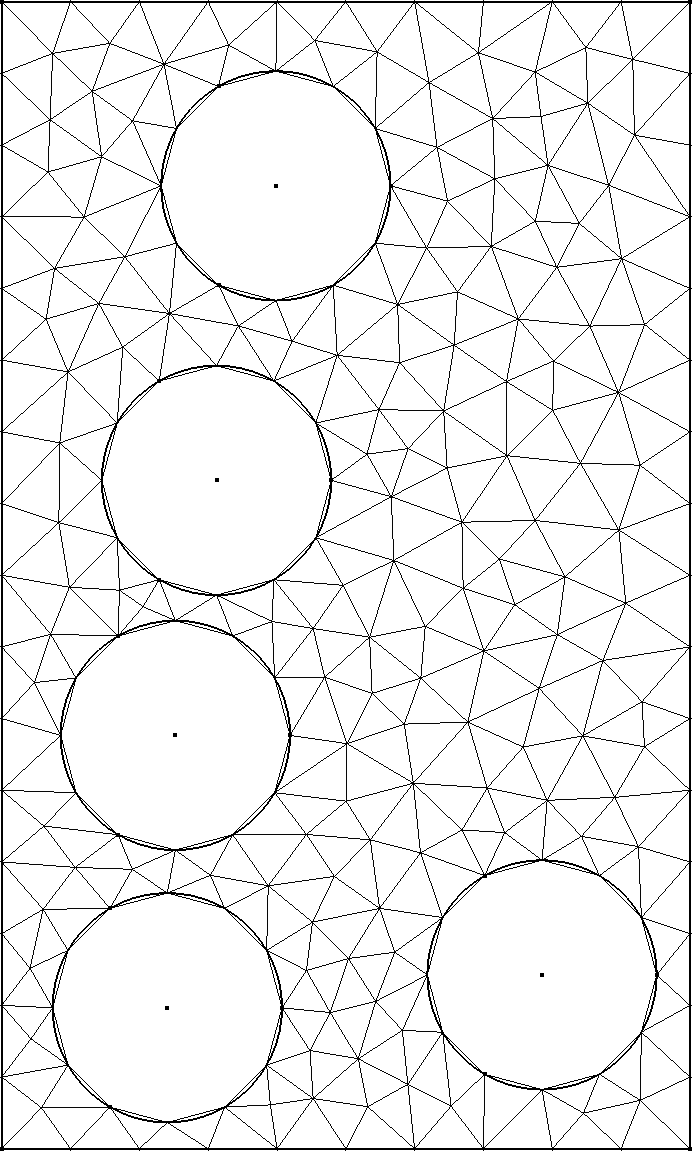}\vspace{0.2em}\\
        \includegraphics[height=1.75cm]{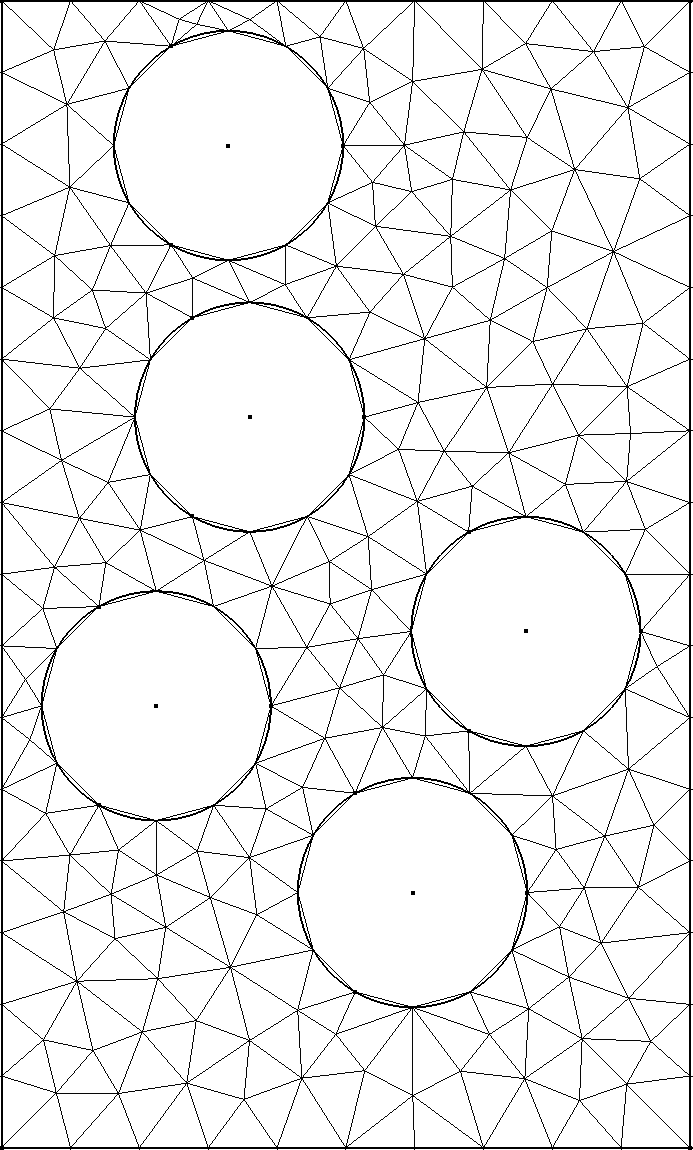}
    \end{subfigure}
    \begin{subfigure}{0.8\textwidth}
        \includegraphics[height=2cm]{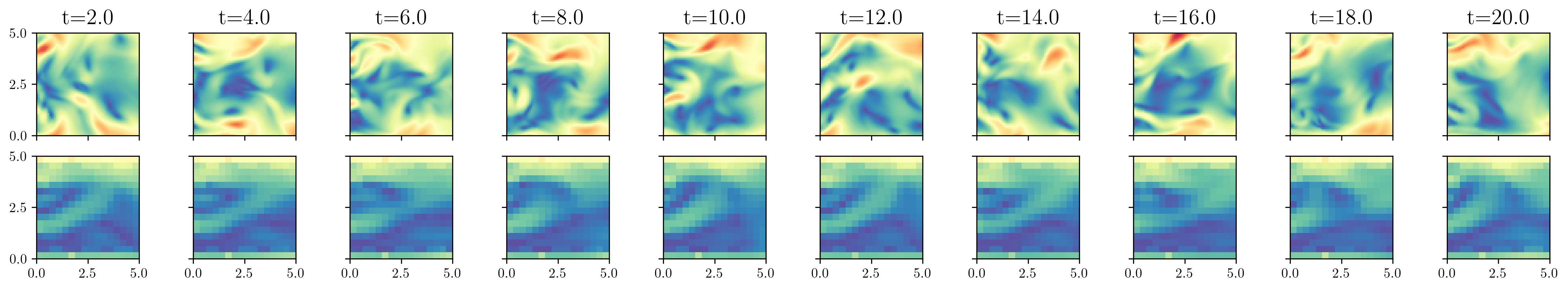}\\
        \includegraphics[height=2cm]{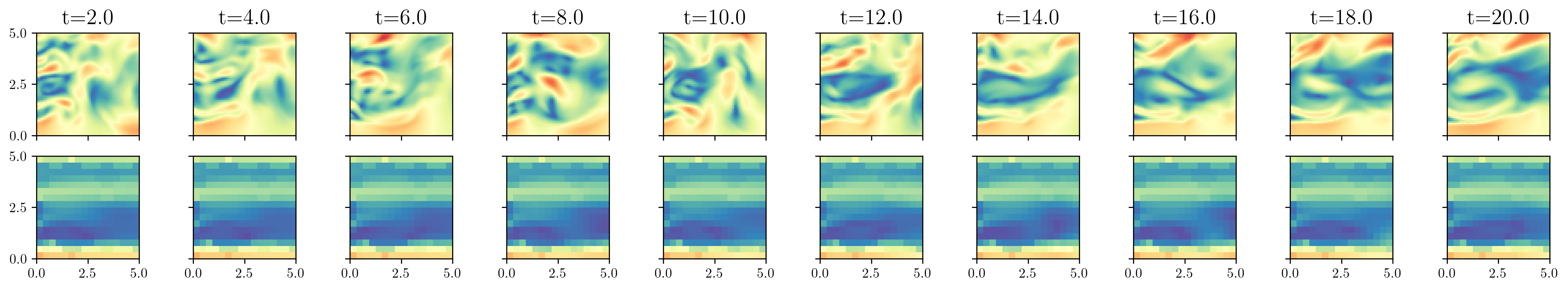}
    \end{subfigure}
    \caption{Velocity magnitude of the low-fidelity and high-fidelity simulations for two different cylinder arrays. (Left to right) Cylinder array configuration and the corresponding (top to bottom) high-fidelity and low-fidelity finite volume simulation results at several time-steps.}
    \label{fig:cylinder-simulation}
\end{figure}

\begin{figure}[H]
    \centering
    \begin{subfigure}{0.48\textwidth}
        \includegraphics[width=\textwidth]{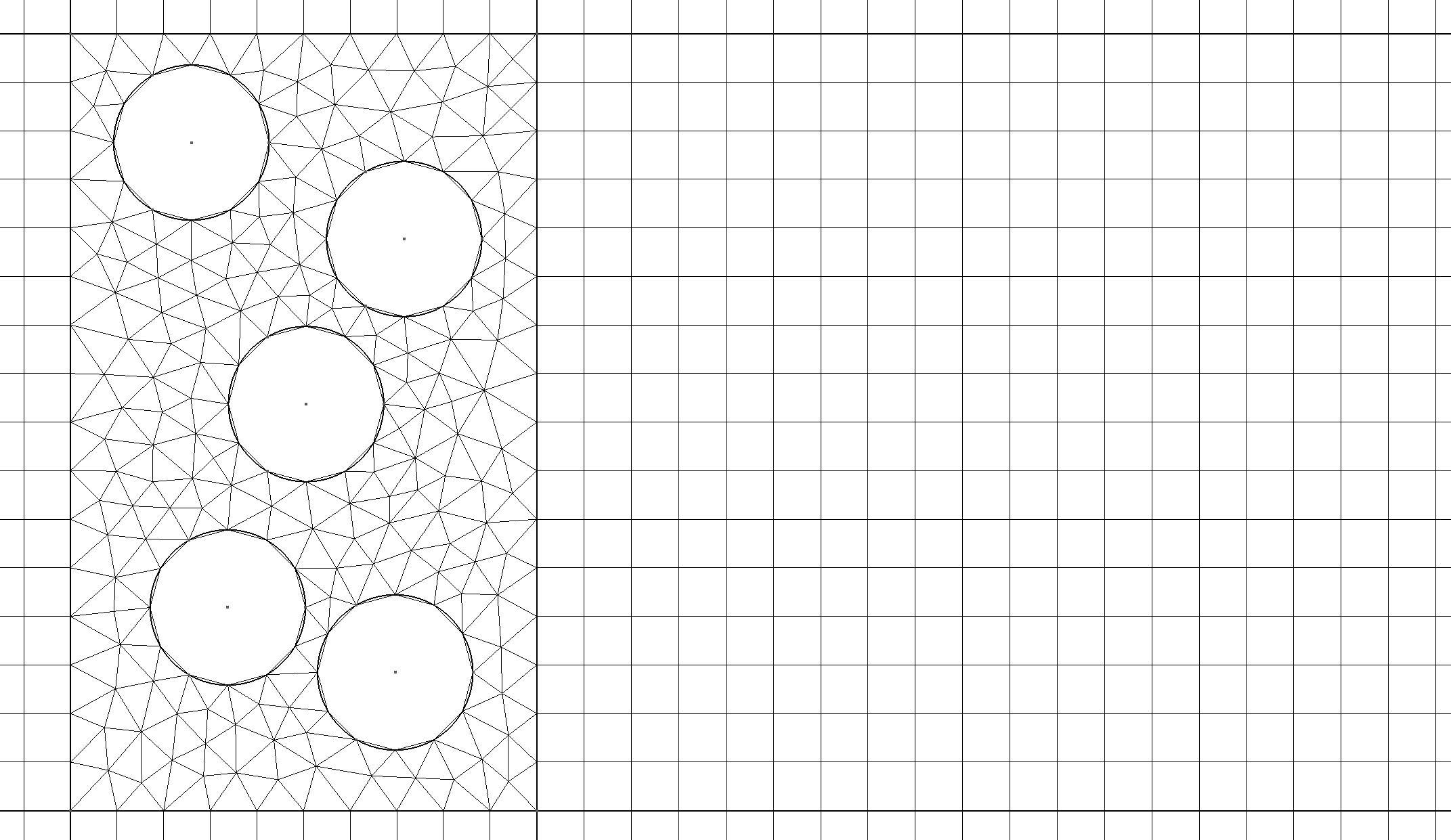}
        \caption{Low-fidelity}
    \end{subfigure}
    ~
    \begin{subfigure}{0.48\textwidth}
        \includegraphics[width=\textwidth]{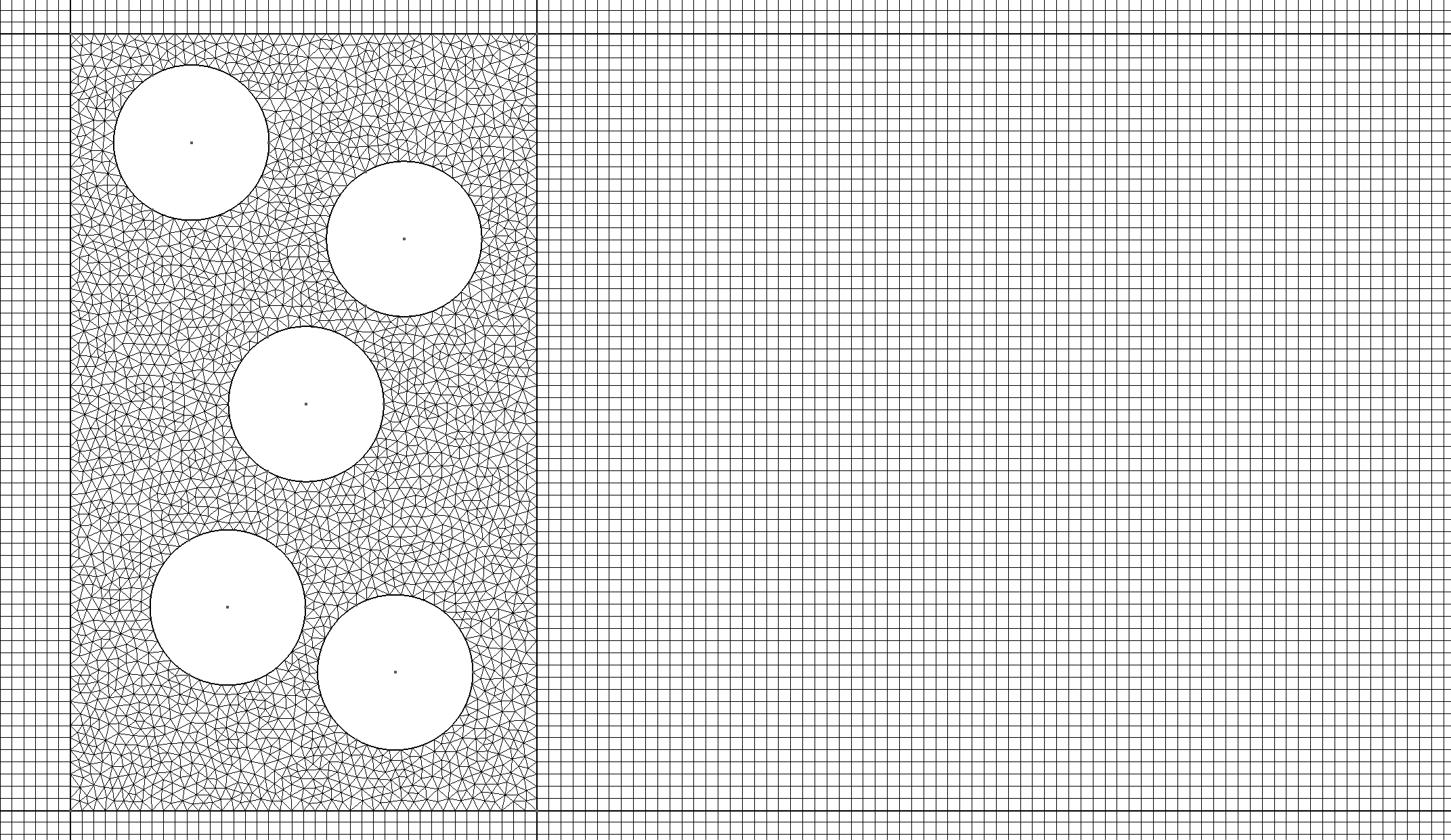}
        \caption{High-fidelity}
    \end{subfigure}
    \caption{Computational mesh around the cylinder array used for the low- and high-fidelity CFD simulations solved with OpenFOAM~\cite{jasak2007openfoam}.}
    \label{fig:cylinder-array-mesh}
\end{figure}

Flow around multiple bluff bodies is important due to its various applications in engineering including: wind flow around urban structures~\cite{tseng2006modeling}, water flow around bridge pylons~\cite{ahmed1998flow, huang2009cfd}, 
wake from an array of wind turbines~\cite{samorani2013wind, gonzalez2010optimization}, modern offshore structures~\cite{patel2013dynamics}, heat transfer applications, etc.
Depicted in Fig.~\ref{fig:cylinder-schem}, in this case study five cylinders are randomly placed within a specified area of a channel with a fixed uniform inlet velocity.
The sub-domain we wish predict is the wake region directly behind the cylinder array in which the majority of the turbulence exists.
Differing from the previous surrogate model where the Reynolds number was varying, the physical boundary of this flow is changing resulting in very different fluid structures in the predictive sub-domain.
The bulk Reynolds number of the flow, set at a constant value $Re=5000$, is governed by the inlet velocity $u_{0}=1$, viscosity $\nu=0.0002$ and the cylinder diameter $d=1$. 
This numerical example is akin to flow optimization problems for which a structure is optimized to yield desired flow properties.
The predicted flow fields for both a low-fidelity and corresponding high-fidelity finite volume simulation are shown in Fig.~\ref{fig:cylinder-simulation} for two different cylinder arrays to demonstrate the difference in the resolved flow features.
The low-fidelity simulator that will be the input of the model has a mesh characteristic resolution of $l_{c} = 5d/16$ and the target high-fidelity field has a characteristic resolution of $l_{c} = 5d/64$ as shown in Fig.~\ref{fig:cylinder-array-mesh}.
The mesh is structured in the wake region allowing for this data to be directly used with our convolutional generative model.
Thus TM-Glow will be provided an input of $\left[16\times 16 \right]$ and predict a field $\left[64\times 64\right]$ both with a time-step size of $\Delta t =0.5$.
The model input for this numerical example is $\bm{x}^{n}=\left\{\bm{u}^{n}_{l}, \bm{p}_{l}^{n}\right\} \in \mathbb{R}^{3,16,16}$ with an output $\bm{y}^{n}=\left\{\bm{u}^{n}_{h}, \bm{p}_{h}^{n}\right\} \in \mathbb{R}^{3,64,64}$.
The full training data set consists of fluid flows with cylinders randomly placed in different configurations.
Just like the previous numerical example, simulations were performed using the OpenFOAM finite volume solver using standard Smagorinsky LES sub-grid scale models~\cite{jasak2007openfoam}.
During training we augment these time-series by splitting them in half into two time-series of $40$ time-steps to artificially create more flow training cases.
Additional details on the computational cost of training of TM-Glow for this numerical example can also be found in Section~\ref{sec:computation}.

\begin{figure}[H]
    \centering
    \includegraphics[width=0.8\textwidth]{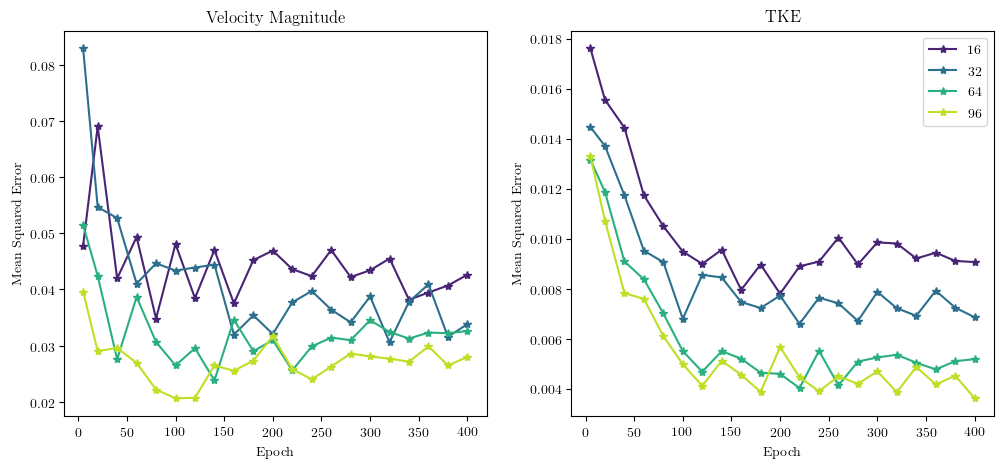}
    \caption{(Left to right) Cylinder array velocity magnitude and turbulent kinetic energy (TKE) error during training of TM-Glow on different data set sizes. Error values were average over five model samples.}
    \label{fig:cylinder-training}
\end{figure}

\begin{figure}[H]
    \centering
    \includegraphics[width=\textwidth]{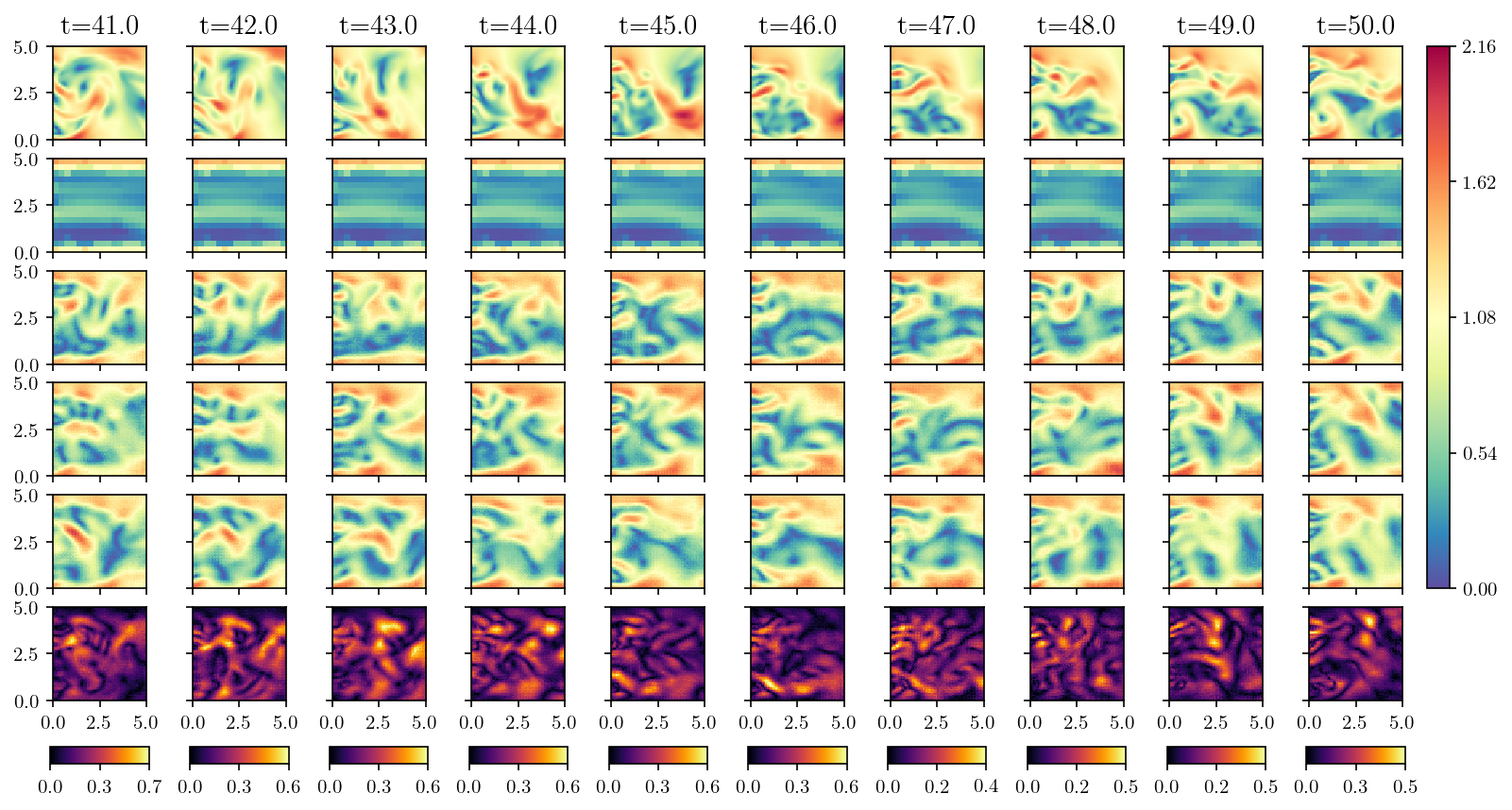}
    \includegraphics[width=\textwidth]{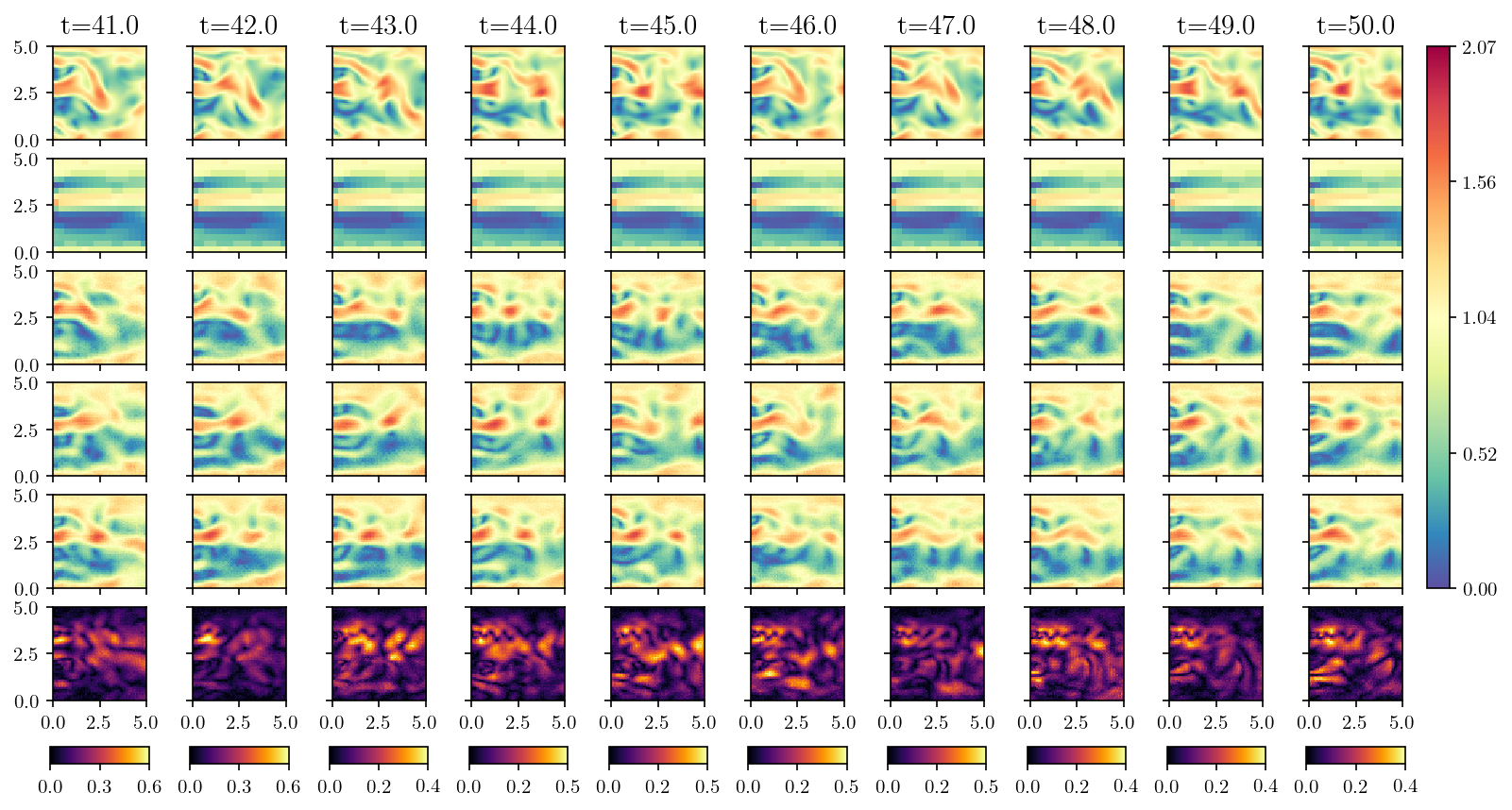}
    \caption{(Top to bottom) Velocity magnitude of the high-fidelity target, low-fidelity input, three TM-Glow samples and standard deviation for two test cases.}
    \label{fig:cylinder-pred-series}
\end{figure}

\begin{figure}[H]
    \centering
    \begin{subfigure}{0.48\textwidth}
        \centering
        \includegraphics[width=\textwidth]{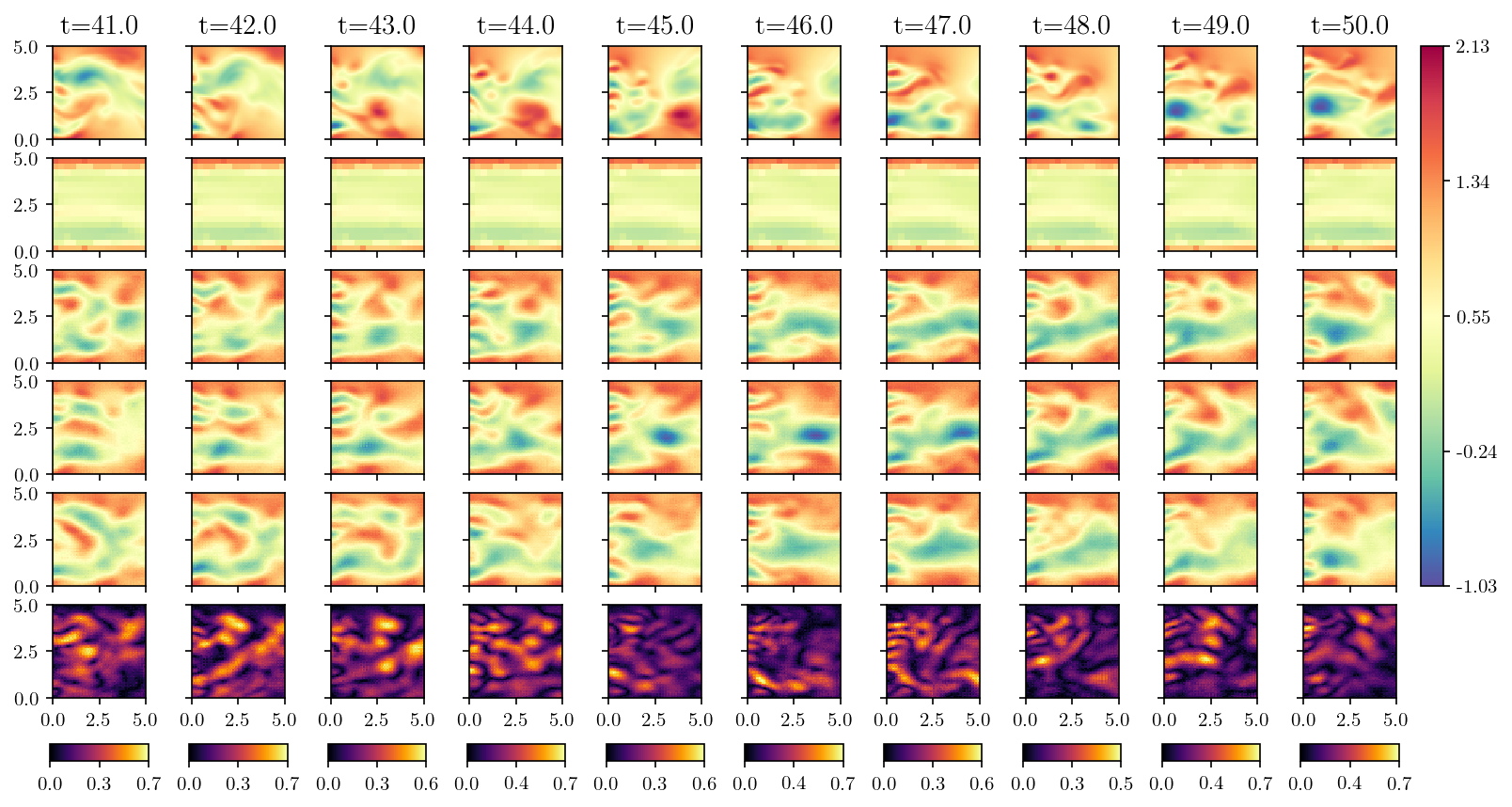}
        \caption{X-velocity}
    \end{subfigure}
    \begin{subfigure}{0.48\textwidth}
        \centering
        \includegraphics[width=\textwidth]{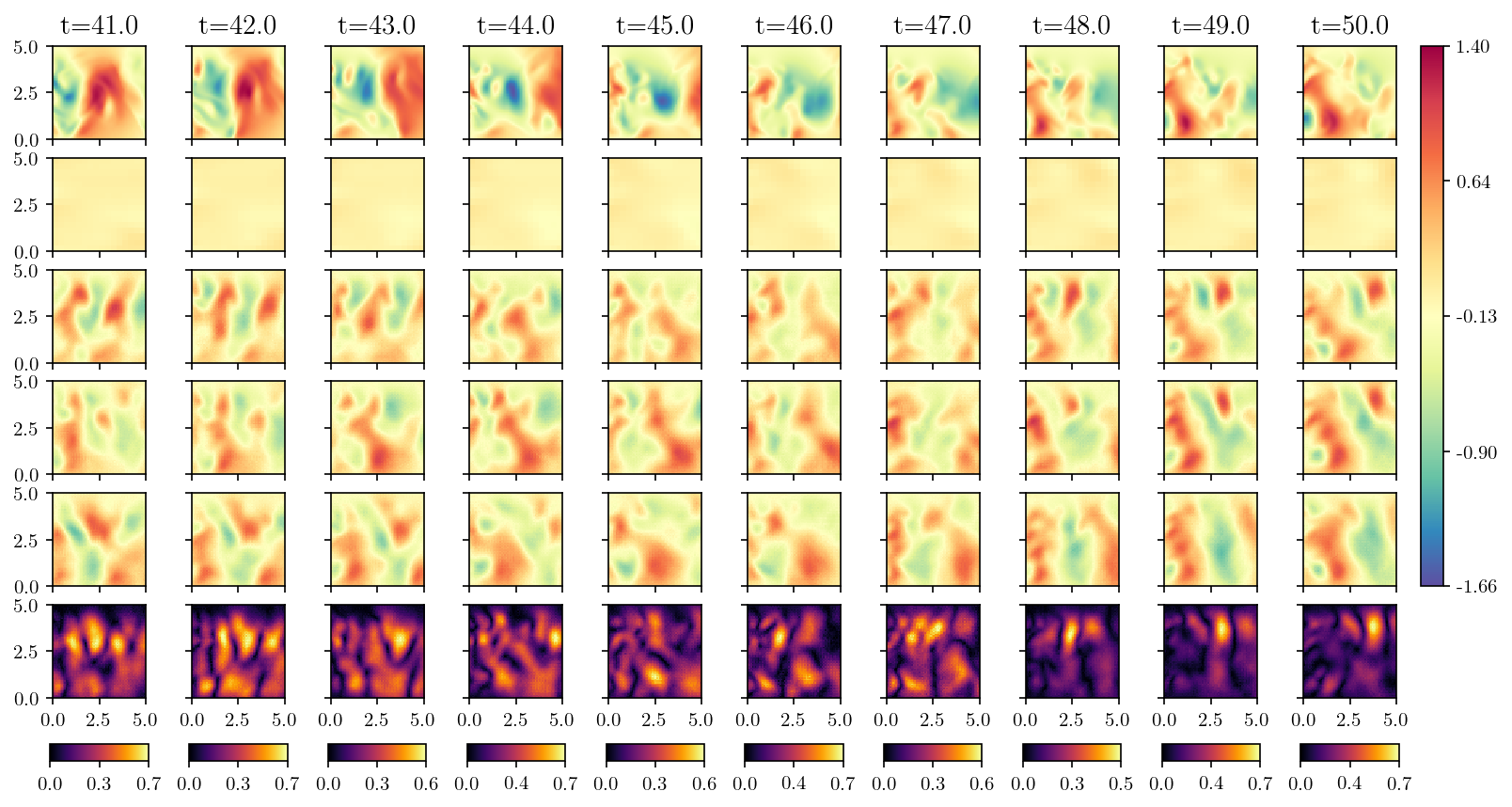}
        \caption{Y-velocity}
    \end{subfigure}\\
    \begin{subfigure}{0.48\textwidth}
        \centering
        \includegraphics[width=\textwidth]{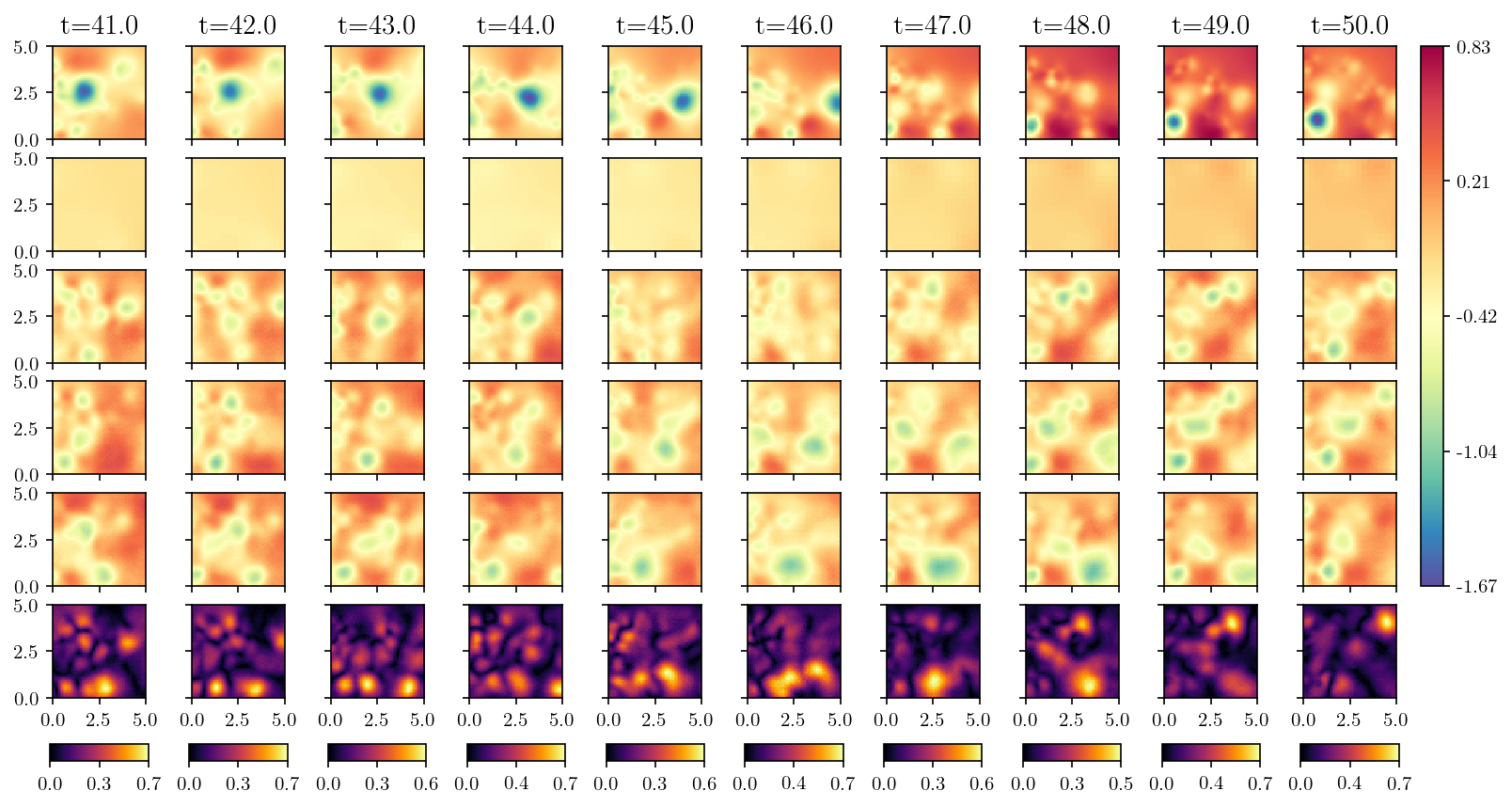}
        \caption{Pressure}
    \end{subfigure}
    \caption{TM-Glow time-series samples of $x-$velocity, $y-$velocity and pressure fields for a cylinder array test case. For each field (top to bottom) the high-fidelity ground truth, low-fidelity input, three TM-Glow samples and the resulting standard deviation are plotted.}
    \label{fig:cylinder-field-sample-1}
\end{figure}

\begin{figure}[H]
    \centering
    \begin{subfigure}{0.48\textwidth}
        \centering
        \includegraphics[width=\textwidth]{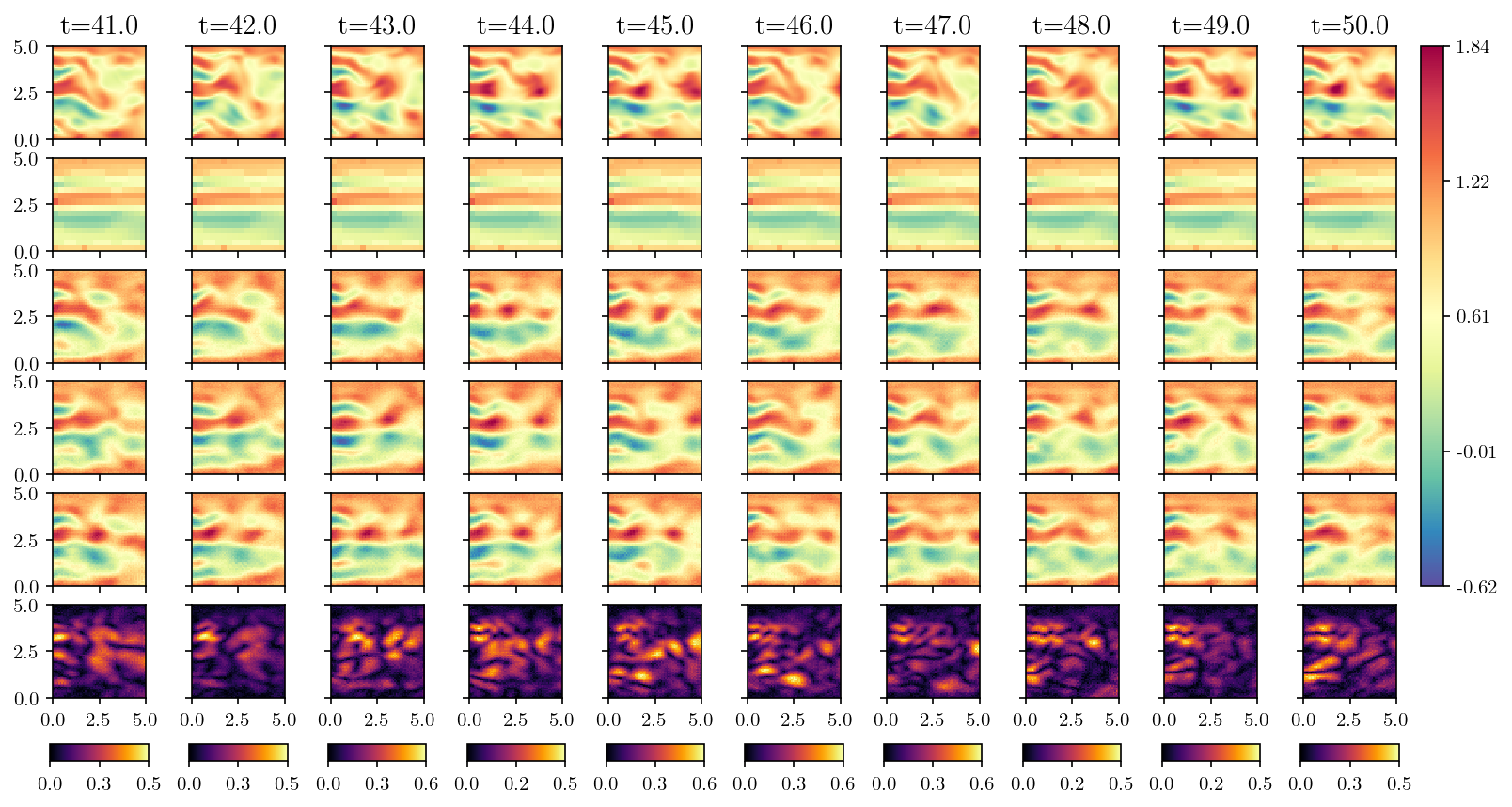}
        \caption{X-velocity}
    \end{subfigure}
    \begin{subfigure}{0.48\textwidth}
        \centering
        \includegraphics[width=\textwidth]{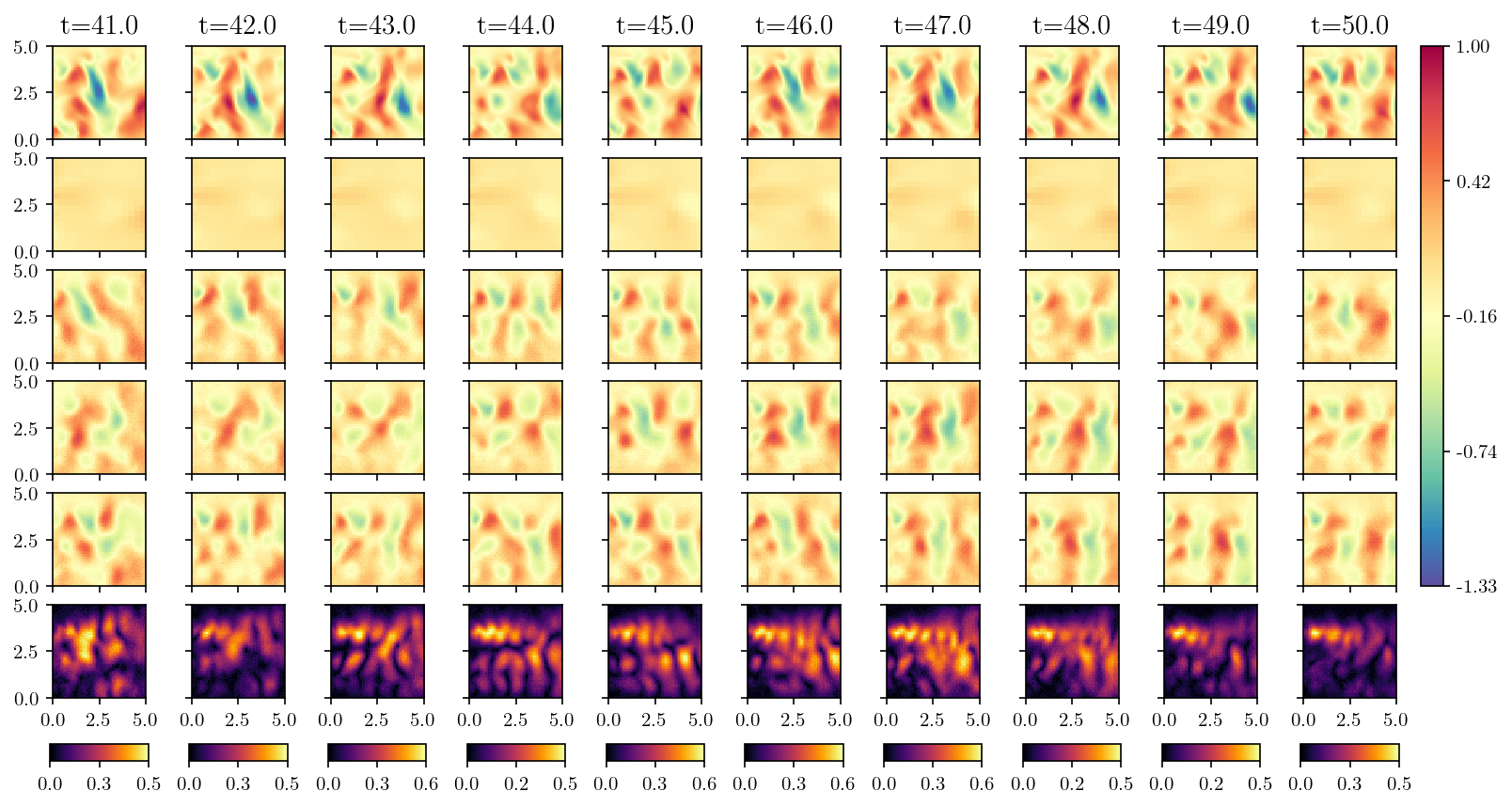}
        \caption{Y-velocity}
    \end{subfigure}\\
    \begin{subfigure}{0.48\textwidth}
        \centering
        \includegraphics[width=\textwidth]{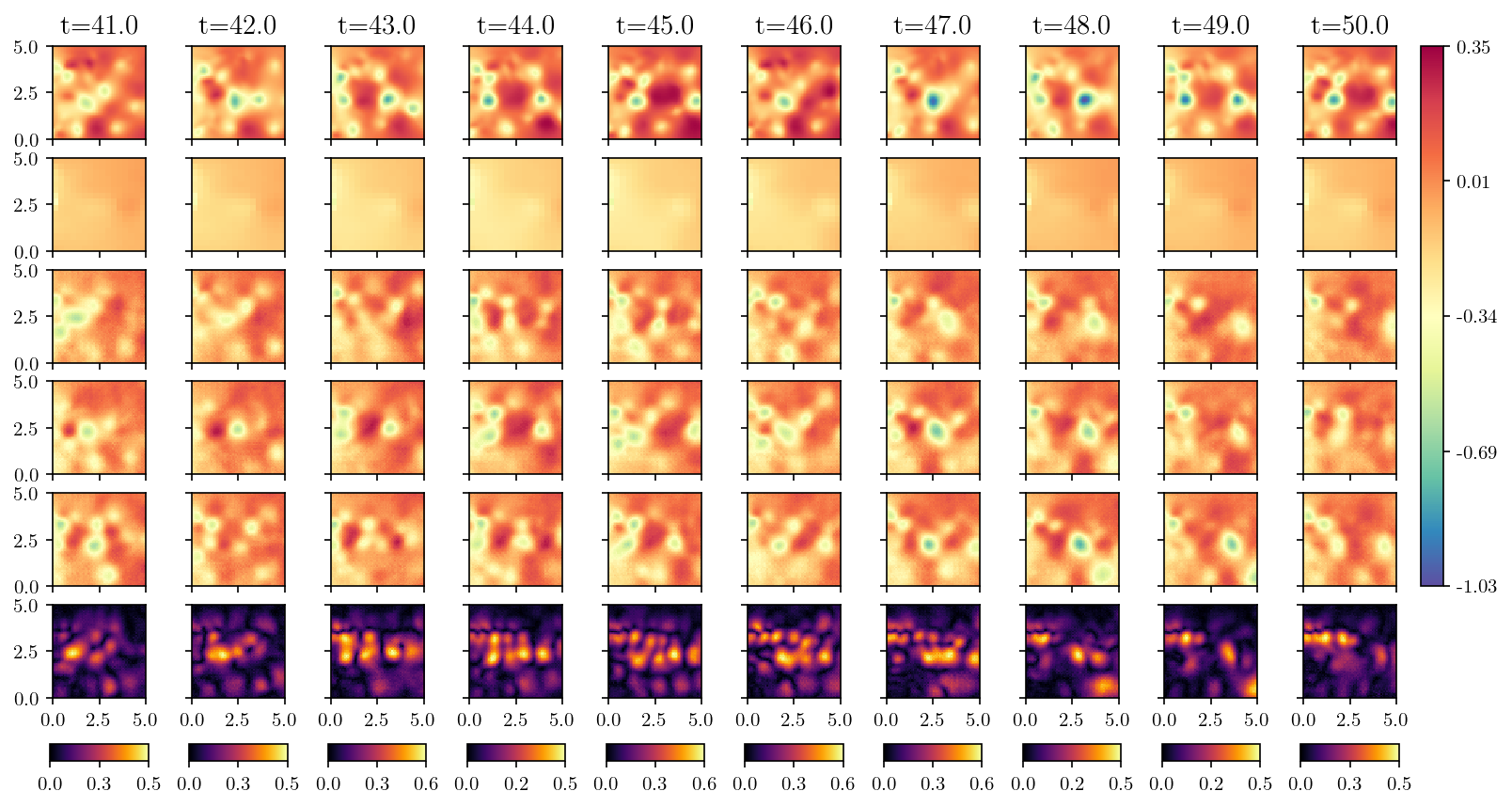}
        \caption{Pressure}
    \end{subfigure}
    \caption{TM-Glow time-series samples of $x-$velocity, $y-$velocity and pressure fields for a cylinder array test case. For each field (top to bottom) the high-fidelity ground truth, low-fidelity input, three TM-Glow samples and the resulting standard deviation are plotted.}
    \label{fig:cylinder-field-sample-2}
\end{figure}

\begin{figure}[H]
    \centering
    \begin{subfigure}{\textwidth}
        \includegraphics[width=\textwidth]{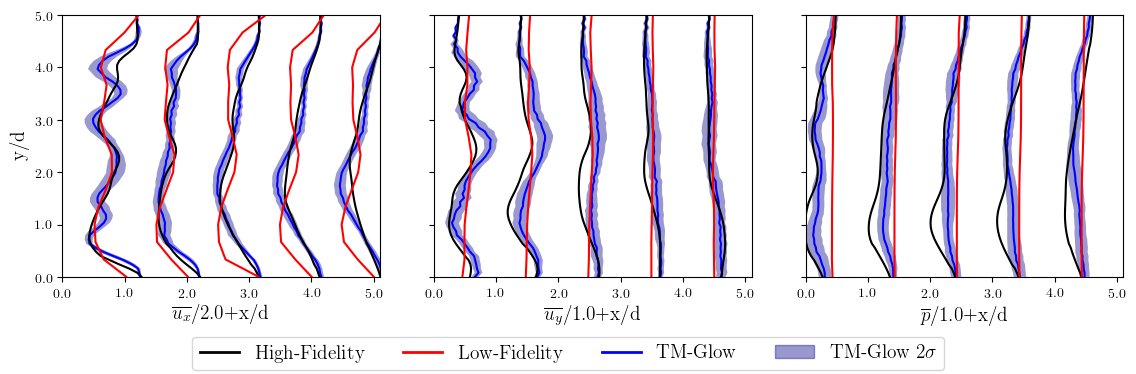}
        \caption{Test-case 1}
    \end{subfigure}\\
    \begin{subfigure}{\textwidth}
        \includegraphics[width=\textwidth]{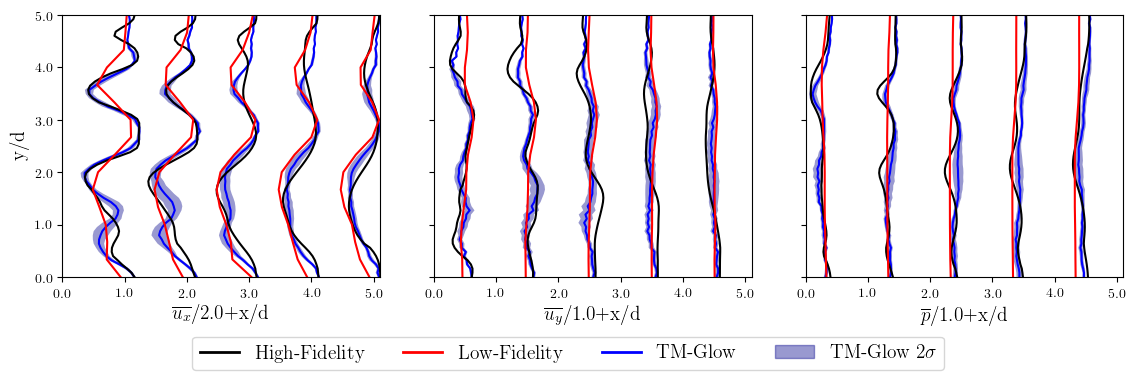}
        \caption{Test-case 2}
    \end{subfigure}
    \caption{Time-averaged flow profiles for two test flows. TM-Glow expectation (TM-Glow) and confidence interval (TM-Glow $2\sigma$) are computed using $20$ time-series samples. }
    \label{fig:cylinder-mean-profiles}
\end{figure}

\begin{figure}[H]
    \centering
    \begin{subfigure}{0.48\textwidth}
        \includegraphics[width=\textwidth]{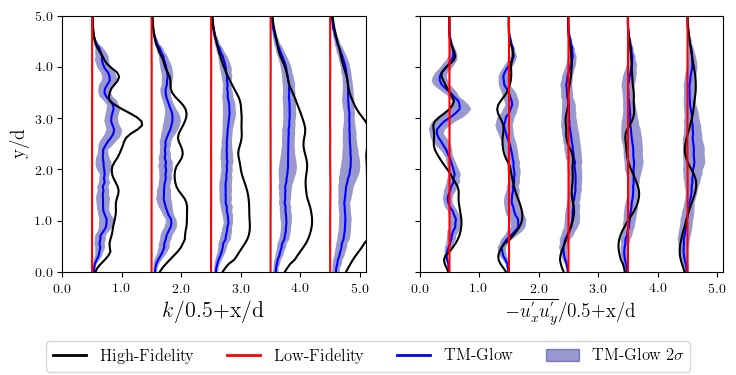}
        \caption{Test-case 1}
    \end{subfigure}
    \begin{subfigure}{0.48\textwidth}
        \includegraphics[width=\textwidth]{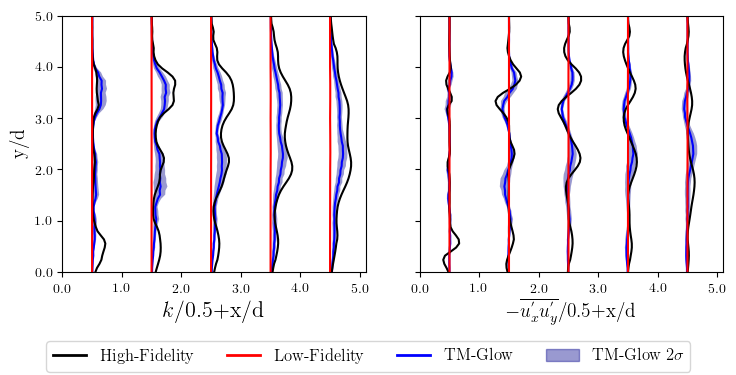}
        \caption{Test-case 2}
    \end{subfigure}
    \caption{Turbulent statistic profiles for two test flows. TM-Glow expectation (TM-Glow) and confidence interval (TM-Glow $2\sigma$) are computed using $20$ time-series samples. }
    \label{fig:cylinder-tke-profiles}
\end{figure}

A test set of $32$ flows, each with a unique cylinder configuration, are used to evaluate the performance of TM-Glow.
Four models are trained on $16$, $32$, $64$ and $96$ flows.
The test MSE error of the velocity magnitude and TKE, defined in~\Eqref{eq:mse-mag} and~\Eqref{eq:mse-tke}, during training are plotted in Fig.~\ref{fig:cylinder-training}.
The test errors of various flow field quantities are listed in Table~\ref{tab:cylinder-error} along with the error obtained from naively interpolating the low-fidelity solution to the high-fidelity mesh.
In general, TM-Glow is able to produce time-average statistics that are far more accurate than the low-fidelity solution.
As the training data set increases, we do see improvements in the flow statistics as we would expect.
We note though, that even on the smallest data set large improvements over the low-fidelity simulation can still easily be obtained.
For the remaining results we will use the model trained on $96$ flows to illustrate the highest accuracy model obtained.
\begin{table}[H]
\caption{Cylinder array test error of various time-averaged flow field quantities of the low-fidelity solution interpolated to the high-fidelity mesh and TM-Glow trained on different training data set sizes. Lower is better. TM-Glow errors were averaged over $20$ samples from the model. The training wall-clock (WC) time of each data set size is also listed.}
\label{tab:cylinder-error}
\resizebox{\textwidth}{!}{%
\begin{tabular}{cccccccc}
  & $MSE\left(\overline{\bm{u}}_{x}\right)$ & $MSE\left(\overline{\bm{u}}_{y}\right)$ & $MSE\left(\overline{\bm{p}}\right)$ & $MSE\left(\sqrt{\overline{\left(\bm{u}^{'}_{x}\right)^{2}}}\right)$ & $MSE\left(\sqrt{\overline{\left(\bm{u}^{'}_{y}\right)^{2}}}\right)$ & $MSE\left(\sqrt{\overline{\left(\bm{p}^{'}\right)^{2}}}\right)$ & WC Hrs. \\ \hline
\multicolumn{1}{c|}{Low-Fidelity} & 0.1033 & 0.0081 & 0.0179 & 0.0655 & 0.0981 & 0.02156 & - \\
\multicolumn{1}{c|}{$16$ Flows} & 0.0461 & 0.0078 & 0.0292 & 0.0116 & 0.0191 & 0.00096 & 4.3 \\
\multicolumn{1}{c|}{$32$ Flows} & 0.0461 & 0.0078 & 0.0166 & 0.0128 & 0.0185 & 0.0093 & 4.9 \\
\multicolumn{1}{c|}{$64$ Flows} & 0.0409 & 0.0062 & 0.0118 & 0.0107 & 0.0172 & 0.0084 & 6.8 \\
\multicolumn{1}{c|}{$96$ Flows} & 0.0386 & 0.0059 & 0.0128 & 0.0100 & 0.0152 & 0.0074 & 10.3
\end{tabular}}
\end{table}
Similar to the previous numerical example, we plot several time-steps of the velocity magnitude for several time-series samples of the model in Fig.~\ref{fig:cylinder-pred-series}.
Additional, samples of each state variable for this numerical test case are illustrated in Figs.~\ref{fig:cylinder-field-sample-1} and~\ref{fig:cylinder-field-sample-2}.
Although the low-fidelity simulation differs significantly from the high-fidelity solution, we can see TM-Glow is able to produce fluid realizations that qualitatively appear similar to the high-fidelity.
In this particular example, the low-fidelity solution exhibits nearly laminar flow due to the coarse discretization used which TM-Glow is able to correct.
Additionally, the fluid flows sampled from TM-Glow appear much more diverse than that seen in the previous numerical example.
Profiles of time-averaged flow quantities and turbulent statistics are plotted in Figs.~\ref{fig:cylinder-mean-profiles} and~\ref{fig:cylinder-tke-profiles} for two test flows.
Indeed we can see that TM-Glow is able to improve both with reasonable uncertainty bounds as well.
However, the model seems to consistently under-predict the turbulent intensity of the flow field.
This could be improved through fine tuning of the loss by weighting the RMS term more heavily.

\section{Computational Cost Analysis}
\label{sec:computation}
\noindent
In the following section, the computational cost associated with the training and prediction of TM-Glow will be discussed.
The cost of a surrogate needs to be low enough to justify its use, which includes the training cost for deep learning models.
To compare processes ran on different hardware and CPU cores, we adopt the measure of a service unit (SU) hour, which is equal to a single CPU core hour or a single GPU hour.
As shown in Table~\ref{tab:hardware}, both the low-fidelity and high-fidelity simulations were ran on CPUs while the deep generative model uses $4$ GPUs in parallel.
Differences between CPU models were neglected since computation of TM-Glow is bottle-necked by the GPU.
The comparison of CPU consumption versus a GPU is not trivial due to the fundamental hardware differences between the two and an in depth investigation using energy consumption or floating point operations is beyond the intended scope of this paper.
Hence, we use this simple definition resembling that used by the Extreme Science and Engineering Discovery Environment (XSEDE).\footnote{For more information regarding XSEDE see: \href{https://www.xsede.org/}{https://www.xsede.org/}.}

For both numerical examples, OpenFOAM finite volume simulator was used due to its extensive validation and efficiency~\cite{jasak2007openfoam}.
Both the low- and high-fidelity simulations used standard  LES Smagorinsky sub-grid scale model~\cite{smagorinsky1963general} with default parameters.
When the high-fidelity simulations were parallelized between CPUs, OpenFOAM's in house domain decomposition algorithm ``scotch'' was used to partition the meshes.
Additionally, the fluid flows are solved between time $t=[0,80]$ for both resolutions but only $t=[40,80]$ is used as training/testing data.
This is done to ensure the flow fields sampled were of fully developed turbulence.
\begin{table}[H]
\caption{Hardware used to run the low-fidelity and high-fidelity CFD simulations as well as the training and prediction of TM-Glow for both numerical examples.}
\label{tab:hardware}
\resizebox{\textwidth}{!}{%
\begin{tabular}{l|ccccc}
                         & CPU Cores & CPU Model  & GPUs & GPU Model         & SU Hour \\ \hline
Low-Fidelity  & 1         & Intel Xeon E5-2680 & -    & -                 & 1       \\
High-Fidelity & 8         & Intel Xeon E5-2680 & -    & -                 & 8       \\
TM-Glow                  & 1         & Intel Xeon Gold 6226 & 4 & NVIDIA Tesla V100 & 8      
\end{tabular}}
\end{table}

\subsection{Turbulent Flow over a Backwards Step}
\noindent
The low-fidelity and high-fidelity simulations for the flow over backwards facing step consisted of a mesh with a resolution of $\Delta x, \Delta y = h/12$ and $\Delta x, \Delta y = h/32$, respectively. 
A sub-section of both meshes are plotted in Fig.~\ref{fig:backward-step-mesh} to illustrate the resolution difference.
This results in the low-fidelity and high-fidelity meshes containing $22$k and $145$k cells, respectively.
A single low-fidelity simulation takes about $4.5$ minutes on a single CPU core while a high-fidelity simulation takes about $42$ minutes on $8$ CPU cores.

In Fig.~\ref{fig:training-computation}, the required SUs needed to train TM-Glow for various data set sizes is plotted.
The majority of the computation is expended on  obtaining  the high-fidelity training data.
In Table~\ref{tab:prediction-computation}, the predictive computational cost of TM-Glow is compared to that of a high-fidelity solver.
We classify the  computational cost of the surrogate's prediction as the cost of the low-fidelity solver as well as the prediction of $20$ model samples.
Here we can see that both in terms of SU hours and raw wall-clock time, the surrogate is significantly more efficient than the high-fidelity solver.
The majority of the computational cost of the surrogate is from the low-fidelity simulation with minimal overhead from TM-Glow.

\subsection{Turbulent Flow around an Array of Cylinders}
\noindent
The low-fidelity and high-fidelity simulations for the flow over a cylinder array consisted of a mesh with a resolution of $\Delta x, \Delta y = 5d/16$ and $\Delta x, \Delta y = 5d/64$, respectively. 
A sub-section of the meshes are plotted in Fig.~\ref{fig:cylinder-array-mesh} to illustrate the resolution difference.
This results in the low-fidelity and high-fidelity meshes containing $9$k and $125$k cells, respectively.
A single low-fidelity simulation takes about $3.1$ minutes on a single CPU core while a high-fidelity simulation takes about $32$ minutes on $8$ CPU cores.

\begin{figure}[H]
    \centering
    \begin{subfigure}{0.48\textwidth}
        \includegraphics[width=\textwidth]{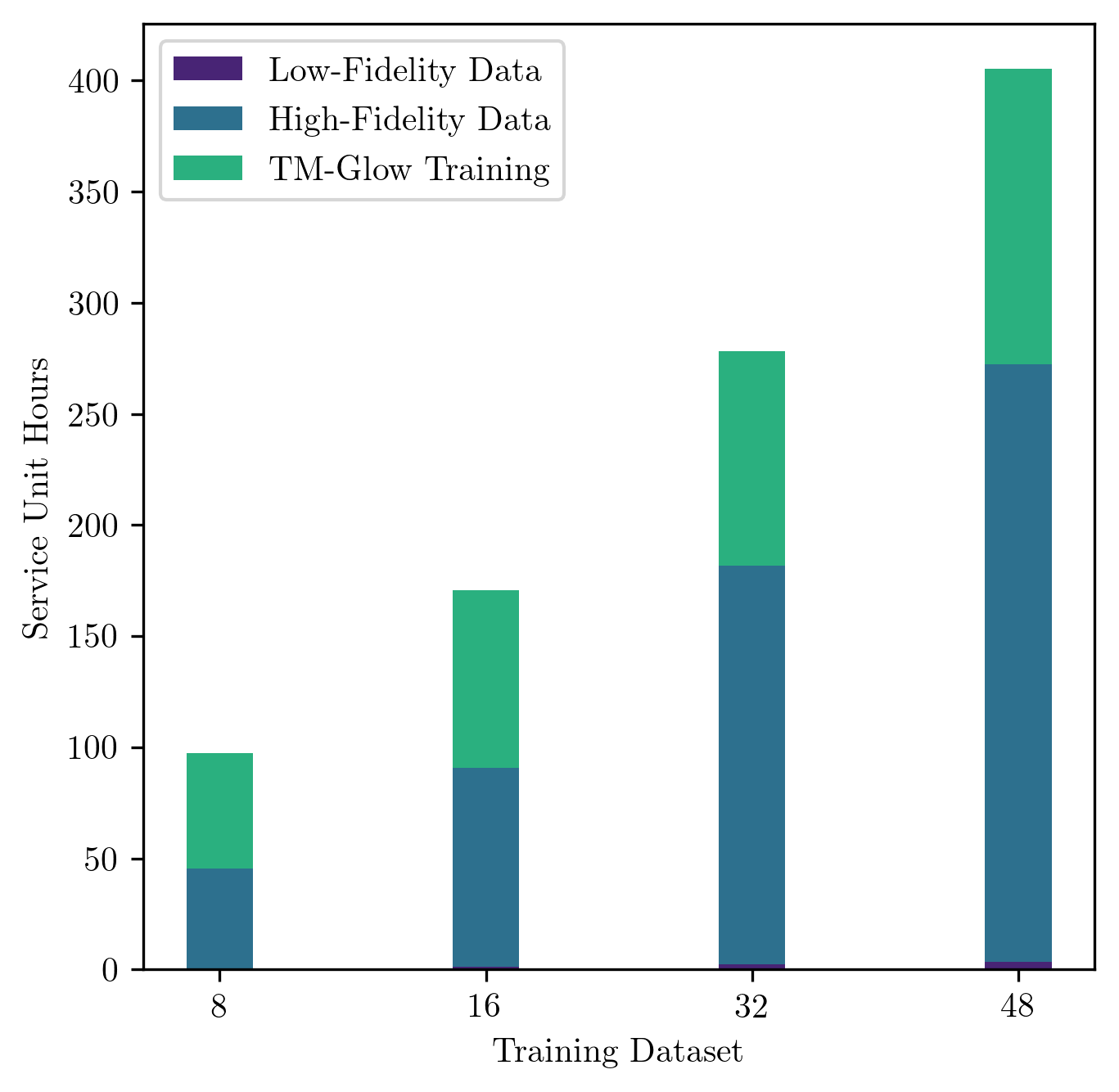}
        \caption{Flow over backwards step}
    \end{subfigure}
    ~
    \begin{subfigure}{0.48\textwidth}
        \includegraphics[width=\textwidth]{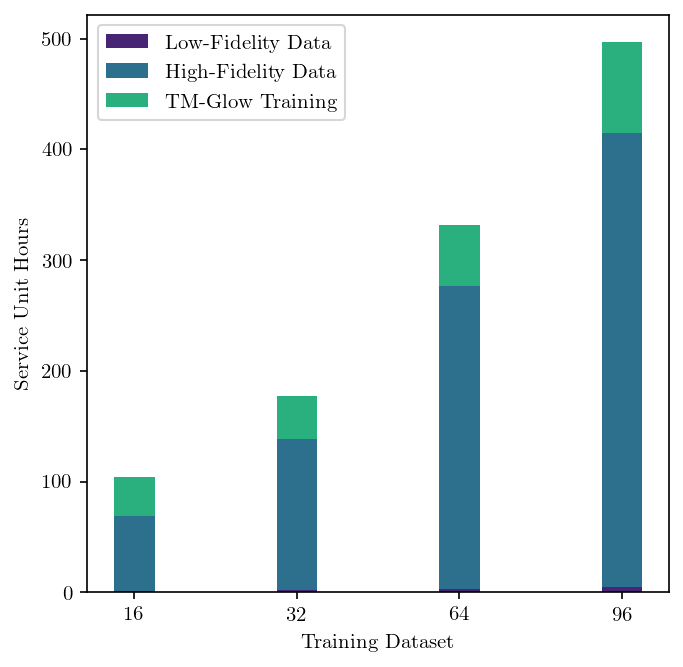}
        \caption{Flow around cylinder array}
    \end{subfigure}
    \caption{Computational requirement for training TM-Glow given training data-sets of various sizes. Computation is quantified using Service Units (SU) defined in Table~\ref{tab:hardware}.}
    \label{fig:training-computation}
\end{figure}

In Fig.~\ref{fig:training-computation}, the required SUs needed to train TM-Glow for various data set sizes is plotted.
Additionally, in Table~\ref{tab:prediction-computation}, the predictive computational cost of TM-Glow is compared to that of a high-fidelity solver.
Again, we classify the  computational cost of the surrogate's prediction as the cost of the low-fidelity solver as well as the prediction of $20$ model samples.
Similar to the flow over a backwards step, we note that the surrogate is much faster than the high-fidelity simulation.
Based on these results, as long as the number of training cases needed to achieve the desired accuracy is reasonable compared to the total number of predictions needed by the surrogate, the use of this model is well justified.
This is of course problem dependent, however for many optimization and inverse settings this will be easily satisfied.
\begin{table}[H]
\caption{Prediction cost of the surrogate compared to the high-fidelity simulator for flow over a backwards step (left) and flow around a cylinder array (right).}
\label{tab:prediction-computation}

\begin{minipage}{0.45\textwidth}
\resizebox{\textwidth}{!}{%
\begin{tabular}{l|cc}
Backwards Step  & SU Hours & Wall-clock (mins) \\ \hline
Low-Fidelity             & 0.06     & 4.5               \\
TM-Glow 20 Samples       & 0.03    & 0.75              \\ \hline
Surrogate Prediction     & 0.09     & 5.25              \\
High-Fidelity Prediction & 5.6      & 42               
\end{tabular}}
\end{minipage}
$\qquad$
\begin{minipage}{0.45\textwidth}
\resizebox{\textwidth}{!}{%
\begin{tabular}{l|cc}
Cylinder Array   & SU Hours & Wall-clock (mins) \\ \hline
Low-Fidelity             & 0.05     & 3.1               \\
TM-Glow 20 Samples       & 0.02    & 0.7              \\ \hline
Surrogate Prediction     & 0.07     & 3.8              \\
High-Fidelity Prediction & 4.27      & 32               
\end{tabular}}
\end{minipage}
\end{table}
%

\section{Conclusion}
\label{sec:conclusion}
\noindent
The application of machine learning methods to CFD requires significant advances to extend such models to realistic problems.
In this work we investigate the prediction of fully-turbulent systems using deep learning.
We proposed a multi-fidelity approach for which a computationally inexpensive low-fidelity solver is used as a conditional input to a deep generative model that predicts fluid realizations at high-fidelity resolution and accuracy.
The model, Transient Multi-fidelity Glow (TM-Glow), is a conditional invertible neural network that allows for the analytical evaluation of the likelihood though the change of variables formula.
TM-Glow is trained using variational backwards KL divergence loss which allows for the seamless combination of data-driven and physics-constraint based learning.
This model was demonstrated on two numerical examples to surrogate model turbulent flow at different Reynolds numbers as well as a stochastic boundary.
With just the low-fidelity solution, TM-Glow was able  to predict diverse samples of turbulent flow time-series that produce accurate mean field/turbulent statistics with error bars for uncertainty quantification.

The multi-fidelity aspect of our model is a key ingredient.
The low-fidelity input provides critical information to the generative model such as information regarding boundary conditions, mean flow properties and general flow field structure.
While this low-fidelity simulation is typically inaccurate, it is a reliable starting point for the model to extrapolate from.
The prediction from low- to high-fidelity is a significantly simpler problem compared to a blind high-fidelity flow prediction allowing for reduced training data-set sizes and training time.
For this reason, we believe that deep learning has significant potential in multilevel/multi-fidelity modeling of a vast number of physical systems where it can be used on even very high-dimensional complex phenomena due to a low-fidelity solver aiding the machine learning model.
In this spirit, future steps to be investigated include the extension of this model to other multi-fidelity physical systems.
Additionally, as the deep learning field evolves, more modern architectures and training techniques could be integrated into the model to increase its predictive capability.
Regardless of potential future directions, TM-Glow demonstrated that modern generative deep learning methods can be used effectively for multi-fidelity modeling of complex dynamical systems.

\section*{Acknowledgements}
\noindent
The authors acknowledge support  from the Defense Advanced Research Projects Agency (DARPA) under the Physics of Artificial Intelligence (PAI) program (contract HR$00111890034$). 
Computing resources were provided by the AFOSR Office of Scientific Research through the DURIP program and by the University of Notre Dame's Center for Research Computing (CRC). The work of NG was also supported by the National Science Foundation (NSF) Graduate Research Fellowship Program grant No. DGE-$1313583$. 

\clearpage
\providecommand{\href}[2]{#2}
\providecommand{\arxiv}[1]{\href{http://arxiv.org/abs/#1}{arXiv:#1}}
\providecommand{\url}[1]{\texttt{#1}}
\providecommand{\urlprefix}{URL }

\bibliographystyle{aims}

\begin{thebibliography}{10}

\bibitem{ahmed1998flow}
\newblock F.~Ahmed and N.~Rajaratnam,
\newblock Flow around bridge piers,
\newblock \emph{Journal of Hydraulic Engineering}, \textbf{124} (1998),
  288--300,
\newblock
  \urlprefix\url{https://ascelibrary.org/doi/abs/10.1061/\%28ASCE\%290733-9429\%281998\%29124\%3A3\%28288\%29}.

\bibitem{ardizzone2019guided}
\newblock L.~Ardizzone, C.~L{\"u}th, J.~Kruse, C.~Rother and U.~K{\"o}the,
\newblock Guided image generation with conditional invertible neural networks,
\newblock \emph{arXiv preprint arXiv:1907.02392}.

\bibitem{bieker2019deep}
\newblock K.~Bieker, S.~Peitz, S.~L. Brunton, J.~N. Kutz and M.~Dellnitz,
\newblock Deep model predictive control with online learning for complex
  physical systems,
\newblock \emph{arXiv preprint arXiv:1905.10094}.

\bibitem{chen2018review}
\newblock L.~Chen, K.~Asai, T.~Nonomura, G.~Xi and T.~Liu,
\newblock A review of {Backward-Facing Step (BFS)} flow mechanisms, heat
  transfer and control,
\newblock \emph{Thermal Science and Engineering Progress}, \textbf{6} (2018),
  194 -- 216,
\newblock
  \urlprefix\url{http://www.sciencedirect.com/science/article/pii/S2451904918300167}.

\bibitem{chung2016hierarchical}
\newblock J.~Chung, S.~Ahn and Y.~Bengio,
\newblock Hierarchical multiscale recurrent neural networks,
\newblock \emph{arXiv preprint arXiv:1609.01704}.

\bibitem{dinh2014nice}
\newblock L.~Dinh, D.~Krueger and Y.~Bengio,
\newblock Nice: Non-linear independent components estimation,
\newblock \emph{arXiv preprint arXiv:1410.8516}.

\bibitem{dinh2016density}
\newblock L.~Dinh, J.~Sohl-Dickstein and S.~Bengio,
\newblock Density estimation using real nvp,
\newblock \emph{arXiv preprint arXiv:1605.08803}.

\bibitem{erturk2008numerical}
\newblock E.~Erturk,
\newblock Numerical solutions of {2-D} steady incompressible flow over a
  backward-facing step, {Part I: High Reynolds} number solutions,
\newblock \emph{Computers and Fluids}, \textbf{37} (2008), 633 -- 655,
\newblock
  \urlprefix\url{http://www.sciencedirect.com/science/article/pii/S0045793007001545}.

\bibitem{geneva2019modeling}
\newblock N.~Geneva and N.~Zabaras,
\newblock Modeling the dynamics of {PDE} systems with physics-constrained deep
  auto-regressive networks,
\newblock \emph{Journal of Computational Physics}, 109056,
\newblock
  \urlprefix\url{http://www.sciencedirect.com/science/article/pii/S0021999119307612}.

\bibitem{geneva2019quantifying}
\newblock N.~Geneva and N.~Zabaras,
\newblock Quantifying model form uncertainty in {Reynolds-averaged} turbulence
  models with {Bayesian} deep neural networks,
\newblock \emph{Journal of Computational Physics}, \textbf{383} (2019), 125 --
  147,
\newblock
  \urlprefix\url{http://www.sciencedirect.com/science/article/pii/S0021999119300464}.

\bibitem{glorot2011deep}
\newblock X.~Glorot, A.~Bordes and Y.~Bengio,
\newblock Deep sparse rectifier neural networks,
\newblock in \emph{Proceedings of the fourteenth international conference on
  artificial intelligence and statistics}, 2011,
\newblock 315--323.

\bibitem{gonzalez2010optimization}
\newblock J.~S. González, A.~G.~G. Rodriguez], J.~C. Mora, J.~R. Santos and
  M.~B. Payan,
\newblock Optimization of wind farm turbines layout using an evolutive
  algorithm,
\newblock \emph{Renewable Energy}, \textbf{35} (2010), 1671 -- 1681,
\newblock
  \urlprefix\url{http://www.sciencedirect.com/science/article/pii/S0960148110000145}.

\bibitem{goodfellow2016deep}
\newblock I.~Goodfellow, Y.~Bengio and A.~Courville,
\newblock \emph{Deep learning},
\newblock MIT press, 2016.

\bibitem{goodfellow2014generative}
\newblock I.~Goodfellow, J.~Pouget-Abadie, M.~Mirza, B.~Xu, D.~Warde-Farley,
  S.~Ozair, A.~Courville and Y.~Bengio,
\newblock Generative adversarial nets,
\newblock in \emph{Advances in neural information processing systems}, 2014,
\newblock 2672--2680.

\bibitem{grathwohl2018ffjord}
\newblock W.~Grathwohl, R.~T. Chen, J.~Betterncourt, I.~Sutskever and
  D.~Duvenaud,
\newblock Ffjord: Free-form continuous dynamics for scalable reversible
  generative models,
\newblock \emph{arXiv preprint arXiv:1810.01367}.

\bibitem{guo2016convolutional}
\newblock X.~Guo, W.~Li and F.~Iorio,
\newblock Convolutional neural networks for steady flow approximation,
\newblock in \emph{Proceedings of the 22Nd ACM SIGKDD International Conference
  on Knowledge Discovery and Data Mining},
\newblock KDD '16, ACM, 2016,
\newblock 481--490,
\newblock \urlprefix\url{http://doi.acm.org/10.1145/2939672.2939738}.

\bibitem{haller2005objective}
\newblock G.~Haller,
\newblock An objective definition of a vortex,
\newblock \emph{Journal of Fluid Mechanics}, \textbf{525} (2005), 1--26.

\bibitem{renkun2019novel}
\newblock R.~Han, Y.~Wang, Y.~Zhang and G.~Chen,
\newblock A novel spatial-temporal prediction method for unsteady wake flows
  based on hybrid deep neural network,
\newblock \emph{Physics of Fluids}, \textbf{31} (2019), 127101,
\newblock \urlprefix\url{https://doi.org/10.1063/1.5127247}.

\bibitem{hennigh2017lat}
\newblock O.~Hennigh,
\newblock Lat-net: compressing lattice {Boltzmann} flow simulations using deep
  neural networks,
\newblock \emph{arXiv preprint arXiv:1705.09036}.

\bibitem{hoffman2006approach}
\newblock J.~Hoffman and C.~Johnson,
\newblock A new approach to computational turbulence modeling,
\newblock \emph{Computer Methods in Applied Mechanics and Engineering},
  \textbf{195} (2006), 2865 -- 2880,
\newblock
  \urlprefix\url{http://www.sciencedirect.com/science/article/pii/S0045782505002306},
\newblock Incompressible CFD.

\bibitem{holgate2019review}
\newblock J.~Holgate, A.~Skillen, T.~Craft and A.~Revell,
\newblock A review of embedded large eddy simulation for internal flows,
\newblock \emph{Archives of Computational Methods in Engineering}, \textbf{26}
  (2019), 865--882.

\bibitem{huang2017densely}
\newblock G.~Huang, Z.~Liu, L.~van~der Maaten and K.~Q. Weinberger,
\newblock Densely connected convolutional networks,
\newblock in \emph{The IEEE Conference on Computer Vision and Pattern
  Recognition (CVPR)}, 2017.

\bibitem{huang2009cfd}
\newblock W.~Huang, Q.~Yang and H.~Xiao,
\newblock {CFD} modeling of scale effects on turbulence flow and scour around
  bridge piers,
\newblock \emph{Computers and Fluids}, \textbf{38} (2009), 1050 -- 1058,
\newblock
  \urlprefix\url{http://www.sciencedirect.com/science/article/pii/S0045793008000704},
\newblock Advances in Computational Fluid Dynamics.

\bibitem{hunt1988eddies}
\newblock J.~C. Hunt, A.~A. Wray and P.~Moin,
\newblock Eddies, streams, and convergence zones in turbulent flows,
\newblock in \emph{Center for Turbulence Research Report}, vol. CTR-S88, 1988,
\newblock \urlprefix\url{https://ntrs.nasa.gov/search.jsp?R=19890015184}.

\bibitem{ioffe2015batch}
\newblock S.~Ioffe and C.~Szegedy,
\newblock Batch normalization: Accelerating deep network training by reducing
  internal covariate shift,
\newblock \emph{arXiv preprint arXiv:1502.03167}.

\bibitem{jacobsen2018revnet}
\newblock J.-H. Jacobsen, A.~Smeulders and E.~Oyallon,
\newblock i-revnet: {Deep} invertible networks,
\newblock \emph{arXiv preprint arXiv:1802.07088}.

\bibitem{jasak2007openfoam}
\newblock H.~Jasak, A.~Jemcov, Z.~Tukovic et~al.,
\newblock {OpenFOAM}: A {C++} library for complex physics simulations,
\newblock in \emph{International workshop on coupled methods in numerical
  dynamics}, vol. 1000,
\newblock IUC Dubrovnik, Croatia, 2007,
\newblock 1--20.

\bibitem{kim2019deep}
\newblock B.~Kim, V.~C. Azevedo, N.~Thuerey, T.~Kim, M.~Gross and
  B.~Solenthaler,
\newblock Deep fluids: A generative network for parameterized fluid
  simulations,
\newblock \emph{Computer Graphics Forum}, \textbf{38} (2019), 59--70,
\newblock
  \urlprefix\url{https://onlinelibrary.wiley.com/doi/abs/10.1111/cgf.13619}.

\bibitem{kingma2014adam}
\newblock D.~P. Kingma and J.~Ba,
\newblock Adam: A method for stochastic optimization,
\newblock \emph{arXiv preprint arXiv:1412.6980}.

\bibitem{kingma2013auto}
\newblock D.~P. Kingma and M.~Welling,
\newblock Auto-encoding variational bayes,
\newblock \emph{arXiv preprint arXiv:1312.6114}.

\bibitem{kingma2018glow}
\newblock D.~P. Kingma and P.~Dhariwal,
\newblock Glow: Generative flow with invertible 1x1 convolutions,
\newblock in \emph{Advances in Neural Information Processing Systems}, 2018,
\newblock 10215--10224.

\bibitem{kumar2019videoflow}
\newblock M.~Kumar, M.~Babaeizadeh, D.~Erhan, C.~Finn, S.~Levine, L.~Dinh and
  D.~Kingma,
\newblock Videoflow: A flow-based generative model for video,
\newblock \emph{arXiv preprint arXiv:1903.01434}.

\bibitem{kumar2019maximum}
\newblock R.~Kumar, S.~Ozair, A.~Goyal, A.~Courville and Y.~Bengio,
\newblock Maximum entropy generators for energy-based models,
\newblock \emph{arXiv preprint arXiv:1901.08508}.

\bibitem{lapeyre2019training}
\newblock C.~J. Lapeyre, A.~Misdariis, N.~Cazard, D.~Veynante and T.~Poinsot,
\newblock Training convolutional neural networks to estimate turbulent sub-grid
  scale reaction rates,
\newblock \emph{Combustion and Flame}, \textbf{203} (2019), 255 -- 264,
\newblock
  \urlprefix\url{http://www.sciencedirect.com/science/article/pii/S0010218019300835}.

\bibitem{lecun2006tutorial}
\newblock Y.~LeCun, S.~Chopra, R.~Hadsell, M.~Ranzato and F.~Huang,
\newblock A tutorial on energy-based learning,
\newblock \emph{Predicting structured data}.

\bibitem{li2018learning}
\newblock C.~Li, J.~Li, G.~Wang and L.~Carin,
\newblock Learning to sample with adversarially learned likelihood-ratio, 2018,
\newblock \urlprefix\url{https://openreview.net/forum?id=S1eZGHkDM}.

\bibitem{ling2016reynolds}
\newblock J.~Ling, A.~Kurzawski and J.~Templeton,
\newblock Reynolds averaged turbulence modelling using deep neural networks
  with embedded invariance,
\newblock \emph{Journal of Fluid Mechanics}, \textbf{807} (2016), 155–166,
\newblock \urlprefix\url{https://doi.org/10.1007/s00162-005-0165-5}.

\bibitem{liu2015multi}
\newblock P.~Liu, X.~Qiu, X.~Chen, S.~Wu and X.-J. Huang,
\newblock Multi-timescale long short-term memory neural network for modelling
  sentences and documents,
\newblock in \emph{Proceedings of the 2015 conference on empirical methods in
  natural language processing}, 2015,
\newblock 2326--2335.

\bibitem{maulik2019subgrid}
\newblock R.~Maulik, O.~San, A.~Rasheed and P.~Vedula,
\newblock Subgrid modelling for two-dimensional turbulence using neural
  networks,
\newblock \emph{Journal of Fluid Mechanics}, \textbf{858} (2019), 122–144.

\bibitem{mitran2001comparison}
\newblock S.~M. Mitran,
\newblock \emph{A comparison of adaptive mesh refinement approaches for large
  eddy simulation},
\newblock Technical report, Washington University, Seattle, Department of
  Applied Mathematics, 2001.

\bibitem{mo2019deep}
\newblock S.~Mo, Y.~Zhu, N.~Zabaras, X.~Shi and J.~Wu,
\newblock Deep convolutional encoder-decoder networks for uncertainty
  quantification of dynamic multiphase flow in heterogeneous media,
\newblock \emph{Water Resources Research}, \textbf{55} (2019), 703--728,
\newblock
  \urlprefix\url{https://agupubs.onlinelibrary.wiley.com/doi/abs/10.1029/2018WR023528}.

\bibitem{mohan2019compressed}
\newblock A.~Mohan, D.~Daniel, M.~Chertkov and D.~Livescu,
\newblock Compressed convolutional lstm: An efficient deep learning framework
  to model high fidelity 3d turbulence,
\newblock \emph{arXiv preprint arXiv:1903.00033}.

\bibitem{patel2013dynamics}
\newblock M.~H. Patel,
\newblock \emph{Dynamics of offshore structures},
\newblock Butterworth-Heinemann, 2013.

\bibitem{pope2001turbulent}
\newblock S.~B. Pope,
\newblock \emph{Turbulent flows},
\newblock Cambridge University Press, Cambridge, 2000.

\bibitem{quemere2002zonal}
\newblock P.~Quéméré and P.~Sagaut,
\newblock Zonal multi-domain rans/les simulations of turbulent flows,
\newblock \emph{International Journal for Numerical Methods in Fluids},
  \textbf{40} (2002), 903--925,
\newblock
  \urlprefix\url{https://onlinelibrary.wiley.com/doi/abs/10.1002/fld.381}.

\bibitem{rabault2019artificial}
\newblock J.~Rabault, M.~Kuchta, A.~Jensen, U.~R{\'e}glade and N.~Cerardi,
\newblock Artificial neural networks trained through deep reinforcement
  learning discover control strategies for active flow control,
\newblock \emph{Journal of Fluid Mechanics}, \textbf{865} (2019), 281–302.

\bibitem{raissi2019physics}
\newblock M.~Raissi, P.~Perdikaris and G.~Karniadakis,
\newblock Physics-informed neural networks: A deep learning framework for
  solving forward and inverse problems involving nonlinear partial differential
  equations,
\newblock \emph{Journal of Computational Physics}, \textbf{378} (2019), 686 --
  707,
\newblock
  \urlprefix\url{http://www.sciencedirect.com/science/article/pii/S0021999118307125}.

\bibitem{raissi2019deep}
\newblock M.~Raissi, Z.~Wang, M.~S. Triantafyllou and G.~E. Karniadakis,
\newblock Deep learning of vortex-induced vibrations,
\newblock \emph{Journal of Fluid Mechanics}, \textbf{861} (2019), 119–137.

\bibitem{sagaut2013multiscale}
\newblock P.~Sagaut,
\newblock \emph{Multiscale and multiresolution approaches in turbulence: LES,
  DES and hybrid RANS/LES methods: applications and guidelines},
\newblock World Scientific, 2013.

\bibitem{samorani2013wind}
\newblock M.~Samorani,
\newblock The wind farm layout optimization problem,
\newblock in \emph{Handbook of Wind Power Systems} (eds. P.~M. Pardalos,
  S.~Rebennack, M.~V.~F. Pereira, N.~A. Iliadis and V.~Pappu),
\newblock Springer Berlin Heidelberg, Berlin, Heidelberg, 2013,
\newblock 21--38,
\newblock \urlprefix\url{https://doi.org/10.1007/978-3-642-41080-2\_2}.

\bibitem{schluter2004large}
\newblock J.~U. Schlüter, H.~Pitsch and P.~Moin,
\newblock Large-eddy simulation inflow conditions for coupling with
  reynolds-averaged flow solvers,
\newblock \emph{AIAA Journal}, \textbf{42} (2004), 478--484,
\newblock \urlprefix\url{https://doi.org/10.2514/1.3488}.

\bibitem{xingjian2015convolutional}
\newblock X.~SHI, Z.~Chen, H.~Wang, D.-Y. Yeung, W.-k. Wong and W.-c. WOO,
\newblock Convolutional {LSTM} network: A machine learning approach for
  precipitation nowcasting,
\newblock in \emph{Advances in Neural Information Processing Systems 28},
\newblock Curran Associates, Inc., 2015,
\newblock 802--810,
\newblock
  \urlprefix\url{http://papers.nips.cc/paper/5955-convolutional-lstm-network-a-machine-learning-approach-for-precipitation-nowcasting.pdf}.

\bibitem{smagorinsky1963general}
\newblock J.~Smagorinsky,
\newblock General circulation experiments with the primitive equations: I. the
  basic experiment,
\newblock \emph{Monthly Weather Review}, \textbf{91} (1963), 99--164,
\newblock
  \urlprefix\url{https://doi.org/10.1175/1520-0493(1963)091<0099:GCEWTP>2.3.CO;2}.

\bibitem{sobel19683x3}
\newblock I.~Sobel and G.~Feldman,
\newblock A 3x3 isotropic gradient operator for image processing,
\newblock \emph{Presented at a talk at the Stanford Artificial Intelligence
  Project}, 271--272.

\bibitem{speziale1997computing}
\newblock C.~G. Speziale,
\newblock Computing non-equilibrium turbulent flows with time-dependent {RANS}
  and {VLES},
\newblock in \emph{Fifteenth International Conference on Numerical Methods in
  Fluid Dynamics},
\newblock Springer, 1997,
\newblock 123--129.

\bibitem{subramaniam2020turbulence}
\newblock A.~Subramaniam, M.~L. Wong, R.~D. Borker, S.~Nimmagadda and S.~K.
  Lele,
\newblock Turbulence enrichment using generative adversarial networks,
\newblock \emph{arXiv preprint arXiv:2003.01907}.

\bibitem{sun2020surrogate}
\newblock L.~Sun, H.~Gao, S.~Pan and J.-X. Wang,
\newblock Surrogate modeling for fluid flows based on physics-constrained deep
  learning without simulation data,
\newblock \emph{Computer Methods in Applied Mechanics and Engineering},
  \textbf{361} (2020), 112732.

\bibitem{tabak2013family}
\newblock E.~G. Tabak and C.~V. Turner,
\newblock A family of nonparametric density estimation algorithms,
\newblock \emph{Communications on Pure and Applied Mathematics}, \textbf{66}
  (2013), 145--164.

\bibitem{tabak2010density}
\newblock E.~G. Tabak, E.~Vanden-Eijnden et~al.,
\newblock Density estimation by dual ascent of the log-likelihood,
\newblock \emph{Communications in Mathematical Sciences}, \textbf{8} (2010),
  217--233.

\bibitem{taghizadeh2020turbulence}
\newblock S.~Taghizadeh, F.~D. Witherden and S.~S. Girimaji,
\newblock Turbulence closure modeling with data-driven techniques: physical
  compatibility and consistency considerations,
\newblock \emph{arXiv preprint arXiv:2004.03031}.

\bibitem{terracol2005hybrid}
\newblock M.~Terracol, E.~Manoha, C.~Herrero, E.~Labourasse, S.~Redonnet and
  P.~Sagaut,
\newblock Hybrid methods for airframe noise numerical prediction,
\newblock \emph{Theoretical and Computational Fluid Dynamics}, \textbf{19}
  (2005), 197--227.

\bibitem{terracol2001multilevel}
\newblock M.~Terracol, P.~Sagaut and C.~Basdevant,
\newblock A multilevel algorithm for large-eddy simulation of turbulent
  compressible flows,
\newblock \emph{Journal of Computational Physics}, \textbf{167} (2001), 439 --
  474,
\newblock
  \urlprefix\url{http://www.sciencedirect.com/science/article/pii/S0021999100966877}.

\bibitem{tompson2017accelerating}
\newblock J.~Tompson, K.~Schlachter, P.~Sprechmann and K.~Perlin,
\newblock Accelerating eulerian fluid simulation with convolutional networks,
\newblock in \emph{Proceedings of the 34th International Conference on Machine
  Learning - Volume 70},
\newblock ICML'17, JMLR.org, 2017,
\newblock 3424--3433,
\newblock \urlprefix\url{http://dl.acm.org/citation.cfm?id=3305890.3306035}.

\bibitem{travin2002physical}
\newblock A.~Travin, M.~Shur, M.~Strelets and P.~R. Spalart,
\newblock Physical and numerical upgrades in the detached-eddy simulation of
  complex turbulent flows,
\newblock in \emph{Advances in LES of Complex Flows} (eds. R.~Friedrich and
  W.~Rodi),
\newblock Springer Netherlands, Dordrecht, 2002,
\newblock 239--254.

\bibitem{tseng2006modeling}
\newblock Y.-H. Tseng, C.~Meneveau and M.~B. Parlange,
\newblock Modeling flow around bluff bodies and predicting urban dispersion
  using large eddy simulation,
\newblock \emph{Environmental Science \& Technology}, \textbf{40} (2006),
  2653--2662,
\newblock \urlprefix\url{https://doi.org/10.1021/es051708m}.

\bibitem{wang2017physics}
\newblock J.-X. Wang, J.-L. Wu and H.~Xiao,
\newblock Physics-informed machine learning approach for reconstructing
  {Reynolds} stress modeling discrepancies based on {DNS} data,
\newblock \emph{Phys. Rev. Fluids}, \textbf{2} (2017), 034603,
\newblock
  \urlprefix\url{https://link.aps.org/doi/10.1103/PhysRevFluids.2.034603}.

\bibitem{wang2018investigations}
\newblock Z.~Wang, K.~Luo, D.~Li, J.~Tan and J.~Fan,
\newblock Investigations of data-driven closure for subgrid-scale stress in
  large-eddy simulation,
\newblock \emph{Physics of Fluids}, \textbf{30} (2018), 125101,
\newblock \urlprefix\url{https://doi.org/10.1063/1.5054835}.

\bibitem{werhahn2019multi}
\newblock M.~Werhahn, Y.~Xie, M.~Chu and N.~Thuerey,
\newblock A multi-pass {GAN} for fluid flow super-resolution,
\newblock \emph{arXiv preprint arXiv:1906.01689}.

\bibitem{wiewel2019latent}
\newblock S.~Wiewel, M.~Becher and N.~Thuerey,
\newblock Latent space physics: Towards learning the temporal evolution of
  fluid flow,
\newblock \emph{Computer Graphics Forum}, \textbf{38} (2019), 71--82,
\newblock
  \urlprefix\url{https://onlinelibrary.wiley.com/doi/abs/10.1111/cgf.13620}.

\bibitem{wang2019reynolds}
\newblock J.~Wu, H.~Xiao, R.~Sun and Q.~Wang,
\newblock {Reynolds-averaged Navier-Stokes} equations with explicit data-driven
  {Reynolds} stress closure can be ill-conditioned,
\newblock \emph{Journal of Fluid Mechanics}, \textbf{869} (2019), 553–586.

\bibitem{xiao2016quantifying}
\newblock H.~Xiao, J.-L. Wu, J.-X. Wang, R.~Sun and C.~Roy,
\newblock {Quantifying and reducing model-form uncertainties in
  {Reynolds-averaged} {Navier}--{Stokes} simulations: A data-driven,
  physics-informed {Bayesian} approach},
\newblock \emph{Journal of Computational Physics}, \textbf{324} (2016),
  115--136,
\newblock
  \urlprefix\url{http://www.sciencedirect.com/science/article/pii/S0021999116303394}.

\bibitem{xiong2018learning}
\newblock W.~Xiong, W.~Luo, L.~Ma, W.~Liu and J.~Luo,
\newblock Learning to generate time-lapse videos using multi-stage dynamic
  generative adversarial networks,
\newblock in \emph{Proceedings of the IEEE Conference on Computer Vision and
  Pattern Recognition}, 2018,
\newblock 2364--2373.

\bibitem{yang2019adversarial}
\newblock Y.~Yang and P.~Perdikaris,
\newblock Adversarial uncertainty quantification in physics-informed neural
  networks,
\newblock \emph{Journal of Computational Physics}, \textbf{394} (2019), 136 --
  152,
\newblock
  \urlprefix\url{http://www.sciencedirect.com/science/article/pii/S0021999119303584}.

\bibitem{zhao2018learning}
\newblock L.~Zhao, X.~Peng, Y.~Tian, M.~Kapadia and D.~Metaxas,
\newblock Learning to forecast and refine residual motion for image-to-video
  generation,
\newblock in \emph{Proceedings of the European Conference on Computer Vision
  (ECCV)}, 2018,
\newblock 387--403.

\bibitem{zhu2018bayesian}
\newblock Y.~Zhu and N.~Zabaras,
\newblock Bayesian deep convolutional encoder–decoder networks for surrogate
  modeling and uncertainty quantification,
\newblock \emph{Journal of Computational Physics}, \textbf{366} (2018), 415 --
  447,
\newblock
  \urlprefix\url{http://www.sciencedirect.com/science/article/pii/S0021999118302341}.

\bibitem{zhu2019physics}
\newblock Y.~Zhu, N.~Zabaras, P.-S. Koutsourelakis and P.~Perdikaris,
\newblock Physics-constrained deep learning for high-dimensional surrogate
  modeling and uncertainty quantification without labeled data,
\newblock \emph{Journal of Computational Physics}, \textbf{394} (2019), 56 --
  81,
\newblock
  \urlprefix\url{http://www.sciencedirect.com/science/article/pii/S0021999119303559}.

\end{thebibliography}

\end{document}